\documentclass{article}

\usepackage{bbold}
\usepackage{amssymb}
\usepackage{amsmath}
\usepackage{mathtools}
\usepackage{braket}
\usepackage{graphicx}
\usepackage{authblk}

\usepackage{pgffor}

\usepackage{etoolbox}

\usepackage{tikz}
\usetikzlibrary{decorations.markings, calc, positioning, decorations.pathmorphing, fixedpointarithmetic,shapes,patterns,shapes.geometric} 
\AtBeginEnvironment{tikzpicture}{\catcode`$=3 } 
\tikzset{alignmid/.style={baseline={([yshift=-.5ex]current bounding box.center)}}} 
\tikzset{every picture/.append style=alignmid}

\usepackage{ifthen}
\usepackage{xparse,pgffor}
\ExplSyntaxOn

\NewDocumentCommand{\Foreach}{ m O{,} m}
 {
  \__foreach_main:nn { #1 } { #2 } { #3 }
 }

\seq_new:N \l__foreach_arg_seq
\clist_new:N \l__foreach_braced_clist

\cs_new_protected:Npn \__foreach_main:nn #1 #2 #3
 {
  \seq_set_split:Nnn \l__foreach_arg_seq { #2 } { #3 }
  \clist_clear:N \l__foreach_braced_clist
  \seq_map_inline:Nn \l__foreach_arg_seq
   {
    \clist_put_right:Nn \l__foreach_braced_clist { { ##1 } }
   }
  \foreach #1 in~\l__foreach_braced_clist
 }
 \ExplSyntaxOff

\newcommand\coords[1]{\Foreach{\x/\y}[|]{#1}{\coordinate (\x) at (\y);}}

\tikzset{enddots/.style= {postaction=decorate,decoration={name=markings,mark=at position 1 with{
    \ifdimless{\pgflinewidth}{1pt}{\newcommand\dotlw{1pt}}{\newcommand\dotlw{\pgflinewidth}}
    \ifdimless{\pgflinewidth}{1.5pt}{\newcommand\dotdist{1.5pt}}{\newcommand\dotdist{\pgflinewidth}}
    \fill (1*\dotdist,0) circle (0.5*\dotlw);
    \fill (2.5*\dotdist,0) circle (0.5*\dotlw);
    \fill (4*\dotdist,0) circle (0.5*\dotlw);
}}}}
\tikzset{startdots/.style= {postaction=decorate,decoration={name=markings,mark=at position 0 with{
   \ifdimless{\pgflinewidth}{1pt}{\newcommand\dotlw{1pt}}{\newcommand\dotlw{\pgflinewidth}}
   \ifdimless{\pgflinewidth}{1.5pt}{\newcommand\dotdist{1.5pt}}{\newcommand\dotdist{\pgflinewidth}}
   \fill (-1*\dotdist,0) circle (0.5*\dotlw);
   \fill (-2.5*\dotdist,0) circle (0.5*\dotlw);
   \fill (-4*\dotdist,0) circle (0.5*\dotlw);
}}}}

\tikzset{
    -|/.style={to path={-| (\tikztotarget)}},
    |-/.style={to path={|- (\tikztotarget)}},
}

\tikzset{midlabeldistance/.store in=\midlabeldistance,midlabeldistance=0.2}
\tikzset{midlabelpos/.store in=\midlabelpos,midlabelpos=0.5}
\tikzset{midlabel/.style={decoration={markings, mark=at position \midlabelpos with {\node at (0,\midlabeldistance) {#1};}}, postaction={decorate}}}
\tikzset{midlabelr/.style={decoration={markings, mark=at position \midlabelpos with {\node at (0,-\midlabeldistance) {#1};}}, postaction={decorate}}}

\usepackage{hyperref}
\definecolor{darkblue}{RGB}{46,48,146}
\hypersetup{colorlinks=true,linkcolor=darkblue,pdfborderstyle={0 0 0}}

\usepackage{amsthm}
\theoremstyle{definition}
\newtheorem{myobs}{Observation}
\newtheorem{myrem}{Remark}
\newtheorem{mycom}{Comment}
\newtheorem{mydef}{Definition}

\newtheorem{myprop}{Proposition}

\newcommand{\tdef}[2]{\hypertarget{#2}{{\bf #1}}} 

\tikzset{enddist/.store in=\enddist,enddist=0.2}
\tikzset{startdist/.store in=\startdist,startdist=0.2}
\tikzset{endlabel/.style={decoration={markings, mark=at position \pgfdecoratedpathlength-0.005cm with {\node at (\enddist,0) {#1};}}, postaction={decorate}}}
\tikzset{startlabel/.style={decoration={markings, mark=at position 0.005cm with {\node at (-\startdist,0) {#1};}}, postaction={decorate}}}

\tikzset{cdirpos/.store in=\cdirpos,cdirpos=0.5}
\tikzset{cdir/.style={postaction=decorate,decoration={name=markings,mark=at position \cdirpos with{
    \draw[solid,line width=0.7] (#1 0.04,0)--(-#1 0.04,0.08) (#1 0.04,0)--(-#1 0.04,-0.08);
  }}}}

\tikzset{ndirpos/.store in=\ndirpos,ndirpos=0.5}
\tikzset{ndir/.style={postaction=decorate,decoration={name=markings,mark=at position \ndirpos with{
    \draw[solid,semithick] (#1 -0.04,-0.08)arc(90-#1 180:90:0.08);
  }}}}

\tikzset{flags/.style={postaction=decorate,decoration={name=markings,mark=at position 0.06cm with{
    \draw[fill,solid,semithick,sharp corners] (-0.06,0)--(0,#1 0.09)--(0.06,0)--cycle;
  }}}}
\tikzset{flage/.style={postaction=decorate,decoration={name=markings,mark=at position \pgfdecoratedpathlength -0.06cm with{
    \draw[fill,solid,semithick,sharp corners] (-0.06,0)--(0,#1 0.09)--(0.06,0)--cycle;
  }}}}

\tikzset{arrs/.style={postaction=decorate,decoration={name=markings,mark=at position 0.04 cm with{
    \fill[sharp corners] (-#1 0.03,0.07)--(#1 0.05,0)--(-#1 0.03,-0.07);
  }}}}
\tikzset{arre/.style={postaction=decorate,decoration={name=markings,mark=at position \pgfdecoratedpathlength -0.04 cm with{
    \fill[sharp corners] (-#1 0.03,0.07)--(#1 0.05,0)--(-#1 0.03,-0.07);
  }}}}

\tikzset{shapels/.style={line width=1.1}}

\newcommand{\piclab}[6][]{
\pic[#1] (#2) at #4 {#3};
\node at ($#4+#5$){#6};
}

\newcommand\pics[3][]{\Foreach{\x/\y}[|]{#3}{\pic[#1](\x) at (\y){#2};}}

\tikzset{crossboxside/.pic={ 
\node[transform shape, minimum width=0.3cm,minimum height=0.3cm,rectangle,draw,shapels] (-c) at (0,0){};
\draw (-c.north west)--(-c.south east) (-c.north east)--(-c.south west);
\fill (-c.north east)--(0,0)--(-c.south east)--cycle;
\coords{-t/0,0.15|-b/0,-0.15|-l/-0.15,0|-r/0.15,0};
}}

\tikzset{bwbox/.pic={ 
\node[transform shape, minimum width=0.3cm,minimum height=0.3cm,rectangle,draw,shapels] (-c) at (0,0){};
\fill (-c.north east)--++(-0.15,0)|-(-c.south east)--cycle;
\coords{-t/0,0.15|-b/0,-0.15|-l/-0.15,0|-r/0.15,0};
}}

\tikzset{bwdiamond/.pic={ 
\draw[shapels] (0,-0.2)--(0.2,0)--(0,0.2)--(-0.2,0)--cycle;
\fill (0,-0.2)--(0.2,0)--(0,0.2)--cycle;
\coords{-t/0,0.2|-b/0,-0.2|-l/-0.2,0|-r/0.2,0};
}}

\tikzset{circ/.pic={ 
\node[transform shape, minimum width=0.3cm,minimum height=0.3cm,circle,draw,shapels,inner sep=0] (-c) at (0,0){};
}}

\tikzset{smallcirc/.pic={ 
\node[transform shape, minimum width=0.18cm,minimum height=0.18cm,circle,draw,shapels,inner sep=0] (-c) at (0,0){};
}}

\tikzset{smallfcirc/.pic={ 
\node[transform shape, minimum width=0.15cm,minimum height=0.15cm,circle,draw,fill,shapels] (-c) at (0,0){};
}}

\tikzset{fcirc/.pic={ 
\node[transform shape, minimum width=0.3cm,minimum height=0.3cm,circle,draw,fill,shapels] (-c) at (0,0){};
}}

\tikzset{chiralcirc/.pic={ 
\node[transform shape, minimum width=0.3cm,minimum height=0.3cm,circle,draw,shapels] (-c) at (0,0){};
\fill (0,-0.15)--(0,0.05)--(0.1,-0.05)--cycle;
}}

\tikzset{chiralfcirc/.pic={ 
\node[transform shape, minimum width=0.3cm,minimum height=0.3cm,circle,draw,fill,shapels] (-c) at (0,0){};
\fill[white] (0,-0.15)--(0,0.05)--(0.1,-0.05)--cycle;
}}

\tikzset{chiraltriangle/.pic={
\coords{-c1/-30:0.2|-c2/90:0.2|-c3/-150:0.2};
\draw[shapels] (-30:0.2)--(90:0.2)--(-150:0.2)--cycle;
\fill (-c1)--(0,-0.5*0.2)--(0,0)--cycle;
}}

\tikzset{wid/.style={minimum width=#1}}
\tikzset{hei/.style={minimum height=#1}}

\tikzset{box/.style={draw,rectangle}}
\tikzset{fbox/.style={draw,rectangle, line width=1.1}}
\tikzset{roundbox/.style={draw,rectangle,rounded corners}}
\tikzset{froundbox/.style={draw,rectangle, rounded corners, line width=1.1}}

\tikzset{normal/.style={}}
\tikzset{star/.style={}}
\tikzset{realifsoff/.store in=\realifsoff,realifsoff=0.1}
\tikzset{realifeoff/.store in=\realifeoff,realifeoff=0.1}
\tikzset{realifs/.style={postaction=decorate,decoration={name=markings,mark=at position \realifsoff cm with{
    \draw[solid,semithick,sharp corners] (-#1 0.04,0.07)--(#1 0.04,0)--(-#1 0.04,-0.07);
  }}}}
\tikzset{realife/.style={postaction=decorate,decoration={name=markings,mark=at position \pgfdecoratedpathlength -\realifeoff cm with{
    \draw[solid,semithick,sharp corners] (-#1 0.04,0.07)--(#1 0.04,0)--(-#1 0.04,-0.07);
  }}}}

\tikzset{complex/.style={}}
\tikzset{classical/.style={}}
\tikzset{quantum/.style={line width=1.4}}

\begin{document}
\title{Quantum mechanics is *-algebras and tensor networks}
\author{A.~Bauer}
\affil{Dahlem Center for Complex Quantum Systems, Freie Universit{\"a}t Berlin,
Arnimallee 14, 14195 Berlin}

\maketitle
\begin{abstract}
\noindent
We provide a systematic approach to quantum mechanics from an information-theoretic perspective using the language of tensor networks. Our formulation needs only a single kind of object, so-called positive *-tensors. Physical models translate experimental setups into networks of these *-tensors, and the evaluation of the resulting networks yields the probability distributions describing measurement outcomes. The idea behind our approach is similar to categorical formulations of quantum mechanics. However, our formulation is mathematically simpler and less abstract. Our presentation of the core formalism is completely self-contained and relies on minimal mathematical prerequesites. Therefore, we hope it is in principle also understandable to people without an extensive mathematical background. Additionally, we show how various types of models, like real-time evolutions or thermal systems can be translated into *-tensor networks.
\end{abstract}

\section{Introduction}
Quantum mechanics can be considered the most important development in physics in the 20th century. The first formulations of quantum mechanics are nearly one hundred years old. The way quantum mechanics is taught to students has not changed much in these hundred years. The traditional way, focusing on pure states, Hamiltonians, the Schr\"odinger equation, the harmonic oscillator or the hydrogen atom, surely has its value. However, we find that it largely obscures a deep and fundamental understanding of the very core of quantum mechanics. This has several reasons:
\begin{itemize}
\item Physics can be separated into two types of questions: The first one is about the bare framework which allows us to express different models and compute their predictions. The second one is about finding models within that framework which yield reasonable descriptions of nature. By the ``core of quantum mechanics'' we mean only the framework itself. Concerning the second question, there are many heuristics for building quantum models inspired from classical intuition, some of which come under the name of ``quantization''. We think that this strict separation yields a lot of clarity, though it is not usually done in traditional introductions.
\item The Schr\"odinger equation is usually taken to be the central equation governing quantum mechanics. However, quantum mechanics can be formulated without a continuous time. The latter comes with technicalities like exponential functions that make it harder to see the essential underlying mathematical structure. Enforcing a contiuous time also means treating time and space on an unequal footing, as the latter is often discrete, e.g. in condensed matter models.
\item Many standard textbook examples of quantum systems deal with infinite (continuous) Hilbert spaces, such as the harmonic oscillator. Those infinities come with technicalities that distract from the core of quantum mechanics. In this text we focus on finite Hilbert spaces.
\item Introductions usually focus on pure states and unitary quantum mechanics. We argue that the mixed state level is conceptually much simpler as it allows to unify classical and quantum components of a model, such that states, time evolution, measurements or their outcome probabilities are all represented by the same kind of object.
\item Causality is a very fundamental property of physics. In our language it corresponds to the presence of a global flow of time and a normalization condition of the tensors with respect to this time direction. However, this normalization is not the key feature of quantum mechanics which makes it different from classical mechanics. The central distinguishing property is a different positivity condition. Our approach separates the positivity and normalization condition, which makes things conceptually easier and allows us to also describe non-causal (such as thermal or ground state) quantum physics.
\item Many properties that are attributed to the ``quantum world'' have direct analogues in classical statistical physics: The principle of superposition exists in every theory with a statistical interpretation, including classical statistics. Classical statistical data can have correlations, which formally aren't as different from quantum entanglement as often stated. The ``no-cloning'' theorem also holds for probability distributions in classical statistics. A qualitative difference between the quantum and classical world for all these concepts only appears when we restrict to extremal points in the convex set of states (``pure states''): This is a discrete set for classical statistics but a continuum in the quantum world. If we work on the level of mixed states instead, classical statistics and quantum mechanics become very similar, the only difference being the positivity condition.
\item The use of complex numbers is traditionally considered one of the most fundamental features of quantum mechanics. However, real quantum mechanics has the same expressive power as the complex version, while being simpler for the obvious reason. Real tensors can emulate complex tensors (in which quantum mechanics is traditionally formulated) via ``realification'', as we'll describe in Sec.~(\ref{sec:complex_tensor}). In our viewpoint, complex quantum mechanics is merely a more compact way to write down certain specific models.
\item Dirac notation is the established notation for dealing with quantum systems. We find that it has many shortcomings: First of all it doesn't display any locality structure, e.g. it is inefficient for representing products of operators acting on different parts of a multi-partite Hilbert space. Also, the only purpose of the distinction between bra and ket is a bookkeeping of complex conjugations with can be done more efficiently by using sub- and superscripts, or arrows in a graphical notation. Furthermore, the notation is often repetitive, as for example when writing $\sum_{ij} H_{ij}|i\rangle \langle j|$ where we have to repeat the letters $i$ and $j$ three times. Also we sometimes have to apply computational tricks like the ``insertion of unity'' which are trivial in other notations.
\end{itemize}

This paper is not meant to replace an introductory course on quantum mechanics. The intention is rather to isolate the very core of quantum mechanics from all unnecessary complications in order to uncover the essential mathematical structure that makes it different from classical physics. We ask the question ``What do we want from a physical model?'', and then see that quantum mechanics fulfills all the requirements very naturally and generally. It will also become apparent that it is the most straight forward generalization of classical statistical physics that one could consider.

Our approach combines different conceptual perspectives of quantum mechanics, which go beyond its original formulation:
\begin{itemize}
\item \emph{Information-theoretical perspective}: We don't need to restrict ourselves to describe some form of ``matter'', like fermions, light or atoms. Instead quantum mechanics is seen as a framework describing abstract degrees of freedom, which might be implemented in the real world in different ways (or not at all).
\item \emph{Circuit formulation}: We do not consider a continuum time evolution but cut time into finite intervals and consider the unitary operator describing the time evolution for one interval. We can directly use these unitaries as building blocks of our models instead of working with Hamiltonians. If there is a space with a non-trivial dimension, we can apply a Trotter decomposition to approximate the continuum time evolution by a quantum circuit.
\item \emph{Tensor-networks}: Instead of having a formalism that differentiates between states, circuits, measurement, and so on, we realize that they are all the same kind of mathematical objects: Tensors, that is arrays of complex (or real) numbers. Note, however, that conceptually we're not talking about ``tensor-network ansatzes for ground states'' (known as MPS or PEPS), but rather aim to represent the whole physical situation (e.g., including time evolution and measurements) by a tensor network.
\item \emph{C*-algebra formulation}: C*-algebras were used in early on in quantum mechanics, going back to the work of von Neumann. It was realized later on that classical statistical physics and quantum mechanics can be unified into one theory by using different C*-algebras.
\end{itemize}

While writing up this paper we found a great deal of similarity between our approach and that of \emph{categorical quantum mechanics} \cite{Abramsky2004}. Category theory can be seen as an abstract language for manipulating graph-like diagrams. Different categories are different data structures that fulfill a given graphical calculus. Using category theory, axiomatic structures can be defined in the graphical calculus for arbitrary such data structures. For example, symmetric monoidal categories allow a graphical calculus whose diagrams are networks of shapes (the morphisms of the category) connected by lines (the objects) together with a flow of time. The adjective ``symmetric'' means that we don't have to worry about how the graph is embedded into the paper plain. For example, in the symmetric monoidal category of sets, the items (morphisms) become multi-valued functions between the sets that are associated to connected lines (objects). Evaluating a diagram means composing the functions.

In any sort of physical framework with a notion of locality, models should consist of building blocks that can be combined arbitrarily, subject only to some local constraints. In a categorical setting these building blocks are given by morphisms and combining the morphisms means composing them into networks. E.g. classical deterministic physics corresponds to the (symmetric monoidal) category of sets, or quantum and classical statistical mechanics are both unified by using the category of CPTP maps \cite{Selinger2007,Coecke2013}.

In this paper we do not use symmetric monoidal categories, but an alternative type of structure that we call \emph{tensor types}. The main difference to the categorical language is that the graphical calculus of tensor types does not have any flow of time. This way we also cover non-causal settings such as ground state or thermal physics. In \cite{tensor_type} we give a detailed formal definition of tensor types, together with many examples that are important for physics applications. In this paper we will not talk much about different tensor types. We will instead focus on the one particular tensor type in which quantum mechanics and classical statistics both live, namely positive *-tensors based on real tensors. However, the graphical calculus we develop would also yield positive *-tensors based on other tensor types.

This paper is structured as follows: In Sec.~(\ref{sec:prerequisites}) we introduce the mathematical structures underlying our approach to quantum mechanics, namely tensors, *-algebras, and positive *-tensors. In Sec.~(\ref{sec:models}) we discuss what one would expect from theories like classical or quantum mechanics, and see how these expectations can be formalized using the language of tensor networks. This section is partly philosophical, and might be skipped by readers only interested in the simple mathematical formulation of quantum mechanics. Sec.~(\ref{sec:star_tensor_theory}) contains a definition of quantum mechanics for both thermal systems and systems with time evolution on less than a page. More precisely, we define *-tensor theory, which is a generalization of both classical statistics and quantum mechanics. In Sec.~(\ref{sec:classical}) and Sec.~(\ref{sec:quantum}) we will show how *-tensor models capture both classical and quantum mechanics. In Sec.~(\ref{sec:other_star_algebras}) we look into *-tensor models beyond classical statistics and quantum mechanics (but still based on real tensors) and see that quantum mechanics is enough to emulate them all.

\section{Mathematical prerequisites}
\label{sec:prerequisites}
In this section we will introduce all mathematical prerequisites in a self-contained manner. As we will see, only very elementary mathematics, such as addition and multiplication of real numbers, is required. A knowledge of linear algebra (i.e. vectors, matrices, higher-dimensional arrays, matrix multiplication, inner product etc.) might help though.

\subsection{Tensors and tensor networks}
The predictions of any statistical theory (such as quantum mechanics) are probability distributions, which are arrays of real numbers. As such, it is no surprise that the fundamental building blocks in our approach are also based on such arrays.

\begin{mydef}
A \tdef{(real) tensor}{real_tensor} is defined with respect to a set of \tdef{indices}{index} $I$, together with a collection of \tdef{bases}{basis} $(B_i)_{i\in I}$. Each basis $B_i$ is a finite set. A tensor $T$ associates to each configuration of bases a real number:
\begin{equation}
T: \quad \bigtimes_{i\in I} B_i \rightarrow \mathbb{R}
\end{equation}

The \tdef{tensor product}{tensor_product} of two tensors $T^{(1)}$ and $T^{(2)}$ is a tensor $T$ defined with respect to the disjoint union of the two index sets $I^{(1)}$ and $I^{(2)}$ and basis collections $B^{(1)}$ and $B^{(2)}$, namely the mutual product of all the real numbers:
\begin{equation}
\begin{gathered}
T\left((b^{(1)}_i)_{i\in I^{(1)}},(b^{(2)}_i)_{i\in I^{(2)}}\right)=\\
T^{(1)}\left((b^{(1)}_i)_{i\in I^{(1)}}\right) \cdot T^{(2)}\left((b^{(2)}_i)_{i\in I^{(2)}}\right)
\end{gathered}
\end{equation}
for all $b^{(1)}_i\in B^{(1)}_i$, $b^{(2)}_i\in B^{(2)}_i$.

If two indices $x,y\in I$ of a tensor $T$ have the same basis $B_x=B_y=B$, one can define the \tdef{contraction}{contraction} of those two indices, which is the following tensor $[T]_{xy}$ with the indices $x$ and $y$ missing:
\begin{equation}
[T]_{xy}((b_i)_{i\in I/\{x,y\}})=\sum_{c\in B} T((b_i)_{i\in I/\{x,y\}},b_x=c,b_y=c)
\end{equation}

A subset of indices $i,j\ldots \in I$ can be \tdef{blocked}{index_blocking} by interpreting it as a single index with basis $B_i\times B_j\times \ldots$.

For every basis $B$ there is a $2$-index \tdef{identity tensor}{itentity_tensor}, where $B$ is the basis of both indices:
\begin{equation}
\mathrm{Id}_B(b_0,b_1)=\delta_{b_0,b_1}=
\begin{cases}
1 &\text{if}\quad b_0=b_1\\
0 &\text{otherwise}
\end{cases}
\end{equation}
It has the property that for any tensor, taking the tensor product with the identity tensor and contracting one tensor index with an identity index yields the same tensor again.
\end{mydef}

\begin{mycom}
Note that the way we use the word ``tensor'' is different from the usage in differential geometry: While we are also talking about multi-dimensional arrays we do not need any metric, manifold, tangent space or transformation properties. Also we use the word ``basis'' in a slightly unconventional way: We mean just a (finite) set $B$ such that the vector space is given by $\mathbb{R}^B$ (a $|B|$-dimensional real vector space) rather than a subset of vectors spanning the vector space.

A tensor with one index whose basis is an $m$-element set is just a vector in an $m$-dimensional real vector space. Similarly, $0$-index tensors are just real numbers, and $2$-index tensors are matrices.
\end{mycom}

\begin{mydef}
When taking the tensor product of multiple tensors, it doesn't matter in which order we take it and how we set brackets. Also, contractions of different index pairs commute with each other and with the tensor product. So any sequence of tensor product and contraction operations is already determined by giving the set of involved tensors and the set of index pairs to contract. Such a network of tensors with contractions between their indices will be called a \tdef{tensor network}{tensor_network}. Indices that are not contracted are referred to as \tdef{open indices}{open_index}. The \tdef{evaluation}{evaluation} of a tensor network is the tensor resulting from taking the tensor product of all involved tensors and then performing all contractions. The indices of the evaluation are the open indices of the tensor network.
\end{mydef}

\begin{mydef}
\label{rem:tensor_network_notation}
As indicated by the name, tensor networks come with a very natural notation, which we will refer to as \tdef{tensor-network notation}{tensor_network_notation}: A tensor is denoted by some shape (like a circle, rectangle, or triangle). When there are different tensors of the same shape we want to distinguish, or if we want to refer to a tensor in the text, we might also put label inside or next to the shape. Each index of the tensor corresponds to a position on the boundary of the shape, where a line is sticking out. The endpoints of these lines carry labels. E.g., a $3$-index tensor $A$ could be denoted by:
\begin{equation}
\begin{gathered}
a)\quad
\begin{tikzpicture}
\node[draw,rectangle,minimum width=1cm] (t) {$A$};
\draw ([xshift=-0.3cm]t.north)edge[endlabel=$a$]++(0,0.3);
\draw ([xshift=0.3cm]t.north)edge[endlabel=$b$]++(0,0.3);
\draw (t.east)edge[endlabel=$c$]++(0.5,0);
\end{tikzpicture}\\
b)\quad
\begin{tikzpicture}
\node[draw,circle,minimum width=0.3cm] (t){};
\node[below =0cm of t]{$A$};
\draw (t.west)edge[endlabel=$a$]++(-0.5,0);
\draw (t.north)edge[endlabel=$b$]++(0,0.3);
\draw (t.east)edge[endlabel=$c$]++(0.5,0);
\end{tikzpicture}\\
c)\quad
\begin{tikzpicture}
\coords{0/0,0};
\draw (0)edge[endlabel=$a$]++(60:0.5);
\draw (0)edge[endlabel=$b$]++(-60:0.5);
\draw (0)edge[endlabel=$c$]++(180:0.5);
\begin{scope}
\draw[thick,fill=white] (0)circle(0.2);
\clip (0)circle(0.2);
\draw[thick] (-0.2,0.2)--(0.2,-0.2) (-0.2,-0.2)--(0.2,0.2);
\end{scope}
\end{tikzpicture}
\end{gathered}
\end{equation}
The labels are important when we equate two tensors, as they tell us how to match up the index sets. Sometimes we might omit them and instead use the position of the endpoints and/or the directions they are pointing at to indicate the latter.

The shape of a tensor might be drawn at arbitrary positions, rotated by arbitrary angles, and even mirrored. There are two types of shapes used for tensors: 1) Generic tensors that do not have specific algebraic properties are represented by a (usually rectangular) box with a label inside. 2) Tensors that appear repeatedly throughout the paper and have specific algebraic properties are represented by a fixed shape reflecting these algebraic properties. Most notably, if the shape has some rotation/reflection symmetries, the tensor is supposed to be symmetric under the corresponding index permutations. E.g., if the shape is round, this indicates that the tensor is invariant under cyclic permutations of its indices (as well as reversing the order of indices). If we want to refer to a tensor in the text, or if we have multiple different tensors with the same algebraic properties, we might put a label next to the shape.

The basis of an index is often indicated by the choice of line style of the corresponding line.

The tensor product of two tensors is denoted by placing two shapes next to each other somewhere in the same picture. E.g., the tensor product of a $3$-index tensor $B$ and a $2$-index tensor $C$ could be depicted by:
\begin{equation}
\begin{gathered}
a)\quad
\begin{tikzpicture}
\node[box] (a) at (0,0) {$B$};
\node[box] (b) at (3,0) {$C$};
\draw (a.west)edge[endlabel=$a$]++(-0.5,0);
\draw (a.north)edge[endlabel=$b$]++(0,0.3);
\draw (a.east)edge[endlabel=$c$]++(0.5,0);
\draw (b.west)edge[endlabel=$d$]++(-0.5,0);
\draw (b.east)edge[endlabel=$e$]++(0.5,0);
\end{tikzpicture}\\
b)\quad
\begin{tikzpicture}
\node[draw,regular polygon, regular polygon sides=3,inner sep=0.5pt] (a) at (0,0) {$B$};
\node[fill,circle,minimum width=0.4cm] (b) at (0,0.8){};
\node[above =0cm of b] {$C$};
\draw (a)edge[endlabel=$a$]++(30:1) (a)edge[endlabel=$b$]++(150:1) (a)edge[endlabel=$c$]++(-90:0.6);
\draw[endlabel=$e$,rounded corners] (b)--++(0.4,0)--++(-60:1);
\draw[endlabel=$f$,rounded corners] (b)--++(-0.4,0)--++(-120:1);
\end{tikzpicture}
\end{gathered}
\end{equation}

Contraction of two indices is denoted by connecting the corresponding lines. E.g., contraction of two indices of a $4$-index tensor $D$ could be depicted by:
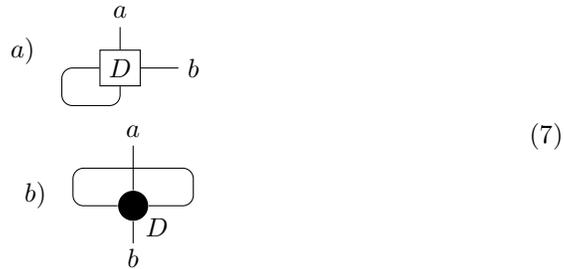
\begin{equation}
\begin{gathered}
a)\quad
\begin{tikzpicture}
\node[draw,rectangle] (t) {$D$};
\draw[rounded corners] (t.west)--++(-0.5,0)--++(0,-0.5)-|(t.south);
\draw (t.north)--++(0,0.3) node[above]{$a$};
\draw (t.east)--++(0.5,0) node[right]{$b$};
\end{tikzpicture}\\
b)\quad
\begin{tikzpicture}
\node[fill,circle,minimum width=0.4cm] (d) at (0,0){};
\node[below right=-0.15cm of d]{$D$};
\draw[rounded corners] (d)--++(0.8,0)--++(0,0.5)--++(-1.6,0)--++(0,-0.5)--(d);
\draw (d)edge[endlabel=$a$]++(0,0.8) (d)edge[endlabel=$b$]++(0,-0.5);
\end{tikzpicture}
\end{gathered}
\end{equation}

If we represent a tensor network with this notation, we obtain graph-like pictures. E.g.
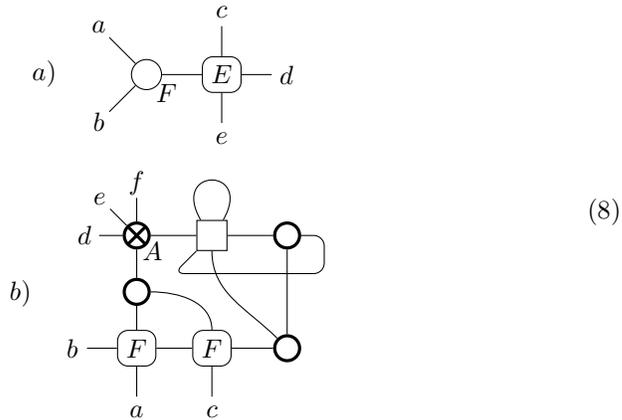
\begin{equation}
\begin{gathered}
a)\quad
\begin{tikzpicture}
\node[draw,rectangle,rounded corners] (e) at (0,0) {$E$};
\node[draw,circle,minimum width=0.4cm] (f) at (-1,0){};
\node[below right=-0.2cm of f]{$F$};
\draw (e)--(f) (f)edge[endlabel=$a$]++(135:0.7) (f)edge[endlabel=$b$]++(-135:0.7) (e.north)edge[endlabel=$c$]++(0,0.4) (e.east)edge[endlabel=$d$]++(0.4,0) (e.south)edge[endlabel=$e$]++(0,-0.4);
\end{tikzpicture}\\
b)\quad
\begin{tikzpicture}
\node[draw,rectangle,rounded corners] (f1) at (0,0) {$F$};
\node[draw,rectangle,rounded corners] (f2) at (1,0) {$F$};
\node[draw,rectangle,minimum width=0.4cm,minimum height=0.4cm] (b) at (1,1.5) {};
\node[draw,circle,minimum width=0.3cm,line width=1.2] (c1) at (0,0.75) {};
\node[draw,circle,minimum width=0.3cm,line width=1.2] (c2) at (2,0) {};
\node[draw,circle,minimum width=0.3cm,line width=1.2] (c3) at (2,1.5) {};
\node[draw,circle,minimum width=0.3cm,line width=1.2] (cr) at (0,1.5) {};
\begin{scope}
\clip (cr)circle(0.15);
\draw[line width=1.2] ($(cr)+(-0.15,0.15)$)--++(0.3,-0.3) ($(cr)+(0.15,0.15)$)--++(-0.3,-0.3);
\end{scope}
\node at ($(cr)+(-45:0.3)$){$A$};
\draw (f1)--(c1) (c1)--(cr) (cr)--(b) (b)--(c3) (c3)--(c2) (c2)--(f2) (f2)--(f1) (f1.south)edge[endlabel=$a$]++(0,-0.4) (f1.west)edge[endlabel=$b$]++(-0.4,0) (f2.south)edge[endlabel=$c$]++(0,-0.4) (cr)edge[endlabel=$d$]++(180:0.5) (cr)edge[endlabel=$e$]++(135:0.5) (cr)edge[endlabel=$f$]++(90:0.5) (b.south)to[out=-90,in=135] (c2) (c1)to[out=0,in=90](f2);
\draw[rounded corners] (b.south west)--++(-0.3,-0.3)--++(2,0)|-(c3);
\draw[rounded corners] ([xshift=-0.15cm]b.north)to[out=120,in=60,looseness=7]([xshift=0.15cm]b.north);
\end{tikzpicture}
\end{gathered}
\end{equation}
In the above, a) shows a shows a very simple tensor network, whose evaluation involves taking the tensor product of two tensors $E$ and $F$ and then contracting one index coming from $E$ with one index coming from $F$. Correspondingly, the representing diagram is obtained by placing $E$ and $F$ next to each other and then connecting two of their index lines. b) shows a more complicated tensor network which involves multiple copies of some tensors.

Note that the tensor network is fully specified by the combinatorics of the underlying graph, i.e., which points on the boundary of the shapes are connected to which others. How we embed those graphs in the paper plain doesn't matter, such that, e.g., crossings of lines do not have any special meaning. E.g., the following two drawings a) and b) represent the same tensor network:
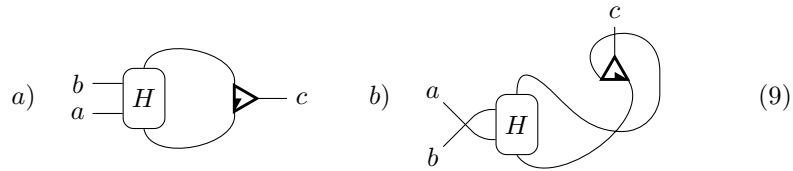
\begin{equation}
a)\quad
\begin{tikzpicture}
\node[roundbox,hei=0.8cm] (h) at (0,0){$H$};
\pic[rotate=-90] (i) at (1.3,0){chiraltriangle};
\draw (h.north)to[bend left=90](i-c3) (h.south)to[bend left=-90](i-c1) ([yshift=-0.2cm]h.west)edge[endlabel=$a$]++(-0.4,0) ([yshift=0.2cm]h.west)edge[endlabel=$b$]++(-0.4,0) (i-c2)edge[endlabel=$c$]++(0.4,0);
\end{tikzpicture}
\qquad
b)\quad
\begin{tikzpicture}
\node[roundbox,hei=0.8cm] (h) at (0,0){$H$};
\pic(i) at (1.3,0.7){chiraltriangle};
\draw (h.north)to[out=90,in=135,looseness=2]++(0.5,0)to[out=-45,in=180]++(0.9,-0.5)to[out=0,in=-90]++(0.5,0.5)--++(0,0.3)to[out=90,in=135,looseness=2.8](i-c3) (h.south)to[bend left=-90](i-c1) (i-c2)edge[endlabel=$c$]++(0,0.4);
\draw[rounded corners,endlabel=$a$] ([yshift=-0.2cm]h.west)--++(-0.2,0)--++(135:0.7);
\draw[rounded corners,endlabel=$b$] ([yshift=0.2cm]h.west)--++(-0.2,0)--++(-135:0.7);
\end{tikzpicture}
\end{equation}

If a set of indices is blocked, this will be denoted by using the concatenation of the labels for the individual indices as the label for the blocked index. E.g., the following shows a $3$-index tensor $J$ which is a blocked version of a $6$-index tensor $K$:
\begin{equation}
\begin{tikzpicture}
\node[roundbox,hei=1cm] (k)at (0,0){$J$};
\draw (k.west)--++(-0.4,0)node[left]{$abc$} ([yshift=-0.3cm]k.east)--++(0.4,0)node[right]{$de$} ([yshift=0.3cm]k.east)edge[endlabel=$f$]++(0.4,0);
\end{tikzpicture}=
\begin{tikzpicture}
\node[roundbox,hei=1cm] (k)at (0,0){$K$};
\draw ([yshift=-0.3cm]k.west)edge[endlabel=$a$]++(-0.4,0) (k.west)edge[endlabel=$b$]++(-0.4,0) ([yshift=0.3cm]k.west)edge[endlabel=$c$]++(-0.4,0) ([yshift=-0.3cm]k.east)edge[endlabel=$d$]++(0.4,0) (k.east)edge[endlabel=$e$]++(0.4,0) ([yshift=0.3cm]k.east)edge[endlabel=$f$]++(0.4,0);
\end{tikzpicture}
\end{equation}

The empty tensor network evaluates to the number $1$, so the latter is consistently represented by some blank space.

The identity tensor is consistently denoted by a free line (one might think of the middle part of the line being the shape and the outer parts being the indices sticking out):
\begin{equation}
\begin{tikzpicture}
\draw (0,0)edge[startlabel=$a$,endlabel=$b$](1,0);
\end{tikzpicture}
\end{equation}
\end{mydef}

\begin{mydef}
The \tdef{direct sum}{direct_sum} of two real tensors $T^{(1)}$ and $T^{(2)}$ with the same index set $I$ but different basis collections $B^{(1)}$ and $B^{(2)}$ is the following real tensor $T$ with basis collection $B_i=B^{(1)}_i\cup B^{(2)}_i$, where the $\cup$ denotes disjoint union:
\begin{equation}
T((b_i))=
\begin{cases}
T^{(1)}((b_i)) & \text{if} \quad b_i\in B^{(1)}_i \quad \forall i\\
T^{(2)}((b_i)) & \text{if} \quad b_i\in B^{(2)}_i \quad \forall i\\
0 & \text{otherwise}
\end{cases}
\end{equation}
Taking the direct sum of tensors commutes with evaluating tensor networks, as long as the latter are connected.
\end{mydef}

\begin{mydef}
A \tdef{gauge transformation}{gauge_transformation} is defined with respect to a set of bases. To each of these bases $B$, it associates another basis $B'$ and a tensor $G_B$ with one $B$ index and one $B'$ index obeying:
\begin{equation}
\label{eq:gauge_isometry}
\begin{gathered}
\begin{tikzpicture}
\piclab{r0}{bwbox}{(0,0.4)}{(90:0.4)}{$G_B$};
\piclab{r1}{bwbox}{(0,-0.4)}{(90:0.4)}{$G_B$};
\draw[rounded corners] (r0-r)--++(0.4,0)|-(r1-r);
\draw (r0-l)edge[endlabel=$a$]++(-0.3,0) (r1-l)edge[endlabel=$b$]++(-0.3,0);
\end{tikzpicture}=
\begin{tikzpicture}
\draw[rounded corners,startlabel=$a$,endlabel=$b$] (0,0.4)--++(0.4,0)|-(0,-0.4);
\end{tikzpicture}\\
\begin{tikzpicture}
\piclab[rotate=180]{r0}{bwbox}{(0,0.4)}{(90:0.4)}{$G_B$};
\piclab[rotate=180]{r1}{bwbox}{(0,-0.4)}{(90:0.4)}{$G_B$};
\draw[rounded corners] (r0-l)--++(0.4,0)|-(r1-l);
\draw (r0-r)edge[endlabel=$a$]++(-0.3,0) (r1-r)edge[endlabel=$b$]++(-0.3,0);
\end{tikzpicture}=
\begin{tikzpicture}
\draw[rounded corners,startlabel=$a$,endlabel=$b$] (0,0.4)--++(0.4,0)|-(0,-0.4);
\end{tikzpicture}
\end{gathered}
\end{equation}
In other words, $G_B$ is an isometry between the vector spaces $\mathbb{R}^B$ and $\mathbb{R}^{B'}$.

A gauge transformation is applied by contracting each index of a tensor with the corresponding tensor $G_B$. E.g. for a tensor $T$ with indices $a,b,c$ having basis $B$, $c,d,e$ having basis $C$, and $f,g$ having basis $D$, we get:
\begin{equation}
\begin{tikzpicture}
\node[roundbox,wid=2.2cm,hei=1cm] (t) {$T$};
\draw ([xshift=-0.9cm]t.north)edge[endlabel=$a$]++(0,0.4);
\draw ([xshift=-0.3cm]t.north)edge[endlabel=$b$]++(0,0.4);
\draw ([xshift=0.3cm]t.north)edge[endlabel=$c$]++(0,0.4);
\draw ([xshift=0.9cm]t.north)edge[endlabel=$d$]++(0,0.4);
\draw ([xshift=-0.6cm]t.south)edge[endlabel=$e$]++(0,-0.4);
\draw ([xshift=0cm]t.south)edge[endlabel=$f$]++(0,-0.4);
\draw ([xshift=0.6cm]t.south)edge[endlabel=$g$]++(0,-0.4);
\end{tikzpicture}\quad \rightarrow\quad
\begin{tikzpicture}
\node[roundbox,wid=2.2cm,hei=1cm] (t) {$T'$};
\draw ([xshift=-0.9cm]t.north)edge[endlabel=$a$]++(0,0.4);
\draw ([xshift=-0.3cm]t.north)edge[endlabel=$b$]++(0,0.4);
\draw ([xshift=0.3cm]t.north)edge[endlabel=$c$]++(0,0.4);
\draw ([xshift=0.9cm]t.north)edge[endlabel=$d$]++(0,0.4);
\draw ([xshift=-0.6cm]t.south)edge[endlabel=$e$]++(0,-0.4);
\draw ([xshift=0cm]t.south)edge[endlabel=$f$]++(0,-0.4);
\draw ([xshift=0.6cm]t.south)edge[endlabel=$g$]++(0,-0.4);
\end{tikzpicture}=
\begin{tikzpicture}
\node[roundbox,wid=2.2cm,hei=1cm] (t) {$T$};
\piclab[rotate=90]{a}{bwbox}{([xshift=-0.9cm,yshift=0.5cm]t.north)}{(-0.3,0.4)}{$G_B$};
\draw ([xshift=-0.9cm]t.north)--(a-l) (a-r)edge[endlabel=$a$]++(0,0.5);
\piclab[rotate=90]{b}{bwbox}{([xshift=-0.3cm,yshift=0.5cm]t.north)}{(-0.3,0.4)}{$G_B$};
\draw ([xshift=-0.3cm]t.north)--(b-l) (b-r)edge[endlabel=$b$]++(0,0.5);
\piclab[rotate=90]{c}{bwbox}{([xshift=0.3cm,yshift=0.5cm]t.north)}{(-0.3,0.4)}{$G_B$};
\draw ([xshift=0.3cm]t.north)--(c-l) (c-r)edge[endlabel=$c$]++(0,0.5);
\piclab[rotate=90]{d}{bwbox}{([xshift=0.9cm,yshift=0.5cm]t.north)}{(-0.3,0.4)}{$G_C$};
\draw ([xshift=0.9cm]t.north)--(d-l) (d-r)edge[endlabel=$d$]++(0,0.5);
\piclab[rotate=-90]{e}{bwbox}{([xshift=-0.6cm,yshift=-0.5cm]t.south)}{(-0.3,-0.4)}{$G_C$};
\draw ([xshift=-0.6cm]t.south)--(e-l) (e-r)edge[endlabel=$e$]++(0,-0.5);
\piclab[rotate=-90]{f}{bwbox}{([xshift=0cm,yshift=-0.5cm]t.south)}{(-0.3,-0.4)}{$G_D$};
\draw ([xshift=0cm]t.south)--(f-l) (f-r)edge[endlabel=$f$]++(0,-0.5);
\piclab[rotate=-90]{g}{bwbox}{([xshift=0.6cm,yshift=-0.5cm]t.south)}{(-0.3,-0.4)}{$G_D$};
\draw ([xshift=0.6cm]t.south)--(g-l) (g-r)edge[endlabel=$g$]++(0,-0.5);
\end{tikzpicture}
\end{equation}

Due to Eq.~(\ref{eq:gauge_isometry}), applying a gauge transformation commutes with contraction of indices and leaves the identity invariant. It also commutes with the tensor product and the trivial tensor.
\end{mydef}

As announced in the introduction, the constructions in the following sections could in principle be formulated for types of tensors other than real tensors. In fact, we could take any sort of data structure defined with respect to a set of indices, with a notion of tensor product and contraction, fulfilling axioms such that the diagrammatic calculus makes sense \cite{tensor_type}. One might think that for doing quantum mechanics one should use complex tensors, which would be another tensor type. However, we will see that in our approach, real tensors are the more natural tensor type to formulate quantum mechanics. In Sec.~(\ref{sec:complex_tensor}) we will see how real tensors can emulate the complex tensors commonly used in quantum mechanics.

\subsection{*-algebras: Definition}
In this section we will provide an axiomatic definition of *-algebras. This definition makes sense for arbitrary tensor types, though readers not interested in this level of abstraction might imagine all tensors are ordinary real tensors as introduced in the section before.

\begin{mydef}
A \tdef{*-algebra}{star_algebra} for a basis $B$ consists of an infinite set of tensors for each number $n$ of indices and each choice of \tdef{*-orientations}{star_orientation} $\{\text{left},\text{right}\}^n$ of the indices. The basis of all indices of all tensors is $B$. In tensor-network notation, we will represent a *-algebra tensor by an empty circle with clockwise or counter-clockwise flags where the index lines meet the circle. The latter indicate whether the *-orientation for this index is left (counter-clockwise) or right (clockwise). This choice of shape implies symmetries of the tensors which are consistent with the permutation axiom and the orientation reversal axiom below. E.g. the *-algebra tensor with $5$ indices $a,b,c,d,e$ and *-orientations $(\text{right},\text{left},\text{right},\text{left},\text{left})$ would be denoted as:
\begin{equation}
\begin{tikzpicture}
\pics{circ}{c/0,0};
\draw[star] (c-c)edge[flags=-,endlabel=$a$]++(0:0.5) (c-c)edge[flags=+,endlabel=$b$]++(72:0.5) (c-c)edge[flags=-,endlabel=$c$]++(144:0.5) (c-c)edge[flags=+,endlabel=$d$]++(216:0.5) (c-c)edge[flags=+,endlabel=$e$]++(-72:0.5);
\end{tikzpicture}
\end{equation}

The set of *-algebra tensors has to obey the following axioms:
\begin{itemize}
\item The \tdef{permutation axiom}{permutation_axiom}: They are invariant under arbitrary cyclic permutations of the indices that leave the *-orientations invariant. E.g.
\begin{equation}
\begin{tikzpicture}
\pics{circ}{0/0,0};
\draw[star] (0-c)edge[flags=-,endlabel=$a$]++(0.5,0) (0-c)edge[flags=-,endlabel=$b$]++(0,0.5) (0-c)edge[flags=-,endlabel=$c$]++(-0.5,0) (0-c)edge[flags=-,endlabel=$d$]++(0,-0.5);
\end{tikzpicture}=
\begin{tikzpicture}
\pics{circ}{0/0,0};
\draw[star] (0-c)edge[flags=-,endlabel=$c$]++(0.5,0) (0-c)edge[flags=-,endlabel=$d$]++(0,0.5) (0-c)edge[flags=-,endlabel=$a$]++(-0.5,0) (0-c)edge[flags=-,endlabel=$b$]++(0,-0.5);
\end{tikzpicture}
\end{equation}
Note that this axiom is already implied by the notation through the choice of a rotation-invariant shape for the tensor.

\item The \tdef{orientation reversal axiom}{orientation_reversal_axiom}: Inverting all *-orientations is the same as inverting the order of indices. E.g., in the following the order is inverted from $a,b,c,d$ to $d,c,b,a$:
\begin{equation}
\begin{tikzpicture}
\pics{circ}{c/0,0};
\draw[star] (c-c)edge[flags=+,endlabel=$a$]++(45:0.5) (c-c)edge[flags=-,endlabel=$b$]++(135:0.5) (c-c)edge[flags=+,endlabel=$c$]++(-135:0.5) (c-c)edge[flags=+,endlabel=$d$]++(-45:0.5);
\end{tikzpicture}=
\begin{tikzpicture}
\pics{circ}{c/0,0};
\draw[star] (c-c)edge[flags=-,endlabel=$a$]++(-45:0.5) (c-c)edge[flags=-,endlabel=$d$]++(45:0.5) (c-c)edge[flags=-,endlabel=$c$]++(135:0.5) (c-c)edge[flags=+,endlabel=$b$]++(-135:0.5);
\end{tikzpicture}
\end{equation}
Also this axiom is implied by the notation.

\item The following \tdef{fusion axiom}{fusion_axiom}: Contracting a right-oriented index of one *-algebra tensor with a left-oriented index of another *-algebra tensor evaluates to the *-algebra tensor with respect to the remaining open indices. E.g.:
\begin{equation}
\begin{gathered}
a)\quad
\begin{tikzpicture}
\pics{circ}{c1/0,0|c2/0.8,0};
\draw[star] (c1-c)edge[flags=+,flage=+](c2-c);
\end{tikzpicture}=
\begin{tikzpicture}
\pics{circ}{c1/0,0};
\end{tikzpicture}\\
b)\quad
\begin{tikzpicture}
\pics{circ}{c1/0,0|c2/0.8,0};
\draw[star] (c1-c)edge[flags=+,flage=+](c2-c) (c2-c)edge[flags=-,endlabel=$a$]++(0.5,0);
\end{tikzpicture}=
\begin{tikzpicture}
\pics{circ}{c1/0,0};
\draw[star] (c1-c)edge[flags=-,endlabel=$a$]++(0.5,0);
\end{tikzpicture}\\
c)\quad
\begin{tikzpicture}
\pics{circ}{c1/0,0|c2/0.8,0};
\draw[star] (c1-c)edge[flags=+,endlabel=$b$]++(135:0.4) (c1-c)edge[flags=+,endlabel=$c$]++(-135:0.4) (c1-c)edge[flags=+,flage=+](c2-c) (c2-c)edge[flags=+,endlabel=$a$]++(45:0.4) (c2-c)edge[flags=+,endlabel=$d$]++(-45:0.4);
\end{tikzpicture}=
\begin{tikzpicture}
\pics{circ}{c/0,0};
\draw[star] (c-c)edge[flags=+,endlabel=$a$]++(45:0.5) (c-c)edge[flags=+,endlabel=$b$]++(135:0.5) (c-c)edge[flags=+,endlabel=$c$]++(-135:0.5) (c-c)edge[flags=+,endlabel=$d$]++(-45:0.5);
\end{tikzpicture}\\
d)\quad
\begin{tikzpicture}
\pics{circ}{c1/0,0|c2/0.8,0};
\draw[star] (c1-c)edge[flags=+,flage=+](c2-c) (c2-c)edge[flags=+,endlabel=$a$]++(-90:0.4) (c2-c)edge[flags=+,endlabel=$b$]++(-45:0.4) (c2-c)edge[flags=+,endlabel=$c$]++(0:0.4) (c2-c)edge[flags=+,endlabel=$d$]++(45:0.4) (c2-c)edge[flags=+,endlabel=$e$]++(90:0.4);
\end{tikzpicture}=
\begin{tikzpicture}
\pics{circ}{c/0,0};
\draw[star]  (c-c)edge[flags=+,endlabel=$a$]++(-90:0.4) (c-c)edge[flags=+,endlabel=$b$]++(-45:0.4) (c-c)edge[flags=+,endlabel=$c$]++(0:0.4) (c-c)edge[flags=+,endlabel=$d$]++(45:0.4) (c-c)edge[flags=+,endlabel=$e$]++(90:0.4);
\end{tikzpicture}
\end{gathered}
\end{equation}

\item The \tdef{support convention}{support_convention}: The *-algebra tensor with two indices with opposite *-orientations equals the identity tensor:
\begin{equation}
\begin{tikzpicture}
\pics{circ}{0/0,0};
\draw[star] (0-c)edge[flags=+,endlabel=$b$]++(0.5,0) (0-c)edge[flags=-,endlabel=$a$]++(-0.5,0);
\end{tikzpicture}=
\begin{tikzpicture}
\draw[star] (0,0)edge[startlabel=$a$,endlabel=$b$](1,0);
\end{tikzpicture}
\end{equation}
\end{itemize}
\end{mydef}

\begin{myobs}
\label{obs:star_algebra_finite}
It might seem that working with *-algebras involves dealing with an infinite set of tensors and an infinite set of axioms which would be quite cumbersome. Luckily, due to the fusion axiom the whole *-algebra is already determined by the *-algebra tensor with two $\text{right}$ and one $\text{left}$ indices:
\begin{equation}
\begin{tikzpicture}
\pics{circ}{c/0,0};
\draw[star]  (c-c)edge[flags=+,endlabel=$c$]++(-90:0.4) (c-c)edge[flags=-,endlabel=$a$]++(30:0.4) (c-c)edge[flags=-,endlabel=$b$]++(150:0.4);
\end{tikzpicture}
\end{equation}
We can view parts of the *-algebra axioms as definitions for all other *-algebra tensors. The $0$-, $1$- and $2$-index *-algebra tensors can be obtained in the following way:
\begin{equation}
\begin{gathered}
\begin{tikzpicture}
\pics{circ}{c1/0,0};
\draw[star] (c1-c)edge[flags=-,endlabel=$a$]++(0.5,0);
\end{tikzpicture}\coloneqq
\begin{tikzpicture}
\pics{circ}{c/0,0};
\draw[star]  (c-c)edge[flags=-,endlabel=$a$]++(0:0.4);
\draw[star,flags=+,flage=+] (c-c)to[out=-120,in=-90,looseness=2]++(-0.5,0)to[out=90,in=120,looseness=2](c-c);
\end{tikzpicture}\\
\begin{tikzpicture}
\pics{circ}{c/0,0};
\draw[star]  (c-c)edge[flags=-,endlabel=$a$]++(0:0.4) (c-c)edge[flags=-,endlabel=$b$]++(180:0.4);
\end{tikzpicture}\coloneqq
\begin{tikzpicture}
\pics{circ}{c1/0,-0.8|c2/0,0};
\draw[star] (c1-c)edge[flags=-,flage=-](c2-c) (c2-c)edge[flags=-,endlabel=$a$]++(30:0.4) (c2-c)edge[flags=-,endlabel=$b$]++(150:0.4);
\end{tikzpicture}\\
\begin{tikzpicture}
\pics{circ}{c1/0,0};
\end{tikzpicture}=
\begin{tikzpicture}
\pics{circ}{c1/0,0|c2/0.8,0};
\draw[star] (c1-c)edge[flags=+,flage=+](c2-c);
\end{tikzpicture}
\end{gathered}
\end{equation}
An $n+2$-index *-algebra tensor can be obtained by fusing $n$ $3$-index tensors. By contracting with the $2$-index tensor above we can invert the *-orientation of individual indices. This way we obtain all *-algebra tensors from only the one above.

Additionally, the infinite set of axioms is generated by a finite subset, consisting essentially of the associativity law below and some conditions involving inverting the *-orientations.
\end{myobs}

\begin{mycom}
The support convention is not essential for the axiomatic structure, and we merely impose it out of convenience. If we don't impose it, there can be a proper sub-vector space of $\mathbb{R}^B$, such that every *-algebra tensor is only supported within this sub-vector space. The $2$-index *-algebra tensor with opposite *-orientations is the projector onto this sub-vector space. By setting it to the identity map we can ensure that the tensors are always fully supported.
\end{mycom}

\begin{mycom}
Multiplying each $n$-index *-algebra tensor by the number $\alpha^{2-n}$ for some $0\neq \alpha\in \mathbb{R}$ leaves all the axioms invariant. Thus, *-algebras come in families related by those normalizations with $\alpha$, and by gauge transformations. Other than that, the set of families is discrete.

We can fix the $\alpha$ normalization (i.e. canonically choose one representative in the corresponding family) in different ways. One of them is to impose that contracting two neighboring indices of a *-algebra tensor with matching *-orientations yields the *-algebra tensor with those indices missing. E.g.:
\begin{equation}
\label{eq:loop_normalization}
\begin{tikzpicture}
  \pics{circ}{v1/0,0};
  \draw[star] (v1-c)edge[flags=-,endlabel=$a$] ++(-120:0.6) (v1-c)edge[flags=+,endlabel=$b$] ++(180:0.6) (v1-c)edge[flags=+,endlabel=$c$] ++(120:0.6);
\draw[star,flags=+,flage=+] (v1-c)to[out=60,in=180] ++(30:0.6) to[out=0,in=0] (-30:0.6) to[out=180,in=-60] (v1-c);
\end{tikzpicture}=
\begin{tikzpicture}
  \pics{circ}{v1/0,0};
  \draw[star] (v1-c)edge[flags=-,endlabel=$a$] ++(-120:0.6) (v1-c)edge[flags=+,endlabel=$b$] ++(180:0.6) (v1-c)edge[flags=+,endlabel=$c$] ++(120:0.6);
\end{tikzpicture}
\end{equation}
This fixes the normalization of the $0$-index *-algebra tensor:
\begin{equation}
\begin{tikzpicture}
\pic (s) at (0,0){circ};
\end{tikzpicture}=
\begin{tikzpicture}
\pic (s) at (0,0){circ};
\draw[star,flags=+,rounded corners] (s-c)--++(0,0.4)--++(0.4,0)--++(0,-0.8)--++(-0.4,0)--(s-c);
\draw[star] (s-c)edge[flags=-]++(0,-0.3);
\end{tikzpicture}=
\begin{tikzpicture}
\coordinate (s-c) at (0,0);
\draw[star,rounded corners] (s-c)--++(0,0.4)--++(0.4,0)--++(0,-0.8)--++(-0.4,0)--(s-c);
\end{tikzpicture}=
|B|
\end{equation}
\end{mycom}

\begin{mycom}
*-algebras as we defined them above are equivalent to what is known as \emph{finite-dimensional unital real C*-algebra} in the literature. 

Such a (finite-dimensional unital real C)*-algebra over a vector space $\mathbb{R}^B$ is given by three linear maps:
\begin{equation}
\begin{aligned}
\textrm{unit } \eta: \mathbb{C}^1 & \longrightarrow \mathbb{C}^B \quad [\eta(1)=\mathbf{1}]\\
\textrm{involution } t: \mathbb{C}^B & \longrightarrow \mathbb{C}^B\quad [t(a)=a^*]\\
\textrm{product } \mu: \mathbb{C}^{B\times B} & \longrightarrow \mathbb{C}^B\quad [\mu(a,b)=ab]
\end{aligned}
\end{equation}
such that the axioms Eqs.~(\ref{eq:cstar_ax1}, \ref{eq:cstar_ax2}, \ref{eq:cstar_ax3}, \ref{eq:cstar_ax4}, \ref{eq:cstar_ax5}) hold.

We will think of those three linear maps as the tensors given by their coefficients in the canonical basis of $\mathbb{R}^B$. Every *-algebra according to our definition yields a *-algebra in the conventional definition via the following identification:
\begin{equation}
\begin{gathered}
\eta_a=
\begin{tikzpicture}
\pics{circ}{x/0,0};
\draw[star](x-c)edge[flags=+,endlabel=$a$]++(0.5,0);
\end{tikzpicture}\\
t_{ab}=
\begin{tikzpicture}
\pics{circ}{x/0,0};
\draw[star](x-c)edge[flags=+,endlabel=$a$]++(-0.5,0);
\draw[star](x-c)edge[flags=+,endlabel=$b$]++(0.5,0);
\end{tikzpicture}\\
\mu^{ab}_c=
\begin{tikzpicture}
\pics{circ}{c/0,0};
\draw[star]  (c-c)edge[flags=+,endlabel=$c$]++(0:0.4) (c-c)edge[flags=-,endlabel=$a$]++(120:0.4) (c-c)edge[flags=-,endlabel=$b$]++(-120:0.4);
\end{tikzpicture}
\end{gathered}
\end{equation}

We will show this by giving a list of all conventional *-algebra axioms and then giving the set of permutation, orientation reversal and fusion axioms from which they follow.
\begin{enumerate}
\item The unit is left invariant by the *-operation:
\begin{equation}
\label{eq:cstar_ax1}
\mathbf{1}^*=\mathbf{1} \qquad t \circ \eta=\eta
\end{equation}
This follows from the following fusion axiom:
\begin{equation}
\begin{tikzpicture}
\pics{circ}{x/0,0|y/1,0};
\draw[star](x-c)edge[flags=+,flage=+](y-c) (y-c)edge[flags=+,endlabel=$a$]++(0.5,0);
\end{tikzpicture}=
\begin{tikzpicture}
\pics{circ}{x/0,0};
\draw[star](x-c)edge[flags=+,endlabel=$a$]++(0.5,0);
\end{tikzpicture}
\end{equation}

\item The \emph{involution} is named after the following property:
\begin{equation}
\label{eq:cstar_ax2}
(a^*)^*=a \qquad t \circ t=\mathbb{1}
\end{equation}
This follows from the following fusion and arrow reversal axiom,
\begin{equation}
\begin{tikzpicture}
\pics{circ}{x/0,0|y/0.7,0};
\draw[star] (x-c)edge[flags=-,flage=-](y-c) (x-c)edge[flags=-,endlabel=$a$]++(-0.5,0) (y-c)edge[flags=+,endlabel=$b$]++(0.5,0);
\end{tikzpicture}=
\begin{tikzpicture}
\pics{circ}{x/0,0};
\draw[star](x-c)edge[flags=-,endlabel=$a$]++(-0.5,0);
\draw[star](x-c)edge[flags=+,endlabel=$b$]++(0.5,0);
\end{tikzpicture},\quad
\begin{tikzpicture}
\pics{circ}{x/0,0};
\draw[star](x-c)edge[flags=+,endlabel=$a$]++(-0.5,0);
\draw[star](x-c)edge[flags=+,endlabel=$b$]++(0.5,0);
\end{tikzpicture}=
\begin{tikzpicture}
\pics{circ}{x/0,0};
\draw[star](x-c)edge[flags=-,endlabel=$a$]++(-0.5,0);
\draw[star](x-c)edge[flags=-,endlabel=$b$]++(0.5,0);
\end{tikzpicture}
\end{equation}
together with the support convention.

\item The defining property of the unit is:
\begin{equation}
\label{eq:cstar_ax3}
a \cdot \mathbf{1}=a=\mathbf{1} \cdot a \qquad \mu \circ (\mathbb{1}\otimes \eta)=\mathbb{1}=\mu \circ (\eta \otimes \mathbb{1})
\end{equation}
It follows from the following fusion and orientation reversal axiom,
\begin{equation}
\begin{tikzpicture}
\pics{circ}{m/0,0|e/-0.7,0.3};
\draw[star] (m-c)edge[flags=+,endlabel=$b$]++(0.5,0);
\draw[star,flags=-,rounded corners,endlabel=$a$] (m-c)--++(-0.3,-0.3)--++(-0.5,0);
\draw[star,flags=-,rounded corners] (m-c)--++(-0.3,0.3)--(e-c);
\end{tikzpicture}=
\begin{tikzpicture}
\pics{circ}{x/0,0};
\draw[star](x-c)edge[flags=-,endlabel=$a$]++(-0.5,0);
\draw[star](x-c)edge[flags=+,endlabel=$b$]++(0.5,0);
\end{tikzpicture}=
\begin{tikzpicture}
\pics{circ}{m/0,0|e/-0.7,-0.3};
\draw[star] (m-c)edge[flags=+,endlabel=$b$]++(0.5,0);
\draw[star,flags=-,rounded corners,endlabel=$a$] (m-c)--++(-0.3,0.3)--++(-0.5,0);
\draw[star,flags=-,rounded corners] (m-c)--++(-0.3,-0.3)--(e-c);
\end{tikzpicture}
\end{equation}
together with the support convention.

\item The \emph{*-property} is given by
\begin{equation}
\label{eq:cstar_ax4}
(ab)^*=b^*a^* \qquad t\circ \mu= \mu \circ (t\otimes t) \circ \mathrm{Swap}
\end{equation}
where $\mathrm{Swap}$ is the operation that exchanges the two arguments.

It follows from the following fusion and orientation reversal axiom:
\begin{equation}
\begin{tikzpicture}
\pics{circ}{m/0,0|t/0.7,0};
\draw[star] (m-c)edge[flags=+,flage=+](t-c)(t-c)edge[flags=-,endlabel=$c$]++(0.5,0);
\draw[rounded corners,star,flags=-,endlabel=$a$] (m-c)--++(-0.3,0.3)--++(-0.4,0);
\draw[rounded corners,star,flags=-,endlabel=$b$] (m-c)--++(-0.3,-0.3)--++(-0.4,0);
\end{tikzpicture}=
\begin{tikzpicture}
\pics{circ}{m/0,0|t1/-0.7,0.3|t2/-0.7,-0.3};
\draw[star] (m-c)edge[flags=+,endlabel=$c$]++(0.5,0) (t1-c)edge[flags=+,endlabel=$b$]++(-0.5,0) (t2-c)edge[flags=+,endlabel=$a$]++(-0.5,0);
\draw[star,flags=+,flage=+,rounded corners] (t1-c)--++(0.4,0)--(m-c);
\draw[star,flags=+,flage=+,rounded corners] (t2-c)--++(0.4,0)--(m-c);
\end{tikzpicture}
\end{equation}

\item \emph{Associativity} is given by
\begin{equation}
\label{eq:cstar_ax5}
(ab)c=a(bc) \qquad \mu \circ (\mu \otimes \mathbb{1})=\mu \circ (\mathbb{1} \otimes \mu)
\end{equation}

It follows from the following fusion axiom:
\begin{equation}
\begin{tikzpicture}[rotate=90]
\pics{circ}{0/0,0|1/0.5,-0.5};
\draw[star] (0-c)edge[flags=+, flage=+](1-c) (0-c)edge[flags=-,endlabel=$a$]++(-0.4,0.4) (0-c)edge[flags=-,endlabel=$b$]++(0.4,0.4) (1-c)edge[flags=-,endlabel=$c$]++(0.4,0.4) (1-c)edge[flags=+,endlabel=$d$]++(0,-0.5);
\end{tikzpicture}=
\begin{tikzpicture}[rotate=90]
\pics{circ}{0/0,0|1/-0.5,-0.5};
\draw[star] (0-c)edge[flags=+, flage=+](1-c) (0-c)edge[flags=-,endlabel=$b$]++(-0.4,0.4) (0-c)edge[flags=-,endlabel=$c$]++(0.4,0.4) (1-c)edge[flags=-,endlabel=$a$]++(-0.4,0.4) (1-c)edge[flags=+,endlabel=$d$]++(0,-0.5);
\end{tikzpicture}
\end{equation}
\end{enumerate}
\end{mycom}

\begin{mydef}
A *-algebra is called \tdef{commutative}{star_commutative}, if all *-algebra tensors are invariant under arbitrary (instead of only cyclic) index permutations that leave the *-orientations invariant. We will represent commutative *-algebras by a smaller circle. E.g.
\begin{equation}
\begin{tikzpicture}
\pics{smallcirc}{0/0,0};
\draw[normal] (0-c)edge[flags=-]node[pos=1,right]{$a$}++(0.5,0) (0-c)edge[flags=-]node[pos=1,above]{$b$}++(0,0.5) (0-c)edge[flags=-]node[pos=1,left]{$c$}++(-0.5,0) (0-c)edge[flags=-]node[pos=1,below]{$d$}++(0,-0.5);
\end{tikzpicture}=
\begin{tikzpicture}
\pics{smallcirc}{0/0,0};
\draw[normal] (0-c)edge[flags=-]node[pos=1,right]{$c$}++(0.5,0) (0-c)edge[flags=-]node[pos=1,above]{$d$}++(0,0.5) (0-c)edge[flags=-]node[pos=1,left]{$b$}++(-0.5,0) (0-c)edge[flags=-]node[pos=1,below]{$a$}++(0,-0.5);
\end{tikzpicture}
\end{equation}
\end{mydef}

\subsection{*-algebras: Examples and classification}
In this section we will give some examples for *-algebras, and state a complete classification. Note that this section is specific to real tensors and does not make sense for arbitrary tensor types.

\begin{mydef}
The \tdef{trivial *-algebra}{trivial_star_algebra} is the *-algebra with the trivial one-element basis whose tensors are all given by the number $1$.
\end{mydef}

\begin{mydef}
For each finite set $B$, the \tdef{delta *-algebra}{delta_star_algebra} is the commutative *-algebra with basis $B$ given by the following tensors:
\begin{equation}
\begin{tikzpicture}
\piclab{s}{smallcirc}{(0,0)}{(-135:0.3)}{$\delta$};
\draw[star] (s-c)edge[endlabel=$a$]++(-90:0.4) (s-c)edge[endlabel=$b$]++(180:0.4) (s-c)edge[endlabel=$c$]++(90:0.4);
\node at ($(s-c)+(0:0.4)$){$\ldots$};
\end{tikzpicture}
\begin{cases}
1 \textrm{ if } a=b=c=\ldots\\
0 \textrm{ otherwise}
\end{cases}.
\end{equation}
The tensors of this *-algebra do not depend on the *-orientations, so we omit the flags to indicate the latter.
\end{mydef}

\begin{mycom}
Let us compute the result $Z$ of contracting two $1$-index tensors $X$ and $Y$ with two indices of the $3$-index *-algebra tensor:
\begin{equation}
\label{eq:algebra_action}
\begin{tikzpicture}
\node[froundbox](z)at(0,0){$Z$};
\draw[star] (z.south)edge[endlabel=$a$]++(-90:0.4);
\end{tikzpicture}=
\begin{tikzpicture}
\piclab{s}{smallcirc}{(0,0)}{(90:0.3)}{$\delta$};
\node[froundbox] (x) at (-0.5,0.7) {$X$};
\node[froundbox] (y) at (0.5,0.7) {$Y$};
\draw[star,rounded corners] (s-c)--++(-0.5,0.2)--(x.south);
\draw[star,rounded corners] (s-c)--++(0.5,0.2)--(y.south);
\draw[star] (s-c)edge[endlabel=$a$]++(-90:0.5);
\end{tikzpicture}
\end{equation}
For the delta *-algebra we obtain
\begin{equation}
Z(b_a)=X(b_a) Y(b_a)
\end{equation}
So we see that the $3$-index *-algebra tensor corresponds to the structure coefficients of the algebra of real functions over the set $B$ under point-wise multiplication.

The $2$-index *-algebra tensor corresponds to the identity linear map. The $1$-index tensor is either the real function that is $1$ for every element of $B$, or, interpreted as a linear functional, the integral/sum over the function.
\end{mycom}

\begin{mydef}
For each finite set $B$, the \tdef{matrix *-algebra}{matrix_star_algebra} is the *-algebra with basis $B\times B$ whose $n$-index tensor with *-orientations $o\in \{\text{right},\text{left}\}^n$ is given by:
\begin{equation}
\begin{gathered}
\begin{tikzpicture}
\piclab{c}{circ}{(0,0)}{(-135:0.4)}{$M$};
\draw[star]  (c-c)edge[flags=+,flags=-]node[pos=1,below]{$a_0b_0$}++(-90:0.5) (c-c)edge[flags=+,flags=-]node[pos=1,left]{$a_1b_1$}++(180:0.5) (c-c)edge[flags=+,flags=-]node[pos=1,above]{$a_2b_2$}++(90:0.5);
\node at ($(s-c)+(0:0.5)$){$\ldots$};
\end{tikzpicture}=
\begin{cases}
1 \textrm{ if } s_0(o_0)=t_1(o_1) \text{ and } \ldots \text{ and } s_{n-1}(o_{n-1})=t_0(o_0)\\
0 \textrm{ otherwise}
\end{cases}\\
\text{ where } s_i(\text{right})=b_i,\quad s_i(\text{left})=a_i, \quad  t_i(\text{left})=b_i,\quad t_i(\text{right})=a_i
\end{gathered}
\end{equation}
Where we put flags on both sides to indicate that the *-orientations are according to $o$.
\end{mydef}

\begin{mycom}
The matrix *-algebra tensor with two right indices and one left index is given by:
\begin{equation}
\begin{tikzpicture}
\piclab{c}{circ}{(0,0)}{(-135:0.4)}{$M$};
\draw[star]  (c-c)edge[flags=+]node[pos=1,below]{$a_2b_2$}++(-90:0.5) (c-c)edge[flags=-]node[pos=1,above,xshift=0.2cm]{$a_1b_1$}++(30:0.5) (c-c)edge[flags=-]node[pos=1,above,xshift=-0.2cm]{$a_0b_0$}++(150:0.5);
\end{tikzpicture}=
\begin{tikzpicture}
\coords{m1/-30:0.2|m2/90:0.2|m3/-150:0.2};
\draw[normal,rounded corners,startlabel=$a_1$,endlabel=$a_2$] ($(m1)+(30:0.8)$)--(m1)--++(-90:0.8);
\draw[normal,rounded corners,startlabel=$b_1$,endlabel=$a_0$] ($(m2)+(30:0.8)$)--(m2)--++(150:0.8);
\draw[normal,rounded corners,startlabel=$b_0$,endlabel=$b_2$] ($(m3)+(150:0.8)$)--(m3)--++(-90:0.8);
\end{tikzpicture}
\end{equation}
If we evaluate Eq.~(\ref{eq:algebra_action}) we get:
\begin{equation}
\begin{gathered}
\begin{tikzpicture}
\piclab{s}{circ}{(0,0)}{(-135:0.4)}{$M$};
\node[froundbox] (a) at (-0.5,0.7) {$X$};
\node[froundbox] (b) at (0.5,0.7) {$Y$};
\draw[star,flags=-,rounded corners] (s-c)--++(-0.5,0.2)--(a.south);
\draw[star,flags=-,rounded corners] (s-c)--++(0.5,0.2)--(b.south);
\draw[star] (s-c)edge[flags=+]node[pos=1,below]{$ab$}++(-90:0.5);
\end{tikzpicture}=
\begin{tikzpicture}
\node[froundbox] (a) at (-0.5,0.7) {$X$};
\node[froundbox] (b) at (0.5,0.7) {$Y$};
\draw[normal,rounded corners] ([xshift=0.1cm]a.south)--++(0,-0.4)-|([xshift=-0.1cm]b.south);
\draw[normal] ([xshift=-0.1cm]a.south)edge[endlabel=$a$]++(0,-0.4) ([xshift=0.1cm]b.south)edge[endlabel=$b$]++(0,-0.4);
\end{tikzpicture}\\
Z((b_a,b_b))=
\sum_x X(b_a,x) Y(x,b_b)
\end{gathered}
\end{equation}
So we see that when we interpret $A$ and $B$ as matrices, then the $3$-index *-algebra tensor is nothing but matrix multiplication.

The $2$-index matrix *-algebra tensor with equal *-orientations is given by:
\begin{equation}
\begin{tikzpicture}
\piclab{x}{circ}{(0,0)}{(-90:0.4)}{$M$};
\draw[star,flags=-](x-c)--++(-0.5,0)node[left]{$a_1b_1$};
\draw[star,flags=-](x-c)--++(0.5,0)node[right]{$a_2b_2$};
\end{tikzpicture}=
\begin{tikzpicture}
\draw[startlabel=$a_1$,endlabel=$b_2$] (0,0)--++(1,0);
\draw[startlabel=$b_1$,endlabel=$a_2$] (0,0.4)--++(1,0);
\end{tikzpicture}
\end{equation}
Interpreted as a superoperator acting on the space of matrices, this is the transposition.

The one-index matrix *-algebra tensor is given by:
\begin{equation}
\begin{tikzpicture}
\piclab{x}{circ}{(0,0)}{(180:0.4)}{$M$};
\draw[star](x-c)edge[flags=+,endlabel=$ab$]++(0.5,0);
\end{tikzpicture}=
\begin{tikzpicture}
\draw[rounded corners,startlabel=$a$,endlabel=$b$] (0,0)--++(-0.7,0)--++(0,0.4)--++(0.7,0);
\end{tikzpicture}
\end{equation}
Interpreted as a matrix, this is the identity matrix. As a linear functional on matrices, this is the trace operation.
\end{mycom}

\begin{mydef}
The \tdef{complex number *-algebra}{cn_star_algebra} is the commutative *-algebra with basis $\{\mathbf{1},\mathbf{i}\}$ whose $n$-index tensor with *-orientations $o\in \{\text{right},\text{left}\}^n$ is given by:
\begin{equation}
\begin{gathered}
\begin{tikzpicture}
\piclab{c}{smallcirc}{(0,0)}{(-135:0.4)}{$\mathbb{C}$};
\draw[star]  (c-c)edge[flags=+,flags=-,endlabel=$a_0$]++(-90:0.5)  (c-c)edge[flags=+,flags=-,endlabel=$a_1$]++(180:0.5)  (c-c)edge[flags=+,flags=-,endlabel=$a_2$]++(90:0.5);
\node at ($(s-c)+(0:0.5)$){$\ldots$};
\end{tikzpicture}
=\chi\left(\sum_{i=1}^n c(o_i,a_i) \mod 4\right),\\
c(\text{right},\mathbf{1})=0 ,\quad c(\text{left},\mathbf{1})=0,\\
c(\text{right},\mathbf{i})=1, \quad c(\text{left},\mathbf{i})=3,\\
\chi(0)=1 \quad \chi(1)=0, \quad \chi(2)=-1 \quad \chi(3)=0.
\end{gathered}
\end{equation}
\end{mydef}

\begin{mycom}
Consider the complex number *-algebra tensor with two right- and one left oriented index:
\begin{equation}
\begin{tikzpicture}
\piclab{c}{smallcirc}{(0,0)}{(-135:0.35)}{$\mathbb{C}$};
\draw[star]  (c-c)edge[flags=+]node[pos=1,below]{$c$}++(-90:0.5) (c-c)edge[flags=-]node[pos=1,above,xshift=0.2cm]{$b$}++(30:0.5) (c-c)edge[flags=-]node[pos=1,above,xshift=-0.2cm]{$a$}++(150:0.5);
\end{tikzpicture}=
\left(\begin{pmatrix}1&0\\0&-1\end{pmatrix}, \begin{pmatrix}0&1\\1&0\end{pmatrix}\right)
\end{equation}
where on the right side $a, b, c$ correspond to row, column, and block, respectively, and $\mathbf{1}$ and $\mathbf{i}$ correspond the first and second entry.

If we evaluate Eq.~(\ref{eq:algebra_action}) we get:
\begin{equation}
\begin{gathered}
Z(\mathbf{1})=\left(X(\mathbf{1})Y(\mathbf{1})-X(\mathbf{i})Y(\mathbf{i})\right)\\
Z(\mathbf{i})=\left(X(\mathbf{1})Y(\mathbf{i})+X(\mathbf{i})Y(\mathbf{1})\right)
\end{gathered}
\end{equation}

One can view a complex number $X$ as a vector in $\mathbb{R}^{\{\mathbf{1},\mathbf{i}\}}$ with real part $X(\mathbf{1})$ and imaginary part $X(\mathbf{i})$. We see that with this identification the $3$-leg *-algebra tensor above corresponds to the structure coefficients of the multiplication of complex numbers.

The $2$-index *-algebra tensor is given by
\begin{equation}
\begin{tikzpicture}
\piclab{x}{smallcirc}{(0,0)}{(-90:0.35)}{$\mathbb{C}$};
\draw[star,flags=-](x-c)--++(-0.5,0)node[left]{$a_1b_1$};
\draw[star,flags=-](x-c)--++(0.5,0)node[right]{$a_2b_2$};
\end{tikzpicture}=
\begin{pmatrix}
1&0\\0&-1
\end{pmatrix}
\end{equation}
which is nothing but complex conjugation.

The $1$-index tensor is the complex number $1$.
\end{mycom}

\begin{mydef}
The \tdef{quaternion *-algebra}{quaternion_star_algebra} is the following *-algebra with basis $\{\mathbf{1},\mathbf{i},\mathbf{j},\mathbf{k}\}$: First define recursively the following function on strings
\begin{equation}
\begin{gathered}
Q: \{\mathbf{1},\mathbf{i},\mathbf{j},\mathbf{k}\}^*\longrightarrow \{+1,0,-1\}\\
Q(\ldots \mathbf{1}\ldots)=Q(\ldots\ldots)\\
Q(\ldots \mathbf{i}\mathbf{j}\ldots)=Q(\ldots \mathbf{k}\ldots),\qquad Q(\ldots \mathbf{j}\mathbf{i}\ldots)=-Q(\ldots \mathbf{k}\ldots)\\
Q(\ldots \mathbf{j}\mathbf{k}\ldots)=Q(\ldots \mathbf{i}\ldots),\qquad Q(\ldots \mathbf{k}\mathbf{j}\ldots)=-Q(\ldots \mathbf{i}\ldots)\\
Q(\ldots \mathbf{k}\mathbf{i}\ldots)=Q(\ldots \mathbf{j}\ldots),\qquad Q(\ldots \mathbf{i}\mathbf{k}\ldots)=-Q(\ldots \mathbf{j}\ldots)\\
Q(\varnothing)=1,\qquad Q(\mathbf{i})=0\\
Q(\mathbf{j})=0,\qquad Q(\mathbf{k})=0
\end{gathered}
\end{equation}
Then the quaternion *-algebra tensor with $n$ indices and *-orientations $o\in \{\text{right},\text{left}\}^n$ is given by:
\begin{equation}
\begin{gathered}
\begin{tikzpicture}
\piclab{c}{circ}{(0,0)}{(-135:0.4)}{$\mathbb{H}$};
\draw[star]  (c-c)edge[flags=+,flags=-,endlabel=$a_0$]++(-90:0.5)  (c-c)edge[flags=+,flags=-,endlabel=$a_1$]++(180:0.5)  (c-c)edge[flags=+,flags=-,endlabel=$a_2$]++(90:0.5);
\node at ($(s-c)+(0:0.5)$){$\ldots$};
\end{tikzpicture}=
Q(a_1a_2a_3\ldots) \prod_{0\leq i<n}S(a_i,o_i),\\
S(x,\text{right})=1 \quad \forall x\in \{\mathbf{1},\mathbf{i},\mathbf{j},\mathbf{k}\}\\
S(\mathbf{1},\text{left})=1, \qquad S(x,\text{left})=-1 \quad \forall x\in \{\mathbf{i},\mathbf{j},\mathbf{k}\}\\
\end{gathered}
\end{equation}
\end{mydef}

\begin{mycom}
Consider the quaternion *-algebra tensor with two right- and one left oriented index:
\begin{equation}
\begin{tikzpicture}
\piclab{c}{circ}{(0,0)}{(-135:0.4)}{$\mathbb{H}$};
\draw[star]  (c-c)edge[flags=+]node[pos=1,below]{$c$}++(-90:0.5) (c-c)edge[flags=-]node[pos=1,above,xshift=0.2cm]{$b$}++(30:0.5) (c-c)edge[flags=-]node[pos=1,above,xshift=-0.2cm]{$a$}++(150:0.5);
\end{tikzpicture}=
\begin{pmatrix}
\mathbf{1}:1 & \mathbf{i}:1 & \mathbf{j}:1 & \mathbf{k}:1\\
\mathbf{i}:1 & \mathbf{1}:-1 & \mathbf{k}:1 & \mathbf{j}:-1\\
\mathbf{j}:1 & \mathbf{k}:-1 & \mathbf{1}:-1 & \mathbf{i}:1\\
\mathbf{k}:1 & \mathbf{j}:1 & \mathbf{i}:-1 & \mathbf{1}:-1
\end{pmatrix}
\end{equation}
Here row and column correspond to $a$ and $b$, respectively, in the order $(\mathbf{1},\mathbf{i},\mathbf{j},\mathbf{k})$. The index $c$ is denoted in sparse notation as a list with items ``index configuration : tensor entry'' containing only items for non-zero tensor entries (in the above case there's only one such non-zero entry always). We see that the $3$-index tensor above equals the structure coefficients of the product of a $4$-dimensional real division algebra known as \emph{quaternions}.
\end{mycom}

\begin{mydef}
The \tdef{direct sum}{star_direct_sum} of two *-algebras is the *-algebra formed by the pairwise direct sum of all involved *-tensors. As all the *-algebra axioms involve only connected tensor-networks on the left and right, the direct sum is a *-algebra again. The \tdef{tensor product}{star_tensor_product} of two *-algebras is the *-algebra formed by the pairwise tensor products of all involved *-tensors.
\end{mydef}

\begin{myprop}
Every *-algebra $A$ is gauge equivalent to the direct sum of components:
\begin{equation}
A=\bigoplus_{i} M_{B_i}\otimes X_i
\end{equation}
Here $M_B$ is the matrix *-algebra for the set $B$, and $X_i$ is either 1) the trivial *-algebra, 2) the complex number *-algebra or 3) the quaternionic *-algebra. To be precise, the matrix, trivial, complex or quaternions above were presented for a particular choice of $\alpha$ normalization. This normalization could be different for each component of the direct sum.

In the conventional study of *-algebras, the equivalent statement would be the following: Every real finite-dimensional *-algebra is the direct sum of irreducible blocks. Each of these blocks is $M_n(X)$, which is the algebra of $n\times n$ matrices whose entries are real numbers, complex numbers, or quaternions \cite{Li2003}.
\end{myprop}

\subsection{*-tensors}
In this section we will give an axiomatic definition of positive *-tensors, the building blocks of both quantum mechanics and classical statistical physics in our approach. Positive *-tensors can be seen as a tensor type on their own. One might use a different tensor type than real tensors as input to the construction in this section, yielding another type of positive *-tensors. Normalized positive *-tensors are the causal version of positive *-tensors and are equivalent to the symmetric monoidal category of CPTP maps. As usual, readers not interested in this level of abstraction might just think of all tensors as ordinary real tensors.

\begin{mydef}
A \tdef{*-tensor}{star_tensor} is a tensor for which each basis is equipped with a *-algebra (for that basis).
\end{mydef}

\begin{mydef}
A *-tensor $T$ is \tdef{positive}{positive} if there exists a \tdef{root tensor}{root_tensor} $R$ with the same index set as $T$ except for one additional \tdef{internal index}{internal_index}, such that $T$ is obtained by the following procedure: 1) Take the tensor product of two copies of $R$. 2) Contract the two internal indices of the two copies of $R$. 3) For each index of $T$, contract the two corresponding indices of each copy of $R$ with a copy of the according $3$-index *-algebra tensor. 4) Evaluate the tensor network. The following shows how a positive *-tensor $T$ with $0$-, $1$- or $3$ indices (in a), b) or c)) is obtained from the corresponding root tensor:
\begin{equation}
\label{eq:star_positive}
\begin{gathered}
a)\quad
\begin{tikzpicture}
\node[froundbox] (t) at (0,0) {$T$};
\end{tikzpicture}=
\begin{tikzpicture}
\node[roundbox](r1) at (0,0.4){$R$};
\node[roundbox](r2) at (0,-0.4){$R$};
\draw[normal,rounded corners] (r1.west)--++(-0.2,0)|-(r2.west);
\end{tikzpicture}\\
b)\quad
\begin{tikzpicture}
\node[froundbox] (t) at (0,0) {$T$};
\draw[star] (t.east)edge[endlabel=$a$]++(0.4,0);
\end{tikzpicture}=
\begin{tikzpicture}
\pics{circ}{s/0.7,0};
\node[roundbox](r1) at (0,0.4){$R$};
\node[roundbox](r2) at (0,-0.4){$R$};
\draw[normal,rounded corners] (r1.west)--++(-0.2,0)|-(r2.west);
\draw[star,rounded corners] (r1.east)edge[flage=+](s-c) (r2.east)edge[flage=+](s-c) (s-c)edge[flags=+,endlabel=$a$]++(0.5,0);
\end{tikzpicture}\\
c)\quad
\begin{tikzpicture}
\node[froundbox] (t) at (0,0) {$T$};
\draw[star] ([yshift=0.1cm]t.east)edge[endlabel=$a$]++(0.4,0.3) (t.east)edge[endlabel=$b$]++(0.4,0) ([yshift=-0.1cm]t.east)edge[endlabel=$c$]++(0.4,-0.3);
\end{tikzpicture}=
\begin{tikzpicture}
\pics{circ}{s1/0.9,0.7|s2/0.9,0|s3/0.9,-0.7};
\node[roundbox](r1) at (0,0.5){$R$};
\node[roundbox](r2) at (0,-0.5){$R$};
\draw[normal,rounded corners] (r1.west)--++(-0.2,0)|-(r2.west);
\draw[star] ([yshift=0.1cm]r1.east)edge[flage=+](s1-c) ([yshift=0.1cm]r2.east)edge[flage=+](s1-c) (s1-c)edge[flags=+,endlabel=$a$]++(0.5,0) (r1.east)edge[flage=+](s2-c) (r2.east)edge[flage=+](s2-c) (s2-c)edge[flags=+,endlabel=$b$]++(0.5,0) ([yshift=-0.1cm]r1.east)edge[flage=+](s3-c) ([yshift=-0.1cm]r2.east)edge[flage=+](s3-c) (s3-c)edge[flags=+,endlabel=$c$]++(0.5,0);
\end{tikzpicture}
\end{gathered}
\end{equation}
\end{mydef}

\begin{myrem}
The root tensor $R$ for a given positive *-tensor $T$ is only unique up to an orthogonal gauge acting on the internal index. In fact it turns out that (for real tensors) we can always choose a gauge such that the internal index is trivial (i.e. one-dimensional).
\end{myrem}

\begin{mycom}
What we call ``positive'' in this document would be usually rather referred to as ``non-negative'', or ``positive semidefinite''.
\end{mycom}

\begin{myobs}
\label{obs:positive_tensor_product}
The tensor product $T$ of two positive *-tensors $T_1$ and $T_2$ is again a positive *-tensor. An obvious root tensor for $T$ is given by the tensor product of the two root tensors $R_1$ and $R_2$ (where we block the two internal indices to one single one):
\begin{equation}
\begin{tikzpicture}
\node[roundbox](r1) at (0,0.5){$R$};
\draw (r1.west)--++(-0.3,0)node[left]{$xy$};
\node[below] at (r1.south){$\ldots$};
\end{tikzpicture}=
\begin{tikzpicture}
\node[roundbox](r1) at (0,0.5){$R_1$};
\draw (r1.west)--++(-0.3,0)node[left]{$x$};
\node[below] at (r1.south){$\ldots$};
\node[roundbox](r1) at (0,1.5){$R_2$};
\draw (r1.west)--++(-0.3,0)node[left]{$y$};
\node[below] at (r1.south){$\ldots$};
\end{tikzpicture}
\end{equation}
\end{myobs}

\begin{myobs}
\label{obs:positive_contraction}
The contraction $T'$ of two indices of a positive *-algebra tensor $T$ is again a positive *-algebra tensor. The root tensor $R'$ of $T'$ can be obtained from the root tensor $R$ of $T$ in the following way:
\begin{equation}
\begin{tikzpicture}
\node[roundbox](r1) at (0,0.5){$R'$};
\draw (r1.west)--++(-0.3,0)node[left]{$xy$};
\node[below] at (r1.south){$\ldots$};
\end{tikzpicture}=
\begin{tikzpicture}
\pics{circ}{s1/0.9,0.5};
\node[roundbox](r1) at (0,0.5){$R$};
\draw (r1.west)edge[endlabel=$x$]++(-0.3,0);
\draw[star] ([yshift=0.1cm]r1.east)edge[bend left,flage=+](s1-c) ([yshift=-0.1cm]r1.east)edge[bend right,flage=-](s1-c);
\draw[star] (s1-c)edge[flags=-,endlabel=$y$]++(0.5,0);
\node[below] at (r1.south){$\ldots$};
\end{tikzpicture}
\end{equation}
With this choice we indeed find:
\begin{equation}
\begin{tikzpicture}
\node[froundbox,minimum height=0.7cm] (t) at (0,0) {$T'$};
\node at (0.6,-0.1){$\ldots$};
\end{tikzpicture}=
\begin{tikzpicture}
\node[froundbox,minimum height=1cm] (t) at (0,0) {$T$};
\draw[star,rounded corners] ([yshift=0.3cm]t.east)--++(0.5,0)|-([yshift=-0.1cm]t.east);
\node at (0.5,-0.4){$\ldots$};
\end{tikzpicture}=
\begin{tikzpicture}
\pics{circ}{s1/0.9,0.5|s3/0.9,-0.5};
\node[roundbox](r1) at (0,0.5){$R$};
\node[roundbox](r2) at (0,-0.5){$R$};
\draw[normal,rounded corners] (r1.west)--++(-0.2,0)|-(r2.west);
\draw[star] ([yshift=0.1cm]r1.east)edge[flage=+](s1-c) ([yshift=0.1cm]r2.east)edge[flage=+](s1-c) ([yshift=-0.1cm]r1.east)edge[flage=+](s3-c) ([yshift=-0.1cm]r2.east)edge[flage=+](s3-c);
\draw[star,rounded corners,flags=+,flage=-] (s1-c)--++(0.6,0)|-(s3-c);
\node[below] at (r1.south){$\ldots$};
\node[below] at (r2.south){$\ldots$};
\end{tikzpicture}=
\begin{tikzpicture}
\pics{circ}{s1/0.9,0.5|s3/0.9,-0.5};
\node[roundbox](r1) at (0,0.5){$R$};
\node[roundbox](r2) at (0,-0.5){$R$};
\draw[normal,rounded corners] (r1.west)--++(-0.2,0)|-(r2.west);
\draw[star] ([yshift=0.1cm]r1.east)edge[bend left,flage=+](s1-c) ([yshift=-0.1cm]r1.east)edge[bend right,flage=-](s1-c) ([yshift=0.1cm]r2.east)edge[bend left,flage=+](s3-c) ([yshift=-0.1cm]r2.east)edge[bend right,flage=-](s3-c);
\draw[star,rounded corners,flags=-,flage=+] (s1-c)--++(0.6,0)|-(s3-c);
\node[below] at (r1.south){$\ldots$};
\node[below] at (r2.south){$\ldots$};
\end{tikzpicture}=
\begin{tikzpicture}
\node[roundbox](r1) at (0,0.5){$R'$};
\node[roundbox](r2) at (0,-0.5){$R'$};
\draw[normal,rounded corners] (r1.west)--++(-0.2,0)|-(r2.west);
\node[below] at (r1.south){$\ldots$};
\node[below] at (r2.south){$\ldots$};
\end{tikzpicture}
\end{equation}
Where we used that
\begin{equation}
\begin{tikzpicture}
\pics{circ}{s1/0,0.6|s2/0,-0.6};
\draw[star,rounded corners,flags=+,flage=-] (s1-c)--++(0.5,0)|-(s2-c);
\draw[star] (s1-c)edge[endlabel=$a$,flags=-]++(150:0.5) (s1-c)edge[endlabel=$b$,flags=-]++(-150:0.5) (s2-c)edge[endlabel=$c$,flags=-]++(150:0.5) (s2-c)edge[endlabel=$d$,flags=-]++(-150:0.5);
\end{tikzpicture}=
\begin{tikzpicture}
\pics{circ}{s1/0,0|s2/0.8,0};
\draw[star,flags=+,flage=+] (s1-c)--(s2-c);
\draw[star] (s1-c)edge[endlabel=$a$,flags=-]++(150:0.5) (s1-c)edge[endlabel=$b$,flags=-]++(-150:0.5) (s2-c)edge[endlabel=$c$,flags=+]++(30:0.5) (s2-c)edge[endlabel=$d$,flags=+]++(-30:0.5);
\end{tikzpicture}=
\begin{tikzpicture}
\pics{circ}{s1/0,0};
\draw[star] (s1-c)edge[endlabel=$a$,flags=-]++(150:0.5) (s1-c)edge[endlabel=$b$,flags=-]++(-150:0.5) (s1-c)edge[endlabel=$c$,flags=+]++(30:0.5) (s1-c)edge[endlabel=$d$,flags=+]++(-30:0.5);
\end{tikzpicture}=
\begin{tikzpicture}
\pics{circ}{s1/0,0.6|s2/0,0};
\draw[star,flags=+,flage=+] (s1-c)--(s2-c);
\draw[star] (s1-c)edge[endlabel=$a$,flags=-]++(150:0.5) (s1-c)edge[endlabel=$c$,flags=+]++(30:0.5) (s2-c)edge[endlabel=$b$,flags=-]++(-150:0.5) (s2-c)edge[endlabel=$d$,flags=+]++(-30:0.5);
\end{tikzpicture}=
\begin{tikzpicture}
\pics{circ}{s1/0,0.6|s2/0,-0.6};
\draw[star,rounded corners,flags=-,flage=+] (s1-c)--++(0.5,0)|-(s2-c);
\draw[star] (s1-c)edge[endlabel=$a$,flags=-]++(150:0.5) (s1-c)edge[endlabel=$c$,flags=+]++(-150:0.5) (s2-c)edge[endlabel=$b$,flags=-]++(150:0.5) (s2-c)edge[endlabel=$d$,flags=+]++(-150:0.5);
\end{tikzpicture}
\end{equation}
by the flip axiom and fusion axiom.
\end{myobs}

\begin{myobs}
\label{obs:positive_tensor_network}
The evaluation of a tensor network formed by positive *-tensors yields a positive *-tensor. This follows simply from applying Obs.~(\ref{obs:positive_tensor_product},\ref{obs:positive_contraction}) multiple times.
\end{myobs}

\begin{myobs}
\label{obs:delta_positivity}
Consider a positive (real) *-tensor where all bases are equipped with delta *-algebras. In this case, Eq.~(\ref{eq:star_positive}) becomes
\begin{equation}
\begin{gathered}
T(a,b,\ldots)=\sum_i R(i,a,b,\ldots) R(i,a,b,\ldots)=\\
\sum_i (R(i,a,b,\ldots))^2\geq 0 \quad \forall a,b,\ldots
\end{gathered}
\end{equation}
Conversely, for every $T$ with
\begin{equation}
T(a,b,\ldots)\geq 0 \quad \forall a,b,\ldots
\end{equation}
we can take
\begin{equation}
R(a,b,\ldots)=\sqrt{T(a,b,\ldots)}
\end{equation}
as a root tensor with trivial internal index. So we found that positive *-tensors with respect to delta *-algebras are nothing but entry-wise positive tensors.
\end{myobs}

\begin{myobs}
\label{obs:matrix_positivity}
Consider a positive (real) *-tensor where all bases are equipped with matrix *-algebras. In this case, Eq.~(\ref{eq:star_positive}) becomes
\begin{equation}
\begin{gathered}
T((a,a'),(b,b'),\ldots)=\\
\sum_{i,a'',b'',\ldots} R(i,(a,a''),(b,b''),\ldots) R(i,(a'',a'),(b'',b'),\ldots)
\end{gathered}
\end{equation}
This implies that $T$ interpreted as a matrix from the indices $a,b,\ldots$ to the indices $a',b',\ldots$ is positive semidefinite.

Conversely, for each $T$ with this property we can take
\begin{equation}
R=\sqrt{T}
\end{equation}
as a root tensor, where the square root is taken of $T$ reshaped into a matrix as above. The internal index is trivial in this case.
\end{myobs}

\begin{mydef}
A \tdef{normalized *-tensor}{normalized_star_tensor} is a *-tensor whose indices are divided into \tdef{in}{in_index} and \tdef{out indices}{out_index}, such that: Contracting each $\text{out}$ index with the $1$-index *-algebra tensor yields the tensor product of one $1$-index *-algebra tensor for each $\text{in}$ index. Consider an open index, or a contracted index pair consisting of an in and an out index in a tensor network. In this case, the corresponding line in tensor-network notation can be associated with a direction going from in to out. We'll refer to the latter as the \tdef{normalization direction}{normalization_direction}, and indicate it by a rounded arrow on the corresponding line. E.g.:
\begin{equation}
\label{eq:normalization}
\begin{gathered}
a)\quad
\begin{tikzpicture}
\pics{circ}{s1/1,0.2|s2/1,-0.2};
\node[froundbox,minimum height=0.8cm] (t) at (0,0) {$T$};
\draw[star] ([yshift=0.3cm]t.west)edge[ndir=-,endlabel=$a$]++(-0.5,0);
\draw[star] (t.west)edge[ndir=-,endlabel=$b$]++(-0.5,0);
\draw[star] ([yshift=-0.3cm]t.west)edge[ndir=-,endlabel=$c$]++(-0.5,0);
\draw[star] ([yshift=0.2cm]t.east)edge[ndir=+,ndirpos=0.4,flage=-](s1-c);
\draw[star] ([yshift=-0.2cm]t.east)edge[ndir=+,ndirpos=0.4,flage=-](s2-c);
\end{tikzpicture}=
\begin{tikzpicture}
\pics{circ}{s1/1,0.4|s2/1,0|s3/1,-0.4};
\draw[star] (s1-c)edge[flags=+,endlabel=$a$,ndir=-]++(-0.8,0) (s2-c)edge[flags=+,endlabel=$b$,ndir=-]++(-0.8,0) (s3-c)edge[flags=+,endlabel=$c$,ndir=-]++(-0.8,0);
\end{tikzpicture}\\
b)\quad
\begin{tikzpicture}
\pics{circ}{s1/1,0};
\node[froundbox] (t) at (0,0) {$T$};
\draw[star] (t.west)edge[ndir=-,endlabel=$a$]++(-0.5,0);
\draw[star] (t.east)edge[ndir=+,ndirpos=0.4,flage=-](s1-c);
\end{tikzpicture}=
\begin{tikzpicture}
\pics{circ}{s1/1,0};
\draw[star] (s1-c)edge[flags=+,endlabel=$a$,ndir=-]++(-0.8,0);
\end{tikzpicture}\\
c)\quad
\begin{tikzpicture}
\node[froundbox] (t) at (0,0) {$T$};
\draw[star] (t.west)edge[ndir=-,endlabel=$a$]++(-0.5,0);
\end{tikzpicture}=
\begin{tikzpicture}
\pics{circ}{s1/1,0};
\draw[star] (s1-c)edge[flags=+,endlabel=$a$,ndir=-]++(-0.8,0);
\end{tikzpicture}\\
d)\quad
\begin{tikzpicture}
\pics{circ}{s1/1,0.2|s2/1,-0.2};
\node[froundbox,minimum height=0.7cm] (t) at (0,0) {$T$};
\draw[star] ([yshift=0.2cm]t.east)edge[ndir=+,ndirpos=0.4,flage=-](s1-c);
\draw[star] ([yshift=-0.2cm]t.east)edge[ndir=+,ndirpos=0.4,flage=-](s2-c);
\end{tikzpicture}=
\hspace{2cm}\\
e)\quad
\begin{tikzpicture}
\node[froundbox] (t) at (0,0) {$T$};
\end{tikzpicture}=
\hspace{2cm}
\end{gathered}
\end{equation}
\end{mydef}

\begin{myobs}
\label{obs:normalized_contraction}
Take the tensor product of two normalized *-tensors $T_1$ and $T_2$ and contract $n$ out indices of $T_1$ with $n$ in indices of $T_2$ (for arbitrary $n\geq 0$). The result is a normalized *-tensor again. E.g.:
\begin{equation}
\begin{tikzpicture}
\pics{circ}{s1/1,0.2|s2/1,-0.2|s3/1,0.8};
\node[froundbox,minimum height=1cm] (t1) at (0,0) {$T_2$};
\node[froundbox,minimum height=1cm] (t2) at (-1,0.4) {$T_1$};
\draw[star] ([yshift=0.4cm]t1.west)edge[ndir=-](t2.east);
\draw[star] (t1.west)edge[ndir=-]([yshift=-0.4cm]t2.east);
\draw[star] ([yshift=-0.4cm]t1.west)edge[ndir=-,endlabel=$b$]++(-1.4,0);
\draw[star] (t2.west)edge[ndir=-,endlabel=$a$]++(-0.4,0);
\draw[star] ([yshift=0.2cm]t1.east)edge[ndir=+,ndirpos=0.4,flage=-](s1-c);
\draw[star] ([yshift=-0.2cm]t1.east)edge[ndir=+,ndirpos=0.4,flage=-](s2-c);
\draw[star] ([yshift=0.4cm]t2.east)edge[ndir=+,ndirpos=0.4,flage=-](s3-c);
\end{tikzpicture}=
\begin{tikzpicture}
\pics{circ}{s1/0,-0.4|s2/0,0|s3/0,0.4|s4/0,0.8};
\node[froundbox,minimum height=1cm] (t2) at (-1,0.4) {$T_1$};
\draw[star] (s1-c)edge[flags=+,ndir=-,endlabel=$b$]++(-1.4,0);
\draw[star] (t2.west)edge[ndir=-,endlabel=$a$]++(-0.4,0);
\draw[star] (t2.east)edge[ndir=+,ndirpos=0.4,flage=-](s3-c);
\draw[star] ([yshift=-0.4cm]t2.east)edge[ndir=+,ndirpos=0.4,flage=-](s2-c);
\draw[star] ([yshift=0.4cm]t2.east)edge[ndir=+,ndirpos=0.4,flage=-](s4-c);
\end{tikzpicture}=
\begin{tikzpicture}
\pics{circ}{s1/1,0.4|s3/1,-0.4};
\draw[star] (s1-c)edge[flags=+,endlabel=$a$,ndir=-]++(-0.8,0) (s3-c)edge[flags=+,endlabel=$b$,ndir=-]++(-0.8,0);
\end{tikzpicture}
\end{equation}
\end{myobs}

\begin{myobs}
\label{obs:normalized_star_tensor_network}
Consider a acyclic tensor network formed by normalized *-tensors, that is, a tensor network where all contracted index pairs consist of an in and an out index, such that there is no loop of cyclic normalization directions. The evaluation of such a tensor network yields again a normalized *-tensor. This follows simply from applying Obs.~(\ref{obs:normalized_contraction}) multiple times.
\end{myobs}

\begin{myobs}
\label{obs:delta_normalization}
Consider a normalized *-tensor $T$ with in indices $i,j,\ldots$ and out indices $x,y,\ldots$, where all bases are equipped with delta *-algebras. In this case, Eq.~(\ref{eq:normalization}) becomes:
\begin{equation}
\sum_{b_x,b_y,\ldots}T(b_i,b_j,\ldots,b_x,b_y,\ldots)=1 \quad\forall b_i,b_j,\ldots
\end{equation}
That is, for any fixed configuration of the output indices, the sum of the entries over all input indices equals $1$.
\end{myobs}

\begin{myobs}
Consider a normalized *-tensor $T$ with in indices $i,j,\ldots$ and out indices $x,y,\ldots$, where all bases are equipped with matrix *-algebras. In this case, Eq.~(\ref{eq:normalization}) becomes:
\begin{equation}
\sum_{c_x,c_y,\ldots} T((c_i,c_i),(c_j,c_j),\ldots,(a_x,b_x),(a_y,b_y),\ldots)= \delta_{a_i,b_i} \delta_{a_j,b_j}\ldots
\end{equation}
I.e., if we trace over all out index pairs, we obtain the trace over the in index pairs. Thus, if we interpret $T$ as a superoperator from all the in index pairs to the out index pairs, it is trace preserving.
\end{myobs}

\subsection{Complex tensors as *-tensors}
\label{sec:complex_tensor}
In the sections above we worked with purely real tensors. Traditionally, quantum mechanics is formulated using complex numbers. In this sections we will discuss how *-tensors (for real *-algebras) can emulate complex tensors, such as e.g. vectors of complex vector spaces or complex-linear maps. The central observation is that complex numbers themselves can be interpreted as a $2$-dimensional real *-algebra.

\begin{mydef}
\tdef{Complex tensors}{complex_tensor} are tensors with complex numbers instead of real numbers as entries. Tensor product and contraction are the same except that we now use the addition and multiplication of the complex numbers.
\end{mydef}

\begin{myobs}
\label{obs:real_representation}
A complex number is a vector consisting of two real numbers. Each complex tensor $T$ has a \tdef{real representation}{real_representation}, that is, a real tensor $T^{(r)}$ with one additional index $x$ whose basis is the $2$-element set $\{\mathbf{1},\mathbf{i}\}$, such that
\begin{equation}
\begin{gathered}
T^{(r)}(\ldots,b_x=\mathbf{1})=\mathrm{Real}(T(\ldots))\\
T^{(r)}(\ldots,b_x=\mathbf{i})=\mathrm{Imag}(T(\ldots))
\end{gathered}
\end{equation}
Graphically, the following shows a $3$-index complex tensor, and its $4$-index real representation:
\begin{equation}
\label{eq:compl_tensor}
\begin{tikzpicture}
\node[box] (t) {$T$};
\draw (t.west)edge[endlabel=$a$]++(-0.4,0);
\draw (t.north)edge[endlabel=$b$]++(0,0.3);
\draw (t.east)edge[endlabel=$c$]++(0.4,0);
\end{tikzpicture}
\quad \longleftrightarrow \quad
\begin{tikzpicture}
\node[box] (t) {$T^{(r)}$};
\draw (t.west)edge[endlabel=$a$]++(-0.4,0);
\draw (t.north)edge[endlabel=$b$]++(0,0.3);
\draw (t.east)edge[endlabel=$c$]++(0.4,0);
\draw[normal] (t.north east)edge[endlabel=$x$]++(45:0.4);
\end{tikzpicture}
\end{equation}
The real representation $(T_1\otimes T_2)^{(r)}$ of a tensor product does not equal the tensor product of real representations $T_1^{(r)}\otimes T_2^{(r)}$, as the latter has two additional $x$-indices instead of one. To emulate the tensor product, we have to fuse the two $x$-indices into one with the $3$-index complex number *-algebra tensor. E.g.,
\begin{equation}
\begin{gathered}
\begin{tikzpicture}
\node[box] (t1) at (0,0) {$T_1$};
\draw (t1.west)--++(-0.4,0) node[left]{$a$};
\draw (t1.north)--++(0,0.3) node[above]{$b$};
\draw (t1.east)--++(0.4,0) node[right]{$c$};
\node[box] (t2) at (0,1.5) {$T_2$};
\draw (t2.west)--++(-0.4,0) node[left]{$d$};
\draw (t2.north)--++(0,0.3) node[above]{$e$};
\end{tikzpicture}=
\begin{tikzpicture}
\node[box,minimum width=1cm] (t) {$T$};
\draw (t.west)edge[endlabel=$a$]++(-0.4,0);
\draw (t.north)edge[endlabel=$b$]++(0,0.3);
\draw ([xshift=-0.3cm]t.north)edge[endlabel=$d$]++(0,0.3);
\draw ([xshift=0.3cm]t.north)edge[endlabel=$e$]++(0,0.3);
\draw (t.east)edge[endlabel=$c$]++(0.4,0);
\end{tikzpicture}\\
\quad \longleftrightarrow \quad
\begin{tikzpicture}
\node[box,minimum width=1cm] (t) {$T^{(r)}$};
\draw (t.west)edge[endlabel=$a$]++(-0.4,0);
\draw (t.north)edge[endlabel=$b$]++(0,0.3);
\draw ([xshift=-0.3cm]t.north)edge[endlabel=$d$]++(0,0.3);
\draw ([xshift=0.3cm]t.north)edge[endlabel=$e$]++(0,0.3);
\draw (t.east)edge[endlabel=$c$]++(0.4,0);
\draw[normal] (t.north east)edge[endlabel=$x$]++(45:0.4);
\end{tikzpicture}=
\begin{tikzpicture}
\piclab{c}{smallcirc}{(1.2,0.9)}{(-60:0.25)}{$\mathbb{C}$};
\node[box] (t1) at (0,0) {$T_1^{(r)}$};
\draw (t1.west)--++(-0.4,0) node[left]{$a$};
\draw (t1.north)--++(0,0.3) node[above]{$b$};
\draw (t1.east)--++(0.4,0) node[right]{$c$};
\node[box] (t2) at (0,1.5) {$T_2^{(r)}$};
\draw (t2.west)--++(-0.4,0) node[left]{$d$};
\draw (t2.north)--++(0,0.3) node[above]{$e$};
\draw[normal] (t1.north east)edge[flage=+](c-c) (c-c)edge[flags=+,endlabel=$x$]++(0.5,0);
\draw[normal,flage=+,rounded corners]  (t2.north east)--++(45:0.4)-|(c-c);
\end{tikzpicture}
\end{gathered}
\end{equation}
An index contraction of a real representation $T^{(r)}$ is equal to the real representation of the contraction of the same indices of $T$. This is because real and imaginary part are contracted separately.
\end{myobs}

\begin{mydef}
The \tdef{realification}{realification} of a complex tensor $T$ with index set $I$ and basis collection $B$ with respect to a choice of \tdef{realification directions}{realification_direction} $s\in \{\text{in},\text{out}\}^I$ is the following real tensor $T^{\mathbb{R}}$:
\begin{itemize}
\item $T^{\mathbb{R}}$ has the same index set $I$ as $T$.
\item For index $i\in I$ take the basis set $B_i \times \{\mathbf{1},\mathbf{i}\}$ (instead of $B_i$ for $T$).
\item Consider the real representation $T^{(r)}$.
\item Consider the complex number *-algebra tensor with index set $I\cup\{x\}$ (where $\cup$ denotes disjoint union). The *-orientation of an index $i\in I$ matches the *-orientation of the index $x$ if $s_i=\text{in}$, otherwise they are opposite.
\item Take the tensor product of the two tensors from the previous points, and contract the $x$-index of $T^{(r)}$ with the $x$ index of the *-algebra tensor.
\item $T^{\mathbb{R}}$ is obtained from the tensor of the previous point by blocking the $i$-index coming from $T^{(r)}$ and the $i$-index coming from the *-algebra tensor into the $i$-index of $T^{\mathbb{R}}$.
\end{itemize}
Consider e.g. the realification of the following $3$-index complex tensor $T$ with $s_a=\text{out}$, $s_b=\text{in}$, $s_a=\text{in}$:
\begin{equation}
\label{eq:realification}
\begin{tikzpicture}
\node[box] (t) {$T$};
\draw (t.west)edge[normal,endlabel=$a$]++(-0.5,0);
\draw (t.north)edge[normal,endlabel=$b$]++(0,0.4);
\draw (t.east)edge[normal,endlabel=$c$]++(0.5,0);
\end{tikzpicture}
\quad\rightarrow\quad
\begin{tikzpicture}
\node[box] (t) {$T^{\mathbb{R}}$};
\draw (t.west)edge[normal,realifs=+,endlabel=$a$]++(-0.5,0);
\draw (t.north)edge[normal,realifs=-,endlabel=$b$]++(0,0.4);
\draw (t.east)edge[normal,realifs=-,endlabel=$c$]++(0.5,0);
\end{tikzpicture}=
\begin{tikzpicture}
\piclab{c}{smallcirc}{(0,-0.8)}{(-90:0.35)}{$\mathbb{C}$};
\node[box] (t) {$T^{(r)}$};
\draw (t.west)--++(-0.8,0) node[left]{$a$};
\draw (t.north)--++(0,0.5) node[above]{$b$};
\draw (t.east)--++(0.8,0) node[right]{$c$};
\draw[normal,rounded corners,flags=+] (c-c)--++(-0.7,0) |- ($(t.west)+(-0.8,-0.1)$);
\draw[normal,rounded corners,flags=-] (c-c)--++(0.7,0) |- ($(t.east)+(0.8,-0.1)$);
\draw[normal,flage=+](t.south)--(c-c);
\draw[normal, rounded corners,flags=-] (c-c)--++(0.6,0.5)--++(0,0.7) -| ($(t.north)+(0.1,0.5)$);
\end{tikzpicture}
\end{equation}
\end{mydef}

\begin{mycom}
The realification of complex tensors is closely related to Dirac notation. The realification directions of an index corresponds to whether it is represented as ket or bra. Changing the realification direction amounts to a complex conjugation, which gives a canonical identification of a vector space with its dual (in the finite-dimensional case) (see Rem.(\ref{rem:linear_map_realification})).
\end{mycom}

\begin{myobs}
The tensor product of the realifications of two complex tensors is in general not equal to the realification of the tensor product of the complex tensors. However, contraction commutes with taking the realification, if the realification directions of the contracted indices are opposite to another. Moreover, also taking a tensor product and then contracting one index of the first tensor with one index of the second tensor, commutes with taking the realification. E.g.,
\begin{equation}
\begin{gathered}
\begin{tikzpicture}
\node[box] at (0,0) (t1) {$T_1^{\mathbb{R}}$};
\node[box] at (1.3,0) (t2) {$T_2^{\mathbb{R}}$};
\draw[normal] (t1.west)edge[realifs=+] node[pos=1,left]{$a$}++(-0.5,0);
\draw[normal] (t1.north)edge[realifs=+] node[pos=1,above]{$b$}++(0,0.4);
\draw[normal] (t1.east)edge[realifs=+,realife=+] (t2.west);
\draw[normal] (t2.east)edge[realifs=+] node[pos=1,right]{$c$}++(0.5,0);
\draw[normal] (t2.north)edge[realifs=+] node[pos=1,above]{$d$}++(0,0.4);
\end{tikzpicture}\\=
\begin{tikzpicture}
\piclab{c1}{smallcirc}{(0,-0.8)}{(-90:0.35)}{$\mathbb{C}$};
\piclab{c2}{smallcirc}{(2,-0.8)}{(-90:0.35)}{$\mathbb{C}$};
\node[box] at (0,0) (t1) {$T_1$};
\node[box] at (2,0) (t2) {$T_2$};
\draw (t1.west)--++(-0.7,0) node[left]{$a$};
\draw (t1.north)--++(0,0.4) node[above]{$b$};
\draw (t1.east)--(t2.west);
\draw (t2.east)--++(0.7,0) node[right]{$c$};
\draw (t2.north)--++(0,0.4) node[above]{$d$};
\draw[normal,rounded corners,flags=+] (c1-c)--++(-0.6,0) |- ($(t1.west)+(-0.7,-0.1)$);
\draw[normal,rounded corners,flags=+,flage=+] (c1-c)--++(0.6,0) |- ($(t1.east)+(0.7,-0.1)$)-|($(c2-c)+(-0.6,0)$)--(c2-c);
\draw[normal,flage=+](t1.south)--(c1-c);
\draw[normal, rounded corners,flags=+] (c1-c)--++(0.4,0.4)--++(0,0.8) -| ($(t1.north)+(0.1,0.4)$);
\draw[normal,rounded corners,flags=+] (c2-c)--++(0.6,0) |- ($(t2.east)+(0.7,-0.1)$);
\draw[normal,flage=+](t2.south)--(c2-c);
\draw[normal, rounded corners,flags=+] (c2-c)--++(0.4,0.4)--++(0,0.8) -| ($(t2.north)+(0.1,0.4)$);
\end{tikzpicture}\\=
\begin{tikzpicture}
\piclab{c1}{smallcirc}{(0.5,-0.7)}{(90:0.3)}{$\mathbb{C}$};
\piclab{c2}{smallcirc}{(0.5,-1.4)}{(-90:0.3)}{$\mathbb{C}$};
\coords{c1/0.5,-0.7|c2/0.5,-1.4};
\node[box] at (0,0) (t1) {$T_1$};
\node[box] at (1,0) (t2) {$T_2$};
\draw (t1.west)--++(-0.7,0) node[left]{$a$};
\draw (t1.north)--++(0,0.4) node[above]{$b$};
\draw (t1.east)--(t2.west);
\draw (t2.east)--++(0.7,0) node[right]{$c$};
\draw (t2.north)--++(0,0.4) node[above]{$d$};
\draw[normal,rounded corners,flage=+] (t1.south)|-(c1-c);
\draw[normal,rounded corners,flage=+] (t2.south)|-(c1-c);
\draw[normal,flage=+,flags=+](c1-c)--(c2-c);
\draw[normal,rounded corners,flags=+] (c2-c)--++(-1.1,0) |- ($(t1.west)+(-0.7,-0.1)$);
\draw[normal, rounded corners,flags=+] (c2-c)--++(-0.9,0.75)|- ($(t1.north)+(-0.1,0.2)$)--($(t1.north)+(-0.1,0.4)$);
\draw[normal,rounded corners,flags=+] (c2-c)--++(1.1,0) |- ($(t2.east)+(0.7,-0.1)$);
\draw[normal, rounded corners,flags=+] (c2-c)--++(0.9,0.75)|- ($(t2.north)+(0.1,0.2)$)--($(t2.north)+(0.1,0.4)$);
\end{tikzpicture}
\end{gathered}
\end{equation}
In that sense, real tensors can emulate complex ones.
\end{myobs}

\begin{myrem}
\label{rem:linear_map_realification}
Consider the realification of a $2$-index complex tensor $T$ with the following realification directions:
\begin{equation}
\begin{gathered}
a)\quad
\begin{tikzpicture}
\node[box] (l){$T^{\mathbb{R}}$};
\draw[normal,realifs=-] (l.west)--++(-0.5,0)node[left]{$a$};
\draw[normal,realifs=+] (l.east)--++(0.5,0)node[right]{$b$};
\end{tikzpicture}\\
b)\quad
\begin{tikzpicture}
\node[box] (l){$T^{\mathbb{R}}$};
\draw[normal,realifs=-] (l.west)--++(-0.5,0)node[left]{$a$};
\draw[normal,realifs=-] (l.east)--++(0.5,0)node[right]{$b$};
\end{tikzpicture}
\end{gathered}
\end{equation}
Both realifications can be interpreted as real-linear maps from the index labeled by $a$ to the index labeled by $b$. However, we can also interpret a) as a complex-linear map, and b) as a complex anti-linear map.
\end{myrem}

\begin{myobs}
The Hermitian conjugate of a complex-linear map written as the realification of a complex tensor as in Rem.~(\ref{rem:linear_map_realification}) is given by exchanging the two indices. This is because reversing the realification directions automatically induces complex conjugation:
\begin{equation}
\begin{gathered}
\begin{tikzpicture}
\node[box] (l){$L^{\mathbb{R}}$};
\draw[normal,realifs=-] (l.west)--++(-0.5,0)node[left]{$b$};
\draw[normal,realifs=+] (l.east)--++(0.5,0)node[right]{$a$};
\end{tikzpicture}=
\begin{tikzpicture}
\piclab{c}{smallcirc}{(0,-0.7)}{(-90:0.3)}{$\mathbb{C}$};
\node[box] (t) {$L^{(r)}$};
\draw (t.west)--++(-0.5,0) node[left]{$b$};
\draw (t.east)--++(0.5,0) node[right]{$a$};
\draw[normal,rounded corners,flags=-] (c-c)--++(-0.5,0) |- ($(t.west)+(-0.5,-0.1)$);
\draw[normal,rounded corners,flags=+] (c-c)--++(0.5,0) |- ($(t.east)+(0.5,-0.1)$);
\draw[normal] (t.south)edge[flage=+](c-c);
\end{tikzpicture}\\=
\begin{tikzpicture}
\piclab{c1}{smallcirc}{(0,-0.7)}{(0:0.3)}{$\mathbb{C}$};
\piclab{c2}{smallcirc}{(0,-1.4)}{(-90:0.3)}{$\mathbb{C}$};
\node[box] (t) {$L^{(r)}$};
\draw (t.west)--++(-0.5,0) node[left]{$b$};
\draw (t.east)--++(0.5,0) node[right]{$a$};
\draw[normal,rounded corners,flags=+] (c2-c)--++(-0.5,0) |- ($(t.west)+(-0.5,-0.1)$);
\draw[normal,rounded corners,flags=-] (c2-c)--++(0.5,0) |- ($(t.east)+(0.5,-0.1)$);
\draw[normal] (t.south)edge[flage=-](c1-c) (c1-c)edge[flags=+,flage=+](c2-c);
\end{tikzpicture}=
\begin{tikzpicture}
\piclab{c}{smallcirc}{(0,-0.7)}{(-90:0.3)}{$\mathbb{C}$};
\node[box] (t) {${L^*}^{(r)}$};
\draw (t.west)--++(-0.5,0) node[left]{$b$};
\draw (t.east)--++(0.5,0) node[right]{$a$};
\draw[normal,rounded corners,flags=+] (c-c)--++(-0.7,0) |- ($(t.west)+(-0.5,-0.1)$);
\draw[normal,rounded corners,flags=-] (c-c)--++(0.7,0) |- ($(t.east)+(0.5,-0.1)$);
\draw[normal] (t.south)edge[flage=+](c-c);
\end{tikzpicture}\\=
\begin{tikzpicture}
\node[box] (l){${L^\dagger}^{\mathbb{R}}$};
\draw[normal,realifs=-] (l.west)--++(-0.5,0)node[left]{$a$};
\draw[normal,realifs=+] (l.east)--++(0.5,0)node[right]{$b$};
\end{tikzpicture}
\end{gathered}
\end{equation}
\end{myobs}

\section{What we expect from a model}
\label{sec:models}
In this section we describe a few basic properties that any sensible physical theory should have. We will see that these very reasonable assumptions can be formalized in terms of so-called tensor-network theories.

\subsubsection*{Theories, models, setups}
Let us start by introducing some terminology:
\begin{mydef}
A \tdef{theory}{theory} consists of 1) a set $\mathcal{S}$ of \tdef{setups}{setup}, 2) a set $\mathcal{M}$ of \tdef{models}{model}, 3) a set of predictions $\mathcal{P}$, and 4) a prescription that associates to each pair of model $M\in \mathcal{M}$ and setup $S\in \mathcal{S}$ a prediction $P(M,S)\in \mathcal{P}$. Thereby, the set $\mathcal{P}$ is allowed to depend on $M$ and $S$.
\end{mydef}

The abstract mathematical sets $\mathcal{S}$, $\mathcal{P}$ and $\mathcal{M}$ have the following intended real-world physical interpretation:
\begin{itemize}
\item The setups correspond to different arrangements of the same kind of materials/components in an experiment, or different physical systems that can be observed in the same physical reality. This can for example include different system sizes, or more generally geometrically different configurations in which different components are connected, different times or positions at which measurements are taken, different states of the switches that control the experiment, and so on.
\item The models correspond to different types of components for which we run the same experiment. For example, if we run the same experiments for different materials, those would be described by different models. Also the kinds of measurements are considered part of the models, e.g., detectors of a different type at the same places would correspond to a different model. There is some freedom in what things we attribute to the model and what to the setup. E.g., if we have two different materials, these might either correspond to different components of the same model or the same kind of component in different model. In the first case, we can realize setups that contain both materials, whereas in the second case, they may never be used together.
\item The predictions for a setup correspond to the data that one collects when performing the experiment with that setup. For a deterministic theory, $\mathcal{P}(M,S)$ is just the set of possible outcomes for all the measurements taking place in $S$. When there are multiple measurements, this set is the cartesian product of all the sets of individual outcomes. In a statistical theory, predictions are probability distributions over the set of measurement outcomes.
\end{itemize}

\begin{mycom}
Note that, in contrast to our terminology, people sometimes call things a ``theory'' that we would call a model. For example, for us, quantum electrodynamics would be a model and not a theory, as it makes specific predictions. Also, what people usually call ``models'' (like the ``Ising model'') typically have some parameters to be varied, and thus would be more something like very restricted theories using our terminology. We will think of a theory as something relatively general, such as ``local statistical mechanics'' or ``local quantum mechanics'', and a model as something like ``the Ising model for $\beta=0.5$ and $h=0$'', or ``the Bose-Hubbard model at $t=1$, $U=0.5$ and $\mu=0$''.
\end{mycom}

\subsubsection*{Combination of setups}
If it is possible to do two different experiments, it also has to be possible to do both of them independently. Doing nothing can always be seen as a trivial experiment. Formally, the combination of experiments, and the trivial experiment, correspond to the following structures on the set of setups:
\begin{equation}
\begin{gathered}
1)\quad\otimes: \mathcal{S}\times\mathcal{S}\quad\rightarrow\quad \mathcal{S}\\
2)\quad 1\in \mathcal{S}
\end{gathered}
\end{equation}
If we have a set of experiments that we want to perform independently, the temporal or spacial order in which we do this doesn't matter. Also, combining any experiment with the trivial experiment is that same experiment again. So the following axioms have to hold (for all $p,q,r\in \mathcal{S}$):
\begin{itemize}
\item \emph{Associativity}:
\begin{equation}
(p\otimes q)\otimes r= p\otimes (q\otimes r)
\end{equation}
\item \emph{Commutativity}:
\begin{equation}
p\otimes q=q\otimes p
\end{equation}
\item \emph{identity element}:
\begin{equation}
1\otimes p=p
\end{equation}
\end{itemize}
Such a set with a product and identity is called a \emph{commutative monoid}.

Any pair of observations from two independent experiments can be combined into a single observation for the combined experiment. Making no observation at all can be seen as a trivial observation. The order in which we combine observations doesn't matter and combining an observation with the trivial observation yields the same observation again. So, also the set $\mathcal{P}$ of predictions forms a commutative monoid.

The combination of experiments yields the combination of observations, and the trivial experiment yields the trivial observation. Thus, formally, we have
\begin{equation}
\begin{gathered}
1)\quad P(M,p\otimes q)=P(M,p)\otimes P(M,q)\\
2)\quad P(M,1)=1
\end{gathered}
\end{equation}
So the function $P(M,\cdot): \mathcal{S}\rightarrow \mathcal{P}$ has to be a \emph{monoid homomorphism}.

\subsubsection*{Statistical theories}
Current first-principle theories such as quantum or classical statistical physics describe nature in an information-theoretic way. That is, $\mathcal{P}(M,S)$ should be the set of probability distributions over the set of possible observations/measurement outcomes of $S$. A model can be tested against reality by performing one experiment with the same measurement setup multiple times and recording the frequencies of measurement outcomes. Those frequencies are supposed to converge to the probabilities predicted by the model for this setup.

Using a statistical description for nature might have multiple reasons. On the one hand, talking about probabilities is simply more general than restricting to models that make deterministic predictions for observations given some setup: Statistical theories contain the deterministic case by restricting to distributions with probabilities $0$ and $1$ only. On the other hand, using statistics is often necessary to describe physical phenomena. First, we might only have limited knowledge of a situation, and have to model unknown parameters by probabilities. Second, physical laws might be themselves intrinsically non-deterministic. In the end, a statistical description should not be regarded as a property of nature, but more as the kind of question we ask about nature.

Mathematically, a probability distribution is a real tensor $P(a,b,\ldots)$. The bases $A,B,\ldots$ of the indices $a,b,\ldots$ correspond to the sets of outcomes for different measurements that are done simultaneously in one setup. Those real tensors must obey two constraints: 1) They have positive entries and 2) they are normalized:
\begin{equation}
\begin{gathered}
1)\quad P(a,b,\ldots)\geq 0 \quad\forall a,b,\ldots\\
2)\quad \sum_{a,b,\ldots}P(a,b,\ldots)=1
\end{gathered}
\end{equation}
According to the previous section, the probability distributions should form a commutative monoid. The product is the tensor product. The trivial prediction is the probability distribution that associates the probability $1$ to the only element in a one-element set. Both are compatible with the positivity and normalization condition.

Often it is convenient to slightly generalize and use conditional probability distributions, depending on some input parameters $x,y,z,\ldots$. Those input parameters correspond to different experimental controls, and could always be absorbed by the set of setups. A conditional probability distribution is still a real tensor $P(a,b|\ldots,x,y,\ldots)$ with the same positivity property, but a slightly modified normalization:
\begin{equation}
2) \quad \sum_{a,b,c,\ldots}P(a,b,\ldots|x,y,\ldots)=1 \quad\forall x,y,\ldots
\end{equation}

\subsubsection*{No overfitting}
The most general theory would be the following: The set of models is the set of all Turing machines $M$, and $P(M,S)$ is the evaluation of $M$ with input $S$. For a meaningful theory however, we have to restrict to a small subset of all of these possible models: Doing physics corresponds to the following procedure: 1) Collect data for some ``small'' accessible amount of measurement setups (e.g. obtaining frequencies of measurement outcomes by repetitive runs of the experiment). 2) Choose a theory, and find a model within this theory whose predictions fit the collected data well enough (or the best). 3) Use this model to make predictions for further, untested setups.

How do we choose the theory? The set of models in a theory should be large enough such that it is possible to find a model that fits the collected data well enough. Conversely, it is important that the number of models in the theory is a smaller quantity than the number of different possible measurement setups. Otherwise it would be too easy to find a model that fits the observed data, and there would be no reason to believe that the model generalizes to other setups. In particular, there might be multiple models fitting the observed data but yielding inconsistent predictions for other setups. This problem is known as ``overfitting''.

In fact, the believe in the existence of physical laws for a certain aspect of nature is nothing but the believe that there is a model of a small complexity producing observations for a large set of setups. Often, the different models in a theory are determined by a finite amount of data, whereas they make predictions for an infinite amount of setups.

\begin{mycom}
The concept of overfitting originates from the field of machine learning. Also in applied physics, we ``learn'' physical models by collecting real-world data, except that the we choose the best model by hand and not in an automated way. Also, opposed to e.g. image recognicion, the set of considerable models in a physical theory is usually labelled by a very small number of parameters, and the theory is applicable in a much more general and fundamental scope.
\end{mycom}

\subsubsection*{Combinatorialism}
A particularly natural and appealing way to construct models generated by finite data for an infinite set of setups is the following principle of ``combinatorialism'': A model consists of a finite number of ``components''. Copies of these components can be combined to larger structures. Thereby each component can only interact with a finite number of other components in a finite number of ways. Every possible combination of the components corresponds to a different setup. A combination can be evaluated by starting with one component and gradually adding more. This evaluation yields the predictions of the model for the corresponding setup.

The restrictions of how the components can be combined are only local. So the different setups instances of a local combinatorial structure, which we formalized under the name \emph{lattice type} in \cite{tensor_lattice}. As the restrictions are purely local, there are typically infinitely many setups of one type, corresponding to a ``thermodynamic limit''.

If one wants to model causality in a physical system it is also necessary to introduce a global restriction to the possible setups: One needs an additional (global) arrow of time consistent with the local combinatorial structure. In other words, we need a partial ordering of all the components and give a direction to certain possible interactions between the components. That is, there can't be any closed loops of cyclic time direction. Setups are then instances of the local combinatorial data structure together with such a partial ordering.

\begin{mycom}
An example for this principle are ground state expectation values of local lattice Hamiltonians: The components are a local Hamiltonian and different local measurements. The setups are different lattices (e.g. square lattices of different size) onto which the Hamiltonian is put, together with different points where the measurements take place. The total combinatorial structure consists of copies of the local Hamiltonian terms all over the lattice, as well as copies of the measurement operators at the corresponding places. The evaluation of such a setup yields the corresponding $n$-point correlation function in the ground state of the Hamiltonian.
\end{mycom}

\subsubsection*{Tensor-network theories}
Combinatorialism can be very naturally formalized using tensor networks.

\begin{mydef}
A \tdef{tensor-network theory}{tensor_network_theory} consists of
\begin{itemize}
\item Some monoid of predictions (e.g. probability distributions for statistical theories).
\item A lattice type, i.e. some type of local combinatorial structure.
\item A tensor type, i.e. some data structure whose entities can be combined into networks and evaluated.
\item A monoid homomorphism $f$ from the tensor type (with multiplication given by the tensor product) to the monoid of predictions.
\end{itemize}
The setups $S\in \mathcal{S}$ are given by instances of the lattice type. A \tdef{model}{tensor_network_model} consists of a set of tensors of the tensor type, $M=\{T^{(i)}\}_i$. $P(M,S)$ is computed by: 1) Associate copies of the tensors $\{T^{(i)}\}_i$ and contractions to different components of $S$, yielding a tensor network $\mathop{TN}(M,S)$. 2) Evaluate the tensor network. 3) Apply the monoid homomorphism $f$. In total we have:
\begin{equation}
P(M,S)=f(\mathop{Eval}(\mathop{TN}(M,S)))
\end{equation}
\end{mydef}


For probabilistic theories the predictions are real tensors (with extra constraints). The easiest and most natural type of tensors we could take for such a theory are real tensors. Then $\mathop{Eval}(\mathop{TN}(M,S))$ is already a real tensor. The central remaining question is: How do we get the resulting real tensors to be also positive and normalized?

Positivity could in principle be imposed in an ad hoc manner: We could take for $f$ to be the entry-wise absolute value of the tensors:
\begin{equation}
f(T)(a,b,\ldots)=|T(a,b,\ldots)|
\end{equation}
More generally, for any $\alpha \geq 0$ we could take
\begin{equation}
f(T)(a,b,\ldots)=|T(a,b,\ldots)|^\alpha
\end{equation}
This seems rather arbitrary and inelegant though. Instead we will try to let $f$ be trivial and ensure positivity by imposing conditions to the tensors themselves. The easiest way of getting (entry-wise) positive tensors $\mathop{Eval}(\mathop{TN}(M,S))$ is to use only entry-wise positive tensors $\{T^{(i)}\}_i$. This corresponds to classical statistical mechanics as described in Sec.~(\ref{sec:classical}). There is a very elegant generalization of this way of ensuring positivity, by using positive *-tensors. This corresponds to *-tensor models, which are essentially equivalent to quantum mechanics, as shown in Sec.~(\ref{sec:quantum}). Both entry-wise positive tensors as well as positive *-tensors are subsets of tensors closed under tensor product and contraction, so they are nothing but tensor types.

Also normalization can be imposed in an ad hoc manner by taking $f$ to be the normalization (see Eq.(\ref{eq:prob_normalization})). This is what we do in non-causal classical or quantum physics, e.g. classical statistical physics or thermal quantum systems. One can also enforce the normalization by imposing conditions to the tensors. To this end we need to introduce a partial ordering, or in other words, a flow of time, yielding a notion of causality. In the classical case, the tensors will be multi-variable stochastic maps with respect to this ordering. Causal *-tensor models are a generalization of this using normalized *-tensors, which can be thought of as a generalization of quantum channels.

\section{*-tensor theory}
\label{sec:star_tensor_theory}
In this section we will describe a very general tensor-network theory which unifies both classical statistics and quantum mechanics. The positivity of the predicted probabilities will we guaranteed in a natural and elegant way by using positive *-tensors. In particular, the function $f$ from the previous section only consists of a normalization (or is even trivial in the causal case).

\begin{mydef}
A \tdef{(non-causal) *-tensor model}{star_tensor_model} consists of
\begin{itemize}
\item A set of bases equipped with *-algebras, some of which are delta *-algebras.
\item A set of positive *-tensors with respect to those bases.
\item A prescription that associates to each setup $S$ a tensor network involving copies of the positive *-tensors such that: 1) For each measurement there is one open index. 2) The index is equipped with a delta *-algebra whose basis set is the set of possible measurement outcomes.
\end{itemize}

According to Obs.~(\ref{obs:positive_tensor_network}) the evaluation of such a tensor network yields another positive *-tensor $T(S)$. As the open indices are equipped with delta *-algebras, $T(S)$ are real tensors with positive entries, according to Obs.~(\ref{obs:delta_positivity}). We can make $T(S)$ into a probability distribution $P(S)$ by normalization:
\begin{equation}
\label{eq:prob_normalization}
P(S)(b_i,b_j,\ldots)=\frac{T(S)(b_i,b_j,\ldots)}{\sum_{b_i,b_j,\ldots}T(S)(b_i,b_j,\ldots)}
\end{equation}
$P(S)(b_i,b_j,\ldots)$ is the predicted probability for the combination of measurement outcomes $b_i,b_j,\ldots$ for the setup $S$.
\end{mydef}

\begin{mydef}
A \tdef{causal *-tensor model}{c_star_tensor_model} is a *-tensor model where all *-tensors are normalized. The tensor networks associated to setups have matching normalization directions which are acyclic, that is, there are no closed loops with cyclic normalization directions. In order to guarantee this, the underlying combinatorial structure must be equipped with a flow of time, i.e. a partial ordering of the components.

According to Obs.~(\ref{obs:normalized_star_tensor_network}), the evaluation of such a tensor network yields a normalized *-tensor again. As the open indices have a delta *-algebra, those tensors are real tensors with positive entries, such that for each configuration of values of the input indices the tensor on the output indices is normalized (see Obs.~(\ref{obs:delta_normalization})). This resulting tensor is the stochastic map (or conditional probability distribution) associated to the setup.
\end{mydef}

\section{Classical statistical physics as *-tensor models}
\label{sec:classical}
In this section we will show that both thermal classical statistical physics models and classical statistical processes are *-tensor models.

\subsection{Classical statistical objects as *-tensors}
Let us start by showing how different components of a classical statistical model are represented by (normalized) positive *-tensors. In classical physics, all the *-algebras (even those for non-open indices) are delta *-algebras. We will refer to such indices equipped with the delta *-algebra as \emph{classical indices}.
\begin{itemize}
\item As we have seen in Sec.~(\ref{sec:star_tensor_theory}), a \emph{probability distribution} $P$ is represented by a normalized positive *-tensor with one classical output index. A probability distribution over a $n$-partite system is a normalized positive *-tensor with $n$ classical output indices. A (deterministic) configuration of a classical system is a probability distribution with probability $1$ for this certain configuration and probability $0$ for all other configurations.
\item As we have seen in Sec.~(\ref{sec:star_tensor_theory}), a \emph{stochastic map} (from a $n$-partite system to a $m$-partite system) is represented by a normalized positive *-tensor with one ($n$) classical input indices and one ($m$) classical output indices. A deterministic function from a set of configurations to another set of configurations is a special case of a stochastic map where for each input configuration there is exactly one output configuration with probability $1$, and all other probabilities are $0$.
\item The $n$-index \emph{delta tensor} is a normalized positive *-tensor with one input index and $(n-1)$ output indices (for any choice of input index). The delta tensor is nothing but the classical copy function that associates to one configuration $(n-1)$ copies of that configuration.
\item A \emph{Boltzmann weight} $e^{-\beta H(c_1,c_2,\ldots)}$ depending on a few degrees of freedom $c_1,c_2,\ldots$ is a positive *-tensor with one index for each degree of freedom.
\end{itemize}

\subsection{Typical setups as *-tensor networks}
Let us consider different setups for classical statistical physics and how they can be translated into *-tensor networks.

\subsubsection*{Causal setups}
\begin{myobs}
Start with a probability distribution $p$, perform a stochastic map $S$, and make some measurements $M$. This is captured by the following normalized *-tensor network:
\begin{equation}
\begin{tikzpicture}
\node[froundbox] (r) at (0,0){$p$};
\node[froundbox] (c) at (0.8,0){$S$};
\node[froundbox] (p) at (1.6,0){$M$};
\draw[star] (r)edge[ndir=+](c) (c)edge[ndir=+](p);
\draw[star] (p.east)edge[ndir=+,endlabel=$i$]++(0.4,0);
\end{tikzpicture}
\end{equation}
\end{myobs}

\begin{myobs}
Consider a discrete-time Markov chain where we want to measure joint probabilities at different times. The probability distribution in a Markov chain after $x$ steps is obtained by applying the process matrix $S$ $x$ times to an initial probability $p$. Then, e.g., the joint probability distribution for the configurations after $5$ and $9$ steps is the evaluation of the following causal *-tensor network:
\begin{equation}
\begin{tikzpicture}
\node[froundbox] (0) at (-0.7,0){$p$};
\foreach \i [count=\ii from 1] in {0,...,4}{
\node[froundbox] (\ii) at (\i*0.8,0) {$S$};
\draw[star,ndir=+] (\i.east)--(\ii.west);
}
\pic (i0) at (4*0.8+0.6,0){smallcirc};
\draw[star,ndir=+] (5.east)--(i0-c);
\node[froundbox] (i1) at (4*0.8+1.2,0) {$S$};
\draw[star,ndir=+] (i0-c)--(i1.west);
\foreach \i [count=\ii from 2] in {1,...,3}{
\node[froundbox] (i\ii) at (\i*0.8+4*0.8+1.2,0) {$S$};
\draw[star,ndir=+] (i\i.east)--(i\ii.west);
}
\pic (ii) at ($(i4)+(0.6,0)$){smallcirc};
\pic (d) at ($(i4)+(1.2,0)$){smallcirc};
\draw[star] (ii-c)edge[ndir=+](d-c);
\draw[star,ndir=+] (i4.east)--(ii-c);
\draw[star] (i0-c)edge[ndir=+,endlabel=$i$]++(0,0.4) (ii-c)edge[ndir=+,endlabel=$j$]++(0,0.4);
\end{tikzpicture}
\end{equation}
\end{myobs}

\begin{myobs}
Consider a Markov process, consisting of 1) a set of degrees of freedom, 2) a sequence of stochastic maps, such that each map only acts non-trivially on a few degrees of freedom inside a small region and 3) a few evaluating stochastic maps $O$ that are applied in the end. Such a process is a *-tensor network. E.g., consider a chain of degrees of freedom initialized each in state $s$, which are acted on pairwisely by a local stochastic map $x$. In the end, the stochastic maps $O$ are just direct read outs of single degrees of freedom. This could be represented by the following tensor network:
\begin{equation}
\begin{tikzpicture}
\foreach \n/\x/\y in {i0/-0.3/0,i1/0.3/0,i2/0.7/0,i3/1.3/0,i5/2.3/0,i6/2.7/0,i7/3.3/0,i8/3.7/0,i9/4.3/0}{
\node[draw,circle,inner sep=0.4pt] (\n) at (\x,\y){$s$};
}
\foreach \n/\x/\y in {00/0/0.5,01/1/0.5,02/2/0.5,03/3/0.5,04/4/0.5,10/0.5/1,11/1.5/1,12/2.5/1,13/3.5/1,20/0/1.5,21/1/1.5,22/2/1.5,23/3/1.5,24/4/1.5,30/0.5/2,31/1.5/2,32/2.5/2,33/3.5/2,41/1/2.5,42/2/2.5,43/3/2.5}{
\node[draw,circle,inner sep=0.4pt] (\n)at (\x,\y){$x$};
}
\foreach \n/\x/\y in {a0/-0.3/2,a1/0.2/2.5,a2/0.7/3,a3/1.7/3,a4/2.3/3,a5/2.7/3,a6/3.8/2.5,a7/4.3/2}{
\pic (\n)at (\x,\y){smallcirc};
}
\draw (i0)edge[ndir=+](00) (i1)edge[ndir=+](00) (i2)edge[ndir=+](01) (i3)edge[ndir=+](01) (i5)edge[ndir=+](02) (i6)edge[ndir=+](03) (i7)edge[ndir=+](03) (i8)edge[ndir=+](04) (i9)edge[ndir=+](04) (00)edge[ndir=+](10) (01)edge[ndir=+](10) (01)edge[ndir=+](11) (02)edge[ndir=+](11) (02)edge[ndir=+](12) (03)edge[ndir=+](12) (03)edge[ndir=+](13) (04)edge[ndir=+](13) (20)edge[ndir=-](10) (21)edge[ndir=-](10) (21)edge[ndir=-](11) (22)edge[ndir=-](11) (22)edge[ndir=-](12) (23)edge[ndir=-](12) (23)edge[ndir=-](13) (24)edge[ndir=-](13)  (20)edge[ndir=+](30) (21)edge[ndir=+](30) (21)edge[ndir=+](31) (22)edge[ndir=+](31) (22)edge[ndir=+](32) (23)edge[ndir=+](32) (23)edge[ndir=+](33) (24)edge[ndir=+](33)  (41)edge[ndir=-](30) (41)edge[ndir=-](31) (42)edge[ndir=-](31) (42)edge[ndir=-](32) (43)edge[ndir=-](32) (43)edge[ndir=-](33) (20)edge[ndir=+](a0-c) (30)edge[ndir=+](a1-c) (41)edge[ndir=+](a2-c) (42)edge[ndir=+](a3-c) (42)edge[ndir=+](a4-c) (43)edge[ndir=+](a5-c) (33)edge[ndir=+](a6-c) (24)edge[ndir=+](a7-c) (00)edge[enddots]++(135:0.25) (20)edge[enddots]++(-135:0.25) (04)edge[enddots]++(45:0.25) (24)edge[enddots]++(-45:0.25);
\draw[red] (41)edge[endlabel=$b$]++(0.3,0.5) (43)edge[endlabel=$c$]++(0.3,0.5) (02)edge[endlabel=$a$]++(-0.3,-0.5);
\end{tikzpicture}
\end{equation}
The evaluation of this tensor network is a stochastic map from the open input $a$ to the outputs $b,c$.
\end{myobs}

\subsubsection*{Non-causal setups}
\begin{myobs}
Consider a (thermal) classical statistical system with Hamiltonian $H(c)$ depending on the configuration $c$ of the system. At inverse temperature $\beta$, the probability distribution over the configurations is proportional to the following Boltzmann weight:
\begin{equation}
W(c)=e^{-\beta H(c)}
\end{equation}
A measurement is given by a stochastic matrix $M(o|c)$. Usually this will be a deterministic function $o(c)$. Often the system consists of multiple degrees of freedom, and $o(c)$ corresponds to restricting to some of them. The outcome probability distribution $P(o)$ of measurement outcomes is proportional to the evaluation of the following *-tensor network:
\begin{equation}
\begin{tikzpicture}
\node[froundbox] (r) at (0,0){$W$};
\node[froundbox] (p) at (1,0){$M$};
\draw[star] (r)--(p);
\draw[star] (p.east)edge[endlabel=$o$]++(0.4,0);
\end{tikzpicture}
\end{equation}
\end{myobs}

\begin{myobs}
Consider a thermal classical statistical model with a set of degrees of freedom distributed over some lattice, and Boltzmann weights $W_p(c_{p^{(0)}},c_{p^{(1)}},\ldots)=\mathop{exp}(-\beta H_p(c_{p^{(0)}},c_{p^{(1)}},\ldots))$ depending on the configuration of the degrees of freedom around a set of places $p\in P_W$. Furthermore, consider some conditional probabilities $O_p(o_p|c_{p^{(0)}},c_{p^{(1)}},\ldots)$ depending on the degrees of freedom around some places $p\in P_O$. Usually, $o$ is a deterministic function of $c$, that is, for any set of $c_{p^{(0)}},c_{p^{(1)}},\ldots$ there is exactly one value for $o$ with $O_p=1$.

A setup for such a model consists of a 1) a lattice (of some size) 2) a few places $P_O$ to which we associate observables. The probability distribution for such a setup can be obtained by the following partition function:
\begin{equation}
\begin{gathered}
Z(\mathbf{o})=\sum_{\mathbf{c}} e^{-\beta\sum_p H_p(\mathbf{c})} \mathbf{O}(\mathbf{o}|\mathbf{c})\\
=\sum_{\mathbf{c}} \prod_{p_W\in P_W} \prod_{p_O\in P_O} W_{p_W}(c_{{p_W}^{(0)}},c_{{p_W}^{(1)}},\ldots) O_{p_O}(c_{{p_O}^{(0)}},c_{{p_O}^{(1)}},\ldots)
\end{gathered}
\end{equation}
Here we used $\mathbf{c}$ and $\mathbf{o}$ for a collection of confgurations or outputs.

In order to obtain a proper probability distribution we have to normalize it. The normalization is the partition function without any observables included:
\begin{equation}
P(\mathbf{o})=\frac{Z(\mathbf{o})}{Z(\{\})}
\end{equation}

Such a function $Z(\mathbf{o})$ can be expressed as a tensor network with only positive-entry tensors like the following:
\begin{itemize}
\item Take for each degree of freedom a copy of the delta-tensor, with one index for each Boltzmann weight and each measurement it is part of. The basis of the delta tensor is the configuration set of the degree of freedom.
\item Replace each Boltzmann weight and each measurement by the corresponding tensor.
\item Contract all indices of all delta-tensors with the corresponding index of a Boltzmann weight or measurement.
\item This yields a tensor network with one open index for each measurement.
\end{itemize}

Consider, for example the following kind of model: There are degrees of freedom on the vertices of a $2$-dimensional square lattice. One Boltzmann weight $x$ is associated to each plaquette, depending on the degrees of freedom in the corners of that plaquatte. The measurements correspond to directly reading out a single degree of freedom. The following shows a patch of the representing tensor network with $3$ measurements (in red):
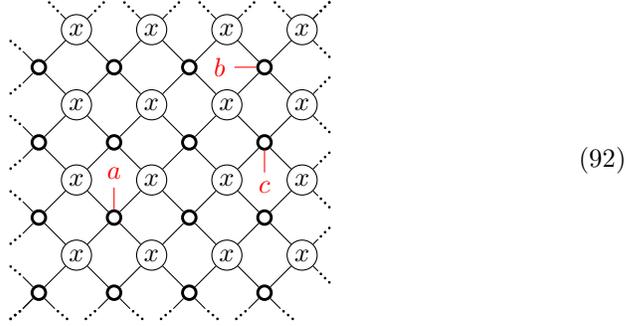
\begin{equation}
\begin{tikzpicture}
\foreach \x in {0,1,2,3}{
\foreach \y in {0,1,2,3}{
\node[draw,circle,minimum width=0.4cm,minimum height=0.4cm,inner sep=1pt] (b\x\y) at (\x+0.5,\y+0.5) {$x$};
\pic (\x\y) at (\x,\y){smallcirc};
}}
\foreach \x in {0,1,2,3}{
\foreach \y in {0,1,2,3}{
\draw (\x\y-c)--(b\x\y);
}}
\foreach[count=\xx from 1] \x in {0,1,2}{
\foreach \y in {0,1,2,3}{
\draw (\xx\y-c)--(b\x\y);
}}
\foreach[count=\xx from 1] \x in {0,1,2,3}{
\foreach[count=\yy from 1] \y in {0,1,2}{
\draw (\x\yy-c)--(b\x\y);
}}
\foreach[count=\xx from 1] \x in {0,1,2}{
\foreach[count=\yy from 1] \y in {0,1,2}{
\draw (\xx\yy-c)--(b\x\y);
}}
\foreach \x in {0,1,2,3}{
\draw (\x0-c)edge[enddots]++(-135:0.3);
\draw (\x0-c)edge[enddots]++(-45:0.3);
\draw (b\x3)edge[enddots]++(135:0.3);
\draw (b\x3)edge[enddots]++(45:0.3);
}
\foreach \y in {0,1,2,3}{
\draw (0\y-c)edge[enddots]++(-135:0.3);
\draw (0\y-c)edge[enddots]++(135:0.3);
\draw (b3\y)edge[enddots]++(-45:0.3);
\draw (b3\y)edge[enddots]++(45:0.3);
}
\draw[red] (11-c)edge[endlabel=$a$]++(0,0.4);
\draw[red] (33-c)edge[endlabel=$b$]++(-0.4,0);
\draw[red] (32-c)edge[endlabel=$c$]++(0,-0.4);
\end{tikzpicture}
\end{equation}

Or, consider another example where the degrees of freedom are on the vertices of a square lattice. The Boltzmann weights are associated to the edges, and only depend on the degrees of freedom at both ends of the edge. Measurements are some function depending on the configuration of two neighboring degrees of freedom. The following shows a patch of the representing tensor network with two measurements (in red):
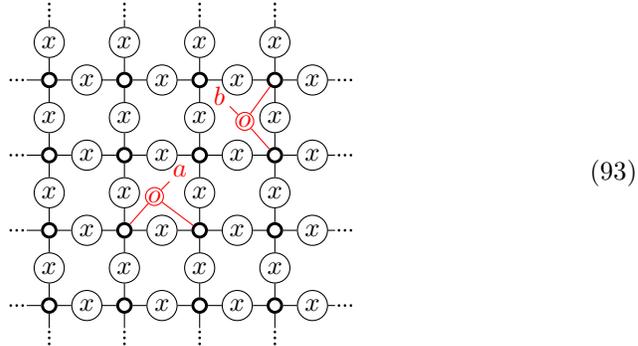
\begin{equation}
\begin{tikzpicture}
\foreach \x in {0,1,2,3}{
\foreach \y in {0,1,2,3}{
\node[draw,circle,minimum width=0.4cm,inner sep=1pt] (a\x\y) at (\x+0.5,\y) {$x$};
\node[draw,circle,minimum width=0.4cm,inner sep=1pt] (b\x\y) at (\x,\y+0.5) {$x$};
\pic (\x\y) at (\x,\y){smallcirc};
}}
\foreach \x in {0,1,2,3}{
\foreach \y in {0,1,2,3}{
\draw (\x\y-c)--(a\x\y);
\draw (\x\y-c)--(b\x\y);
}}
\foreach \x in {0,1,2,3}{
\foreach[count=\yy from 1] \y in {0,1,2}{
\draw (\x\yy-c)--(b\x\y);
}}
\foreach \y in {0,1,2,3}{
\foreach[count=\xx from 1] \x in {0,1,2}{
\draw (\xx\y-c)--(a\x\y);
}}
\foreach \x in {0,1,2,3}{
\draw (\x0-c)edge[enddots]++(-90:0.3);
\draw (0\x-c)edge[enddots]++(180:0.3);
\draw (b\x3)edge[enddots]++(90:0.3);
\draw (a3\x)edge[enddots]++(0:0.3);
}
\node[red,draw,circle,inner sep=0.1pt] (o1) at (1.4,1.45){$o$};
\draw[red] (11-c)--(o1) (21-c)--(o1) (o1)edge[endlabel=$a$]++(0.2,0.2);
\node[red,draw,circle,inner sep=0.1pt] (o1) at (2.6,2.45){$o$};
\draw[red] (33-c)--(o1) (32-c)--(o1) (o1)edge[endlabel=$b$]++(-0.2,0.2);

\end{tikzpicture}
\end{equation}
\end{myobs}

\section{Quantum physics as *-tensor models}
\label{sec:quantum}
In this section we will show that quantum models are a subset of *-algebra models.

\subsection{Quantum mechanical objects as *-tensors}
Let us start by representing different components of a quantum model as (normalized) positive *-tensors. In quantum mechanics, some of the *-algebras are the product of a matrix *-algebra with the complex number *-algebra. We will refer to this *-algebra as the \emph{quantum *-algebra}, and label it with a $Q$. We will refer to indices with the quantum *-algebra as \emph{quantum indices}. Quantum mechanical objects are represented by *-tensors with quantum and classical indices. For clarity we will use thicker lines for the quantum indices in this section.

\begin{itemize}
\item A \emph{density matrix} $\rho$ is a positive complex $n\times n$ matrix such that $\mathop{Tr}(\rho)=1$. It can be represented by a normalized positive *-tensor with one quantum output index. To this end, we take the real representation of $\rho$ (where we called the $x$-index from Obs.(\ref{obs:real_representation}) $c$), and then block the ket, bra and complex index:
\begin{equation}
\begin{tikzpicture}
\node[froundbox](0) at (0,0) {$\rho$};
\draw[quantum,ndir=+] (0)--++(0.5,0)node[right]{$abc$};
\end{tikzpicture}=
\begin{tikzpicture}
\node[roundbox,minimum height=0.6cm] (p) at (0,0) {$\rho$};
\draw ([yshift=0.15cm]p.east)edge[endlabel=$a$]++(0.3,0) ([yshift=-0.15cm]p.east)edge[endlabel=$b$]++(0.3,0);
\draw[complex] (p.north)edge[endlabel=$c$]++(0,0.3);
\end{tikzpicture}
\end{equation}
As we saw in Obs.~(\ref{obs:matrix_positivity}), the matrix being positive is in one-to-one correspondence with the *-tensor being positive. The normalization condition becomes:
\begin{equation}
\begin{tikzpicture}
\piclab{1}{circ}{(0.9,0)}{(0:0.4)}{$Q$};
\node[froundbox](0) at (0,0){$\rho$};
\draw[quantum] (0)edge[flage=-,ndir=+](1-c);
\end{tikzpicture}=
\begin{tikzpicture}
\piclab{c}{smallcirc}{(0,0.7)}{(0:0.3)}{$\mathbb{C}$};
\node[roundbox,minimum height=0.6cm] (p) at (0,0) {$\rho$};
\draw[rounded corners] ([yshift=0.15cm]p.east)--++(0.5,0)|-([yshift=-0.15cm]p.east);
\draw[complex] (p.north)edge[flage=-](c-c);
\end{tikzpicture}
=\mathop{Tr}(\rho)=
\hspace{1cm}
\end{equation}
i.e. the *-tensor is normalized.

If our physical system is multi-partite with $n$ degrees of freedom we can represent it as a normalized positive *-tensor with $n$ quantum output indices. To this end we first block the pairs of ket and bra index for each degree of freedom, and then take the realification of this complex tensor. This way each *-tensor index gets its own copy of the complex index. E.g. for a $3$-partite density matrix we get:
\begin{equation}
\begin{tikzpicture}
\node[froundbox,minimum height=1.5cm](r){$\rho$};
\draw[quantum,ndir=+] ([yshift=0.5cm]r.east)--++(0.5,0)node[right]{$aa'\alpha$};
\draw[quantum,ndir=+](r.east)--++(0.5,0)node[right]{$bb'\beta$};
\draw[quantum,ndir=+] ([yshift=-0.5cm]r.east)--++(0.5,0)node[right]{$cc'\gamma$};
\end{tikzpicture}=
\begin{tikzpicture}
\piclab{c}{smallcirc}{(0,1.1)}{(45:0.35)}{$\mathbb{C}$};
\node[roundbox,minimum height=1.5cm](r){$\rho$};
\draw ([yshift=0.6cm]r.east)--++(0.4,0)node[right]{$a$} ([yshift=0.4cm]r.east)--++(0.5,0)node[right]{$a'$} ([yshift=0.1cm]r.east)--++(0.4,0)node[right]{$b$} ([yshift=-0.1cm]r.east)--++(0.5,0)node[right]{$b'$} ([yshift=-0.4cm]r.east)--++(0.4,0)node[right]{$c$} ([yshift=-0.6cm]r.east)--++(0.5,0)node[right]{$c'$};
\draw[complex] (r.north)edge[flage=+](c-c) (c-c)edge[flags=+,endlabel=$\alpha$]++(180:0.4) (c-c)edge[flags=+,endlabel=$\beta$]++(90:0.4) (c-c)edge[flags=+,endlabel=$\gamma$]++(0:0.4);
\end{tikzpicture}
\end{equation}

A \emph{pure state} is a complex $n$-dimensional vector $\ket{\phi}$ that is normalized $\braket{\phi|\phi}=1$. Every pure state yields a density matrix and therefore a normalized positive *-tensor via:
\begin{equation}
\label{eq:pure_state}
\begin{tikzpicture}
\node[froundbox](0)at(0,0){$\rho$};
\draw[quantum,ndir=+] (0.east)--++(0.3,0)node[right]{$aa'\alpha$};
\end{tikzpicture}=
\begin{tikzpicture}
\piclab{c}{smallcirc}{(-0.6,1)}{(-40:0.3)}{$\mathbb{C}$};
\node[roundbox](0) at (0,0){$\phi$};
\node[roundbox](1) at (0,1){$\phi$};
\draw[rounded corners,complex,flage=+] (0.north)--++(0,0.2)-|(c-c);
\draw[rounded corners,complex,flage=-] (1.north)--++(0,0.2)-|(c-c);
\draw[complex] (c-c)edge[flags=+,endlabel=$\alpha$]++(-0.5,0);
\draw (0)edge[endlabel=$a'$]++(0.5,0) (1)edge[endlabel=$a$]++(0.5,0);
\end{tikzpicture}
\end{equation}

\item An \emph{ensemble} is a set of probabilities $p_i$ together with a set of density matrices $E_i$. If we regard $i$ as an additional index then the collection $E_i$ defines a $3$-index complex tensor. Its real representation yields a normalized positive *-tensor with one classical input and one quantum output index:
\begin{equation}
\begin{tikzpicture}
\node[froundbox](0)at (0,0){$E$};
\draw[quantum,ndir=+] (0.east)--++(0.3,0)node[right]{$abc$};
\draw[star] (0.west)edge[ndir=-,endlabel=$i$]++(-0.3,0);
\end{tikzpicture}=
\begin{tikzpicture}
\node[roundbox,minimum height=0.6cm] (p) at (0,0) {$E$};
\draw ([yshift=0.15cm]p.east)edge[endlabel=$a$]++(0.3,0) ([yshift=-0.15cm]p.east)edge[endlabel=$b$]++(0.3,0);
\draw[complex] (p.north)edge[endlabel=$c$]++(0,0.3);
\draw (p.west)edge[endlabel=$i$]++(-0.3,0);
\end{tikzpicture}
\end{equation}
This is a *-tensor with a classical index $i$ and the other three indices composing to one quantum index. The condition that all the $\rho_i$ are positive individually corresponds to the *-tensor being positive. The condition that the $\rho_i$ all have trace $1$ translates to:
\begin{equation}
\begin{gathered}
\begin{tikzpicture}
\piclab{1}{circ}{(0.9,0)}{(0:0.4)}{$Q$};
\node[froundbox](0) at (0,0){$E$};
\draw[] (0.east)edge[quantum,ndirpos=0.3,ndir=+,flage=-](1-c) (0.west)edge[ndir=-,endlabel=$i$,classical]++(-0.3,0);
\end{tikzpicture}=
\begin{tikzpicture}
\piclab{c}{smallcirc}{(0,0.7)}{(0:0.3)}{$\mathbb{C}$};
\node[roundbox,minimum height=0.6cm] (p) at (0,0) {$E$};
\draw[rounded corners] ([yshift=0.15cm]p.east)--++(0.5,0)|-([yshift=-0.15cm]p.east) (p.west)edge[endlabel=$i$]++(-0.3,0);
\draw[complex] (p.north)edge[flage=-](c-c);
\end{tikzpicture}=
\mathop{Tr}(E_i)\\
=1\forall i=
\begin{tikzpicture}
\piclab{0}{smallcirc}{(0,0)}{(0:0.3)}{$\delta$};
\draw (0-c)edge[endlabel=$i$]++(-0.5,0);
\end{tikzpicture}
\end{gathered}
\end{equation}
I.e. the *-tensor $E$ is normalized.

According to the previous section, the probability distribution $p_i$ defines a normalized positive *-tensor with one classical output index. If we contract this index with the classical input index of $E$ we obtain the density matrix $\rho$ corresponding to the ensemble:
\begin{equation}
\begin{tikzpicture}
\node[froundbox](0)at(0,0){$\rho$};
\draw[star] (0)edge[quantum,ndir=+,endlabel=$x$]++(0.5,0);
\end{tikzpicture}=
\begin{tikzpicture}
\node[froundbox](p)at(0,0){$p$};
\node[froundbox](e)at(0.8,0){$E$};
\draw[star] (p.east)edge[ndir=+](e.west) (e.east)edge[ndir=+,quantum,endlabel=$x$]++(0.3,0);
\end{tikzpicture}
\end{equation}

\item The \emph{trace} is the unique normalized (positive) *-algebra tensor with only one quantum input index.
\begin{equation}
\begin{tikzpicture}
\node[froundbox](0)at (0,0){$\mathop{Tr}$};
\draw[quantum,ndir=-] (0.west)--++(-0.3,0)node[left]{$aa'\alpha$};
\end{tikzpicture}=
\begin{tikzpicture}
\piclab{0}{circ}{(0,0)}{(0:0.4)}{$Q$};
\draw[quantum,flags=+] (0-c)--++(-0.5,0)node[left]{$aa'\alpha$};
\end{tikzpicture}=
\begin{tikzpicture}
\piclab{0}{smallcirc}{(0.5,-0.4)}{(0:0.3)}{$\mathbb{C}$};
\draw[rounded corners,startlabel=$a$,endlabel=$a'$] (0,0)--++(0.5,0)--++(0,0.4)--++(-0.5,0);
\draw[endlabel=$\alpha$,flags=+] (0-c)--++(-0.5,0);
\end{tikzpicture}
\end{equation}
Note that the first equation is exactly the normalization condition for a normalized *-tensor with only one output index.

\item A \emph{quantum channel} is a \emph{completely positive trace-preserving} linear map from density matrices to density matrices. It can represented by a normalized positive *-tensor with one quantum input index and one quantum output index by realification:
\begin{equation}
\begin{tikzpicture}
\node[froundbox](0)at (0,0){$C$};
\draw[quantum,ndir=+] (0.east)--++(0.3,0)node[right]{$bb'\beta$};
\draw[quantum,ndir=-] (0.west)--++(-0.3,0)node[left]{$aa'\alpha$};
\end{tikzpicture}=
\begin{tikzpicture}
\piclab{c}{smallcirc}{(0,0.7)}{(90:0.3)}{$\mathbb{C}$};
\node[roundbox,minimum height=0.6cm] (p) at (0,0) {$C$};
\draw ([yshift=0.15cm]p.east)edge[endlabel=$b$]++(0.3,0) ([yshift=-0.15cm]p.east)edge[endlabel=$b'$]++(0.3,0);
\draw ([yshift=0.15cm]p.west)edge[endlabel=$a$]++(-0.3,0) ([yshift=-0.15cm]p.west)edge[endlabel=$a'$]++(-0.3,0);
\draw[complex] (p.north)edge[flage=+](c-c) (c-c)edge[flags=+,endlabel=$\alpha$]++(-0.4,0) (c-c)edge[flags=-,endlabel=$\beta$]++(0.4,0);
\end{tikzpicture}
\end{equation}
Complete positivity means that for all positive (density) matrices $\rho$ on an extended Hilbert space $H\otimes H'$, $\widetilde{\rho}=(C\otimes \mathbb{1})\rho$ is positive. In particular this holds also for $\rho$ being the maximally entangled state:
\begin{equation}
\begin{gathered}
\rho=\frac{1}{\mathop{dim}(H)} \sum_{i,j} \ket{i}\bra{j} \otimes \ket{i}\bra{j}\\
\begin{tikzpicture}
\node[roundbox,minimum height=1.2cm](r){$\rho$};
\draw ([yshift=0.4cm]r.east)--++(0.4,0)node[right]{$a$} ([yshift=0.2cm]r.east)--++(0.5,0)node[right]{$a'$} ([yshift=-0.2cm]r.east)--++(0.4,0)node[right]{$b$} ([yshift=-0.4cm]r.east)--++(0.5,0)node[right]{$b'$};
\draw[complex] (r.north)edge[endlabel=$\alpha$]++(0,0.3);
\end{tikzpicture}=
\left(\mathop{dim}(H)^{-1}\right)
\begin{tikzpicture}
\piclab{c}{smallcirc}{(-0.5,-0.4)}{(180:0.3)}{$\mathbb{C}$};
\draw[rounded corners,startlabel=$b'$,endlabel=$a'$] (0,0)--++(-0.5,0)--++(0,0.5)--++(0.5,0);
\draw[rounded corners,startlabel=$b$,endlabel=$a$] (-0.1,0.2)--++(-0.5,0)--++(0,0.5)--++(0.5,0);
\draw[complex] (c-c)edge[flags=-,endlabel=$\alpha$]++(0.5,0);
\end{tikzpicture}
\end{gathered}
\end{equation}
We find
\begin{equation}
\begin{tikzpicture}
\node[roundbox,minimum height=1.2cm](r)at(0,0){$\widetilde{\rho}$};
\draw ([yshift=0.4cm]r.east)--++(0.4,0)node[right]{$a$} ([yshift=0.2cm]r.east)--++(0.5,0)node[right]{$a'$} ([yshift=-0.2cm]r.east)--++(0.4,0)node[right]{$b$} ([yshift=-0.4cm]r.east)--++(0.5,0)node[right]{$b'$};
\draw[complex] (r.north)edge[endlabel=$\alpha$]++(0,0.3);
\end{tikzpicture}=
\begin{tikzpicture}
\piclab{co}{smallcirc}{(0.8,0.9)}{(180:0.3)}{$\mathbb{C}$};
\node[roundbox,minimum height=1.2cm](r)at(0,0){$\rho$};
\node[roundbox,minimum height=0.5cm](c)at(0.8,0.3){$C$};
\draw ([yshift=0.1cm]c.east)--++(0.4,0)node[right]{$a$} ([yshift=-0.1cm]c.east)--++(0.5,0)node[right]{$a'$};
\draw ([yshift=0.4cm]r.east)--([yshift=0.1cm]c.west) ([yshift=0.2cm]r.east)--([yshift=-0.1cm]c.west) ([yshift=-0.2cm]r.east)--++(0.4,0)node[right]{$b$} ([yshift=-0.4cm]r.east)--++(0.5,0)node[right]{$b'$};
\draw[complex] (c.north)edge[flage=+](co-c) (co-c)edge[flags=-,endlabel=$\alpha$]++(0.4,0);
\draw[complex,rounded corners,flage=-] (r.north)--++(0,0.7)-|(co-c);
\end{tikzpicture}=
\left(\mathop{dim}(H)^{-1}\right)
\begin{tikzpicture}
\node[roundbox,minimum height=1cm] (p) at (0,0) {$C$};
\draw ([yshift=0.2cm]p.east)edge[endlabel=$a$]++(0.3,0) ([yshift=-0.2cm]p.east)edge[endlabel=$a'$]++(0.3,0);
\draw ([yshift=0.2cm]p.west)edge[endlabel=$b$]++(-0.3,0) ([yshift=-0.2cm]p.west)edge[endlabel=$b'$]++(-0.3,0);
\draw[complex] (p.north)edge[endlabel=$\alpha$]++(0,0.3);
\end{tikzpicture}
\end{equation}
I.e., $C$ is positive as a complex matrix with indices $(a,b)$ and $(a',b')$. So complete positivity implies that the *-tensor $C$ is positive. On the other hand, $C$ being a positive *-tensor immediately implies complete positivity due to Obs.~(\ref{obs:positive_tensor_network}).

The trace-preserving property of the channel $C$ implies that the corresponding *-tensor is normalized (with one in and one out index):
\begin{equation}
\begin{gathered}
\begin{tikzpicture}
\piclab{s}{circ}{(0.9,0)}{(0:0.4)}{$Q$};
\node[froundbox](0)at (0,0){$C$};
\draw[quantum,ndir=+,ndirpos=0.3,flage=-] (0.east)--(s-c);
\draw[quantum,ndir=-] (0.west)--++(-0.3,0)node[left]{$aa'\alpha$};
\end{tikzpicture}=
\begin{tikzpicture}
\piclab{c}{smallcirc}{(0,0.7)}{(90:0.3)}{$\mathbb{C}$};
\piclab{c2}{smallcirc}{(0.7,0.7)}{(0:0.3)}{$\mathbb{C}$};
\node[roundbox,minimum height=0.6cm] (p) at (0,0) {$C$};
\draw[rounded corners] ([yshift=0.15cm]p.east)--++(0.5,0)|-([yshift=-0.15cm]p.east);
\draw ([yshift=0.15cm]p.west)edge[endlabel=$a$]++(-0.3,0) ([yshift=-0.15cm]p.west)edge[endlabel=$a'$]++(-0.3,0);
\draw[complex] (p.north)edge[flage=+](c-c) (c-c)edge[flags=+,endlabel=$\alpha$]++(-0.4,0) (c-c)edge[flags=-,flage=-](c2-c);
\end{tikzpicture}\\=
\begin{tikzpicture}
\node[roundbox,minimum height=0.6cm] (p) at (0,0) {$C$};
\draw[rounded corners] ([yshift=0.15cm]p.east)--++(0.5,0)|-([yshift=-0.15cm]p.east);
\draw ([yshift=0.15cm]p.west)edge[endlabel=$a$]++(-0.3,0) ([yshift=-0.15cm]p.west)edge[endlabel=$a'$]++(-0.3,0);
\draw[complex] (p.north)edge[endlabel=$\alpha$]++(0,0.3);
\end{tikzpicture}=
\begin{tikzpicture}
\piclab{0}{smallcirc}{(0.5,-0.4)}{(0:0.3)}{$\mathbb{C}$};
\draw[rounded corners,startlabel=$a$,endlabel=$a'$] (0,0)--++(0.5,0)--++(0,0.4)--++(-0.5,0);
\draw[endlabel=$\alpha$,flags=+] (0-c)--++(-0.5,0);
\end{tikzpicture}=
\begin{tikzpicture}
\piclab{0}{circ}{(0,0)}{(0:0.4)}{$Q$};
\draw[quantum,flags=+] (0-c)--++(-0.5,0)node[left]{$aa'\alpha$};
\end{tikzpicture}
\end{gathered}
\end{equation}

\item A \emph{unitary operator} $U$ is a $2$-index complex tensor yielding a quantum channel via $C=U\otimes U^\dagger$. The *-tensor corresponding to $C$ can be built from $U^{(r)}$ in the following way:
\begin{equation}
\label{eq:unitary}
\begin{tikzpicture}
\node[froundbox](0)at (0,0){$C$};
\draw[quantum,ndir=+] (0.east)--++(0.3,0)node[right]{$bb'\beta$};
\draw[quantum,ndir=-] (0.west)--++(-0.3,0)node[left]{$aa'\alpha$};
\end{tikzpicture}=
\begin{tikzpicture}
\piclab{c}{smallcirc}{(-1,0.9)}{(-135:0.3)}{$\mathbb{C}$};
\node[roundbox](0) at (0,0){$U$};
\node[roundbox](1) at (0,1.2){$U$};
\draw (0.east)edge[endlabel=$b'$]++(0.2,0) (1.east)edge[endlabel=$b$]++(0.2,0);
\draw (0.west)edge[endlabel=$a'$]++(-0.2,0) (1.west)edge[endlabel=$a$]++(-0.2,0);
\draw[rounded corners,complex,flage=+] (0.north)--++(0,0.2)-|(c-c);
\draw[rounded corners,complex,flage=+] (1.north)--++(0,0.2)-|(c-c);
\draw[complex] (c-c)edge[flags=+,endlabel=$\beta$]++(0.4,0) (c-c)edge[flags=-,endlabel=$\alpha$]++(-0.4,0);
\end{tikzpicture}
\end{equation}

\item A \emph{POVM} (projection operator-valued measurement) is a set of complex positive matrices $P_i$ that sum to the identity, $\sum_i P_i=\mathbb{1}$. We can view a POVM as a tensor with two matrix indices, one index labeling $i$ and one complex index. After blocking, we get a *-tensor with one quantum and one classical index:
\begin{equation}
\begin{tikzpicture}
\node[froundbox](0) at (0,0){$P$};
\draw[quantum,ndir=-] (0.west)--++(-0.4,0)node[left]{$aa'\alpha$};
\draw[star,classical] (0.east)edge[ndir=+,endlabel=$i$]++(0.4,0);
\end{tikzpicture}=
\begin{tikzpicture}
\node[roundbox,minimum height=0.6cm] (p) at (0,0) {$P$};
\draw ([yshift=0.15cm]p.west)edge[endlabel=$a$]++(-0.3,0) ([yshift=-0.15cm]p.west)edge[endlabel=$a'$]++(-0.3,0);
\draw[complex] (p.north)edge[endlabel=$\alpha$]++(0,0.3);
\draw (p.east)edge[endlabel=$i$]++(0.3,0);
\end{tikzpicture}
\end{equation}
According to Obs.~(\ref{obs:matrix_positivity}), the condition that all the $P_i$ are positive matrices individually corresponds to the *-tensor being positive. The condition that the $P_i$ sum to $\mathbb{1}$ yields that the *-tensor is normalized with one quantum in index and one classical out index:
\begin{equation}
\begin{gathered}
\begin{tikzpicture}
\piclab{1}{smallcirc}{(0.9,0)}{(0:0.3)}{$\delta$};
\node[froundbox](0) at (0,0){$P$};
\draw[star,classical] (0.east)edge[ndirpos=0.3,ndir=+](1-c);
\draw[quantum,ndir=-] (0.west)--++(-0.3,0)node[left]{$aa'\alpha$};
\end{tikzpicture}=
\begin{tikzpicture}
\piclab{0}{smallcirc}{(0.6,0)}{(0:0.3)}{$\delta$};
\node[roundbox,minimum height=0.6cm] (p) at (0,0) {$P$};
\draw ([yshift=0.15cm]p.west)edge[endlabel=$a$]++(-0.3,0) ([yshift=-0.15cm]p.west)edge[endlabel=$a'$]++(-0.3,0);
\draw[complex] (p.north)edge[endlabel=$\alpha$]++(0,0.3);
\draw (p.east)--(0-c);
\end{tikzpicture}\\
=
\begin{tikzpicture}
\piclab{0}{smallcirc}{(0.5,-0.4)}{(0:0.3)}{$\mathbb{C}$};
\draw[rounded corners,startlabel=$a$,endlabel=$a'$] (0,0)--++(0.5,0)--++(0,0.4)--++(-0.5,0);
\draw[endlabel=$\alpha$,flags=+] (0-c)--++(-0.5,0);
\end{tikzpicture}=
\begin{tikzpicture}
\piclab{0}{circ}{(0,0)}{(0:0.4)}{$Q$};
\draw[quantum,flags=+] (0-c)--++(-0.5,0)node[left]{$aa'\alpha$};
\end{tikzpicture}
\end{gathered}
\end{equation}

\item A \emph{projective measurement} of an observable $O$ is a special case of POVM where the $P_i$ are the complete set of orthogonal projectors onto the eigenspaces of the hermitian matrix $O$. We can diagonalize $O=U\Lambda U^\dagger$ for a diagonal matrix $\Lambda$ and a unitary $U$. In this case the *-tensor is given by:
\begin{equation}
\label{eq:proj_measurement}
\begin{tikzpicture}
\node[froundbox](0) at (0,0){$P$};
\draw[quantum,ndir=-] (0.west)--++(-0.4,0)node[left]{$aa'\alpha$};
\draw[star,classical] (0.east)edge[ndir=+,endlabel=$i$]++(0.4,0);
\end{tikzpicture}=
\begin{tikzpicture}
\piclab{c}{smallcirc}{(-0.9,0.9)}{(0:0.3)}{$\mathbb{C}$};
\piclab{d}{smallcirc}{(0.5,0.5)}{(180:0.3)}{$\delta$};
\node[roundbox](0) at (0,0){$U$};
\node[roundbox](1) at (0,1.2){$U$};
\draw (0.west)edge[endlabel=$a'$]++(-0.2,0) (1.west)edge[endlabel=$a$]++(-0.2,0);
\draw[rounded corners,complex,flage=+] (0.north)--++(0,0.2)-|(c-c);
\draw[rounded corners,complex,flage=+] (1.north)--++(0,0.2)-|(c-c);
\draw[complex] (c-c)edge[flags=-,endlabel=$\alpha$]++(-0.4,0);
\draw[rounded corners] (0.east)-|(d-c) (1.east)-|(d-c) (d-c)edge[endlabel=$i$]++(0.4,0);
\end{tikzpicture}
\end{equation}

\item An \emph{instrument} is a set of completely positive maps $I_i$, whose sum $C=\sum_i I_i$ is trace-preserving. It is represented by a real tensor similar a quantum channel, just that it has the additional index $i$. It can be represented by a *-tensor in the following way:
\begin{equation}
\begin{tikzpicture}
\node[froundbox,minimum height=0.8cm](0)at (0,0){$I$};
\draw[quantum,ndir=+] ([yshift=-0.2cm]0.east)--++(0.3,0)node[right]{$bb'\beta$};
\draw[star,ndir=+,classical] ([yshift=0.2cm]0.east)--++(0.3,0)node[right]{$i$};
\draw[quantum,ndir=-] (0.west)--++(-0.3,0)node[left]{$aa'\alpha$};
\end{tikzpicture}=
\begin{tikzpicture}
\piclab{c}{smallcirc}{(0,1)}{(90:0.3)}{$\mathbb{C}$};
\node[roundbox,minimum height=1cm] (p) at (0,0) {$I$};
\draw ([yshift=0.05cm]p.east)edge[endlabel=$b$]++(0.3,0) ([yshift=-0.35cm]p.east)edge[endlabel=$b'$]++(0.3,0) ([yshift=0.35cm]p.east)edge[endlabel=$i$]++(0.3,0);
\draw ([yshift=0.15cm]p.west)edge[endlabel=$a$]++(-0.3,0) ([yshift=-0.15cm]p.west)edge[endlabel=$a'$]++(-0.3,0);
\draw[complex] (p.north)edge[flage=+](c-c) (c-c)edge[flags=+,endlabel=$\alpha$]++(-0.4,0) (c-c)edge[flags=-,endlabel=$\beta$]++(0.4,0);
\end{tikzpicture}
\end{equation}
The *-tensor is positive as all the $C_i$ are positive complex matrices. The *-tensor is normalized with one quantum input index, one quantum output index, and one classical output index.

\item A \emph{controlled operation} is a set of completely positive trace-preserving maps $O_i$. Just as for instruments, such a collection defines a *-tensor by:
\begin{equation}
\begin{tikzpicture}
\node[froundbox,minimum height=0.8cm](0)at (0,0){$O$};
\draw[quantum,ndir=+] (0.east)--++(0.3,0)node[right]{$bb'\beta$};
\draw[star,ndir=-,classical] ([yshift=0.2cm]0.west)--++(-0.3,0)node[left]{$i$};
\draw[quantum,ndir=-] ([yshift=-0.2cm]0.west)--++(-0.3,0)node[left]{$aa'\alpha$};
\end{tikzpicture}=
\begin{tikzpicture}
\piclab{c}{smallcirc}{(0,1)}{(90:0.3)}{$\mathbb{C}$};
\node[roundbox,minimum height=1cm] (p) at (0,0) {$O$};
\draw ([yshift=0.15cm]p.east)edge[endlabel=$b$]++(0.3,0) ([yshift=-0.15cm]p.east)edge[endlabel=$b'$]++(0.3,0);
\draw ([yshift=0.05cm]p.west)edge[endlabel=$a$]++(-0.3,0) ([yshift=-0.35cm]p.west)edge[endlabel=$a'$]++(-0.3,0) ([yshift=0.35cm]p.west)edge[endlabel=$i$]++(-0.3,0);
\draw[complex] (p.north)edge[flage=+](c-c) (c-c)edge[flags=+,endlabel=$\alpha$]++(-0.4,0) (c-c)edge[flags=-,endlabel=$\beta$]++(0.4,0);
\end{tikzpicture}
\end{equation}
Again, this *-tensor is positive as all the $O_i$ are positive individually. It is normalized with one quantum input index, one quantum output index, and one classical input index.

\item An \emph{anti-unitary} $A$ is an anti-linear operation over a complex vector space such that $AA^\dagger=\mathbb{1}$. For a fixed basis, every anti-unitary can be written as $A=\kappa \overline{A}$, where $\overline{A}$ is a unitary matrix and $\kappa$ is entry-wise complex conjugation with respect to the fixed basis. Given this unitary matrix $\overline{A}$ one can construct the following *-tensor:
\begin{equation}
\begin{tikzpicture}
\node[froundbox](0)at (0,0){$A$};
\draw[quantum,ndir=+] (0.east)--++(0.3,0)node[right]{$bb'\beta$};
\draw[quantum,ndir=-] (0.west)--++(-0.3,0)node[left]{$aa'\alpha$};
\end{tikzpicture}=
\begin{tikzpicture}
\piclab{c}{smallcirc}{(-1,0.9)}{(135:0.3)}{$\mathbb{C}$};
\node[roundbox](0) at (0,0){$\overline{A}$};
\node[roundbox](1) at (0,1.2){$\overline{A}$};
\draw (0.east)edge[endlabel=$b'$]++(0.2,0) (1.east)edge[endlabel=$b$]++(0.2,0);
\draw (0.west)edge[endlabel=$a'$]++(-0.2,0) (1.west)edge[endlabel=$a$]++(-0.2,0);
\draw[rounded corners,complex,flage=+] (0.north)--++(0,0.2)-|(c-c);
\draw[rounded corners,complex,flage=+] (1.north)--++(0,0.2)-|(c-c);
\draw[complex] (c-c)edge[flags=+,endlabel=$\beta$]++(0.4,0) (c-c)edge[flags=+,endlabel=$\alpha$]++(-0.4,0);
\end{tikzpicture}
\end{equation}
Note that the input and output indices $\alpha$ and $\beta$ have equal complex *-orientations, corresponding to the anti-linear behavior. Analogous to the unitary case, this *-tensor is positive and normalized with one quantum in and one quantum out index.

\item A \emph{quantum Boltzmann weight} for a Hamiltonian $H$ at temperature $\beta$ is the positive complex matrix $e^{-\beta H}$. We can represent it by a *-tensor:
\begin{equation}
\begin{tikzpicture}
\node[froundbox](0) at (0,0) {$W$};
\draw[quantum] (0.east)--++(0.4,0)node[right]{$(a,a',\alpha)$};
\end{tikzpicture}=
\begin{tikzpicture}
\node[roundbox,minimum height=0.6cm] (p) at (0,0) {$e^{-\beta H}$};
\draw ([yshift=0.15cm]p.east)edge[endlabel=$a$]++(0.3,0) ([yshift=-0.15cm]p.east)edge[endlabel=$a'$]++(0.3,0);
\draw[complex] (p.north)edge[endlabel=$\alpha$]++(0,0.3);
\end{tikzpicture}
\end{equation}
Positivity of the matrix yields that this *-tensor is positive. Note that this *-tensor is not normalized.
\end{itemize}

\begin{mycom}
Traditionally, complete positivity of superoperators is defined via positivity: A superoperator is positive if it maps density matrices to density matrices. It is complete positive if it is positive after tensoring with the identity acting on an arbitrary auxiliary system. We see that this is a very unfortunate way if defining it as it suggests that complete positivity is a somewhat more advanced concept than positivity. But really, complete positivity is just positivity of the representing tensor shaped into a matrix in the right way, whereas positivity alone is a quite unnatural property that is unconstructive and hard to assert in addition.
\end{mycom}

\subsection{Typical setups as *-tensor networks}
Let us consider different setups for quantum physics and how they can be translated into *-tensor networks.
\subsubsection*{Causal setups}
\begin{myobs}
A typical setup for a quantum models looks like the following: First we prepare an initial state $\ket{\phi}$, then we let it evolve for some time $t$ under a Hamiltonian $H$, and then we measure some observable $O$. Such a setup can be easily represented by a causal *-tensor setup: Via Eq.~(\ref{eq:pure_state}) we get a *-tensor $\rho$ from $\ket{\phi}$, via Eq.~(\ref{eq:unitary}) we get a *-tensor $C$ from the unitary time evolution $U=e^{itH}$, and via Eq.~(\ref{eq:proj_measurement}) we get a *-tensor $P$ from the projective measurement of the observable $O$. The corresponding *-tensor network is given by:
\begin{equation}
\begin{tikzpicture}
\node[froundbox] (r) at (0,0){$\rho$};
\node[froundbox] (c) at (0.8,0){$C$};
\node[froundbox] (p) at (1.6,0){$P$};
\draw[star,quantum] (r)edge[ndir=+](c) (c)edge[ndir=+](p);
\draw[star,classical] (p.east)edge[ndir=+,endlabel=$i$]++(0.4,0);
\end{tikzpicture}
\end{equation}
The evaluation of this tensor-network yields the probability distribution of measurement outcomes.
\end{myobs}

\begin{myobs}
Let us consider a slightly more complex setup: Before measuring the observable $O$ we perform a controlled operation $U$. Then we get the following causal *-tensor setup:
\begin{equation}
\begin{tikzpicture}
\node[froundbox] (r) at (0,0){$\rho$};
\node[froundbox] (c) at (0.8,0){$C$};
\node[froundbox] (u) at (1.6,0){$U$};
\node[froundbox] (p) at (2.4,0){$P$};
\draw[star,quantum] (r)edge[ndir=+](c) (c)edge[ndir=+](u) (u)edge[ndir=+](p);
\draw[star,classical] (p.east)edge[ndir=+,endlabel=$i$]++(0.4,0) (u.north)edge[ndir=-,endlabel=$j$]++(0,0.4);
\end{tikzpicture}
\end{equation}
The evaluation of this *-tensor network yields a stochastic map from $j$ to $i$.
\end{myobs}

\begin{myobs}
Next consider the following setup: Prepare one state of an ensemble, perform a measurement, perform an operation dependent on that measurement outcome, and then perform another measurement. We get the following causal *-tensor setup:
\begin{equation}
\begin{tikzpicture}
\piclab{d}{smallcirc}{(1.2,0.5)}{(45:0.3)}{$\delta$};
\node[froundbox] (e) at (0,0){$E$};
\node[froundbox] (i) at (0.8,0){$I$};
\node[froundbox] (u) at (1.6,0){$U$};
\node[froundbox] (p) at (2.4,0){$P$};
\draw[star,quantum] (e)edge[ndir=+](i) (i)edge[ndir=+](u) (u)edge[ndir=+](p);
\draw[star,classical,rounded corners] (e.west)edge[ndir=-,endlabel=$i$]++(-0.4,0) (p.east)edge[ndir=+,endlabel=$j$]++(0.4,0) (d-c)edge[ndir=+,endlabel=$k$]++(0,0.4) (i)edge[|-,ndir=+](d-c) (d-c)edge[-|,ndir=+](u);
\end{tikzpicture}
\end{equation}
The evaluation of this *-tensor network is a classical stochastic map from $i$ to $k$ and $j$. Physically it corresponds to the joint probability distributions over the measurement outcomes $k$ and $j$ for the different input states $i$.
\end{myobs}

\begin{myobs}
Consider the setup of a Bell test: We start with one state $\ket{\phi}$ of a 2-partite system, then we perform a controlled measurement in each of the two systems:
\begin{equation}
\begin{tikzpicture}
\node[froundbox,minimum height=1cm](0)at (0,0){$\rho$};
\node[froundbox](a) at (1,0.8){$M_1$};
\node[froundbox](b) at (1,-0.8){$M_2$};
\draw[star,quantum] ([yshift=0.2cm]0.east)edge[ndir=+]([yshift=-0.1cm]a.west) ([yshift=-0.2cm]0.east)edge[ndir=+]([yshift=0.1cm]b.west);
\draw[star,classical] ([yshift=0.1cm]a.west)edge[ndir=-,endlabel=$i$]++(-0.4,0) (a.east)edge[ndir=+,endlabel=$a$]++(0.4,0) ([yshift=-0.1cm]b.west)edge[ndir=-,endlabel=$j$]++(-0.4,0) (b.east)edge[ndir=+,endlabel=$b$]++(0.4,0);
\end{tikzpicture}
\end{equation}
The evaluation of this *-tensor network yields a classical stochastic map from $i,j$ to $a,b$. Not all possible stochastic maps are the evaluation of such a tensor network for some $\rho$, $M_1$ and $M_2$. For example, due to causality, after tracing out $a$, $i$ and $b$ are uncorrelated. The Bell inequality shows that if we replace all quantum indices by classical ones and take a classical probability distribution instead of $\rho$, the set of resulting stochastic maps will be smaller than in the quantum case.
\end{myobs}

\begin{myobs}
A quantum circuit is a *-tensor network. Therefore we just replace every state preparation, quantum channel and measurement by the corresponding tensor.
\end{myobs}

\begin{myobs}
A continuous time evolution of a quantum system can be turned into a *-tensor network by picking an ``atomic'' time interval $\delta t$ and the according quantum channel $\delta$ performing the time evolution for the unitary $\delta U=e^{i \delta t H}$. Then, e.g., evolviong a state $\rho$ and performing measurements by an instrument $I$ after $9\delta t$ and $12\delta t$ corresponds to the following causal *-tensor network:
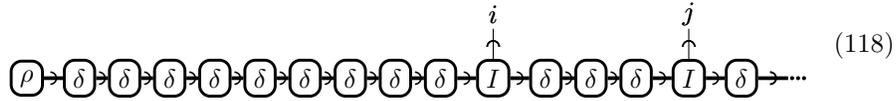
\begin{equation}
\begin{tikzpicture}
\node[froundbox] (0) at (-0.7,0){$\rho$};
\foreach \i [count=\ii from 1] in {0,...,8}{
\node[froundbox] (\ii) at (\i*0.6,0) {$\delta$};
\draw[star,quantum,ndir=+] (\i.east)--(\ii.west);
}
\node[froundbox] (i0) at (8*0.6+0.7,0){$I$};
\draw[star,quantum,ndir=+] (9.east)--(i0.west);
\foreach \i [count=\ii from 1] in {0,...,2}{
\node[froundbox] (i\ii) at (\i*0.6+8*0.6+1.4,0) {$\delta$};
\draw[star,quantum,ndir=+] (i\i.east)--(i\ii.west);
}
\node[froundbox] (ii) at ($(i3)+(0.7,0)$) {$I$};
\node[froundbox] (d) at ($(ii)+(0.7,0)$) {$\delta$};
\draw[star,quantum,ndir=+] (i3.east)--(ii.west);
\draw[star,quantum,ndir=+] (ii.east)--(d.west);
\draw[star,quantum,ndir=+,enddots] (d.east)--++(0.4,0);
\draw[star,classical] (i0.north)edge[ndir=+,endlabel=$i$]++(0,0.4) (ii.north)edge[ndir=+,endlabel=$j$]++(0,0.4);
\end{tikzpicture}
\end{equation}
\end{myobs}

\begin{myobs}
\label{obs:hamiltonian_discretisation}
A continuous (imaginary) time evolution of a quantum many-body system with a local translation-invariant Hamiltonian can be approximated by a tensor network using \emph{Trotterization}. To this end, we perform the following steps:
\begin{enumerate}
\item Pick an ``atomic'' time interval $\Delta t$ and a spacial unit cell.
\item Divide the Hamiltonian terms into a constant (system-size independent) number of non-overlapping sets $I_l$. The Hamiltonian terms within each of those sets commute, such that the time evolution with respecto to a single subset $I_l$ is given by the product
\begin{equation}
U_l(t)=e^{it H_l}=\prod_{i\in I_l} e^{it H_i}=\prod_{i\in I_l} U_i(t)
\end{equation}
\item As the layers overlap they do not in general commute with each other, so $U(\Delta t)\neq \prod_l U_l(\Delta t)$. The same becomes true approximately for small time steps though. So we can approximate the time evolution by $\Delta t$ by dividing it into $n$ steps of size $\Delta t/n$:
\begin{equation}
U(\Delta t)\overset{n\rightarrow\infty}{=} \left(\prod_{l}U_l(\Delta t/n)\right)^n+\mathcal{O}(1/n)= \left(\prod_{l}\prod_{i\in I_l} U_i(\Delta t/n)\right)^n+\mathcal{O}(1/n)
\end{equation}
\item For a fixed $n$, the product of operators on the right hand side yields a tensor network over space, whose combinatorial thickness in time direction is of order $n$.
\item Choose a way to cut this tensor network into patches with the size of the spacial unit cell times $\Delta t$. Evaluate this patch of tensor network and block the indices along the time direction. Regard the sequence of such tensors for larger and larger $n$. \label{itm:block}
\item The blocked indices of those tensors will have a basis that increases exponentially with $n$. We can truncate this basis by projecting to a sub-vector space. We can chose this projection such that the approximation error is as small as possible. This minimal approximation error will decrease when we increase the aimed basis size. We conjecture the following: For a fixed basis size the approximation error stays under a constant threshold, independent of $n$. This threshold converges to $0$ as we increase the basis size. (Practically we often find that it converges to $0$ exponentially quickly.)
\item The tensor network consisting of the truncated blocked tensors for high values of $n$ approximates the continuous (imaginary) time evolution.
\item This tensor network describes time evolution on a pure-state level. If we double the tensor network we get a positive *-tensor network describing the time evolution on the density matrix level.
\end{enumerate}
Let's consider as a simple example a $1$-dimensional quantum spin chain with a translation-invariant nearest-neighbor Hamiltonian. We can divide the Hamiltonian term into two non-overlapping sets, consisting of the terms at the odd and even sites, respectively. Choose a spacial unit cell consisting of two sites, and take as patch all the $U_i(\Delta t/n)$ acting one of the two sites and the one to its right. Then evaluate this tensor network:
\begin{equation}
\begin{tikzpicture}
\node[roundbox,minimum height=0.6cm,minimum width=0.6cm] (t) at (0,0){$P_n$};
\draw (t.north)--++(0,0.4) node[above]{$v'w'$};
\draw (t.south)--++(0,-0.4)node[below]{$vw$};
\draw (t.west)--++(-0.4,0)node[rotate=90,above]{$abcdef$};
\draw (t.east)--++(0.4,0)node[rotate=90,below]{$a'b'c'd'e'f'$};
\end{tikzpicture}=
\begin{tikzpicture}
\tikzset{xtens/.style={roundbox,minimum width=0.9cm,minimum height=0.4cm}};
\node[xtens] (t0) at (0,0){$x$};
\node[xtens] (t1) at (0.5,0.6){$x$};
\node[xtens] (t2) at (0,1.2){$x$};
\node[xtens] (t3) at (0.5,1.8){$x$};
\node[xtens] (t4) at (0,2.4){$x$};
\node[xtens] (t5) at (0.5,3){$x$};
\draw ([xshift=0.25cm]t0.north)--([xshift=-0.25cm]t1.south);
\draw ([xshift=0.25cm]t2.north)--([xshift=-0.25cm]t3.south);
\draw ([xshift=0.25cm]t4.north)--([xshift=-0.25cm]t5.south);
\draw ([xshift=-0.25cm]t1.north)--([xshift=0.25cm]t2.south);
\draw ([xshift=-0.25cm]t3.north)--([xshift=0.25cm]t4.south);
\draw[rounded corners,endlabel=$a$] ([xshift=-0.25cm]t0.north)|-++(-0.3,0.2);
\draw[rounded corners,endlabel=$c$] ([xshift=-0.25cm]t2.north)|-++(-0.3,0.2);
\draw[rounded corners,endlabel=$e$] ([xshift=-0.25cm]t4.north)|-++(-0.3,0.2);
\draw[rounded corners,endlabel=$b$] ([xshift=-0.25cm]t2.south)|-++(-0.3,-0.2);
\draw[rounded corners,endlabel=$d$] ([xshift=-0.25cm]t4.south)|-++(-0.3,-0.2);

\draw[rounded corners,endlabel=$b'$] ([xshift=0.25cm]t1.north)|-++(0.3,0.2);
\draw[rounded corners,endlabel=$d'$] ([xshift=0.25cm]t3.north)|-++(0.3,0.2);
\draw[rounded corners,endlabel=$f'$] ([xshift=0.25cm]t5.north)|-++(0.3,0.2);
\draw[rounded corners,endlabel=$a'$] ([xshift=0.25cm]t1.south)|-++(0.3,-0.2);
\draw[rounded corners,endlabel=$c'$] ([xshift=0.25cm]t3.south)|-++(0.3,-0.2);
\draw[rounded corners,endlabel=$e'$] ([xshift=0.25cm]t5.south)|-++(0.3,-0.2);

\draw[rounded corners,endlabel=$w$] ([xshift=0.25cm]t0.south)--++(0,-0.2);
\draw[rounded corners,endlabel=$v$] ([xshift=-0.25cm]t0.south)--++(0,-0.2);
\draw[rounded corners,endlabel=$w'$] ([xshift=-0.25cm]t5.north)--++(0,0.2);

\draw[rounded corners,startlabel=$v'$,endlabel=$f$] ([xshift=-0.75cm,yshift=0.2cm]t5.north)|-++(-0.3,-0.2);
\end{tikzpicture}
\end{equation}
Here, $x=e^{itH_2}$, where $H_2$ is the nearest-neighbor Hamiltonian term, and $P_n$ is the blocked tensor network patch after step \ref{itm:block} above.

Then we truncate using some isometries $I_1$ and $I_2$:
\begin{equation}
\label{eq:trotter_truncation}
\begin{gathered}
\begin{tikzpicture}
\node[roundbox,minimum height=0.6cm,minimum width=0.6cm] (t) at (0,0){$P_n^{\text{trunc}}$};
\draw (t.north)--++(0,0.4) node[above]{$v'$};
\draw (t.south)--++(0,-0.4)node[below]{$v$};
\draw (t.west)--++(-0.4,0)node[left]{$a$};
\draw (t.east)--++(0.4,0)node[right]{$a'$};
\end{tikzpicture}=
\begin{tikzpicture}
\node[roundbox,minimum height=0.6cm,minimum width=0.6cm] (t) at (0,0){$P_n$};
\node[roundbox] (i0) at (0.8,0) {$I_1$};
\node[roundbox] (i1) at (-0.8,0) {$I_1^\dagger$};
\node[roundbox] (i2) at (0,0.8) {$I_2$};
\node[roundbox] (i3) at (0,-0.8){$I_2^\dagger$};
\draw (t)--(i0) (t)--(i1) (t)--(i2) (t)--(i3);
\draw (i2.north)--++(0,0.3) node[above]{$v'$};
\draw (i3.south)--++(0,-0.3)node[below]{$v$};
\draw (i1.west)--++(-0.3,0)node[left]{$a$};
\draw (i0.east)--++(0.3,0)node[right]{$a'$};
\end{tikzpicture}\\
\begin{tikzpicture}
\node[roundbox,minimum height=0.6cm,minimum width=0.6cm] (t) at (0,0){$P_n$};
\draw (t.north)--++(0,0.4) node[above]{$v'w'$};
\draw (t.south)--++(0,-0.4)node[below]{$vw$};
\draw (t.west)--++(-0.4,0)node[rotate=90,above]{$abcdef$};
\draw (t.east)--++(0.4,0)node[rotate=90,below]{$a'b'c'd'e'f'$};
\end{tikzpicture}\approx
\begin{tikzpicture}
\node[roundbox,minimum height=0.6cm,minimum width=0.6cm] (t) at (0,0){$P_n^{\text{trunc}}$};
\node[roundbox] (i0) at (1.1,0) {$I_1^\dagger$};
\node[roundbox] (i1) at (-1.1,0) {$I_1$};
\node[roundbox] (i2) at (0,0.8) {$I_2^\dagger$};
\node[roundbox] (i3) at (0,-0.8){$I_2$};
\draw (t)--(i0) (t)--(i1) (t)--(i2) (t)--(i3);
\draw (i2.north)--++(0,0.3) node[above]{$v'w'$};
\draw (i3.south)--++(0,-0.3)node[below]{$vw$};
\draw (i1.west)--++(-0.3,0)node[above,rotate=90]{$abcdef$};
\draw (i0.east)--++(0.3,0)node[below,rotate=90]{$a'b'c'd'e'f'$};
\end{tikzpicture}
\end{gathered}
\end{equation}
Now pick a sufficient Trotterization level $n$ and basis $B$ for the truncation, yielding a tensor $P$. If we double $P$ we get a positive *-tensor (where we omitted the fact that $P$ is a complex tensor and we need to use realification):
\begin{equation}
\begin{tikzpicture}
\node[froundbox,minimum height=0.6cm,minimum width=0.6cm] (t) at (0,0){$P$};
\draw[quantum] (t.north)--++(0,0.4) node[above]{$v'\overline{v'}$};
\draw[quantum] (t.south)--++(0,-0.4)node[below]{$v\overline{v}$};
\draw[quantum] (t.west)--++(-0.4,0)node[left]{$a\overline{a}$};
\draw[quantum] (t.east)--++(0.4,0)node[right]{$a'\overline{a'}$};
\end{tikzpicture}=
\begin{tikzpicture}
\node[roundbox,minimum height=0.6cm,minimum width=0.6cm] (t) at (0,0){$P$};
\draw (t.north)--++(0,0.3) node[above]{$v'$};
\draw (t.south)--++(0,-0.3)node[below]{$v$};
\draw (t.west)--++(-0.3,0)node[left]{$a$};
\draw (t.east)--++(0.3,0)node[right]{$a'$};
\node[roundbox,minimum height=0.6cm,minimum width=0.6cm] (t1) at (2.3,0){$P$};
\draw (t1.north)--++(0,0.3) node[above]{$\overline{v'}$};
\draw (t1.south)--++(0,-0.3)node[below]{$\overline{v}$};
\draw (t1.west)--++(-0.3,0)node[left]{$\overline{a}$};
\draw (t1.east)--++(0.3,0)node[right]{$\overline{a'}$};
\end{tikzpicture}
\end{equation}
With this choice an approximation of the (imaginary) time evolution by a *-tensor network is given by:
\begin{equation}
\begin{tikzpicture}
\foreach \x in {0,...,5}{
\foreach \y in {0,...,3}{
\node[froundbox] (\x\y) at (\x*0.8,\y*0.8){$P$};
}}
\foreach \y in {0,...,3}{
\foreach \x [count=\xx from 1] in {0,...,4}{
\draw[quantum] (\x\y)--(\xx\y);
}
\draw[quantum,enddots] (0\y.west)--++(-0.3,0);
\draw[quantum,enddots] (5\y.east)--++(0.3,0);
}
\foreach \x in {0,...,5}{
\foreach \y [count=\yy from 1] in {0,...,2}{
\draw[quantum] (\x\y)--(\x\yy);
}
\draw[quantum,enddots] (\x0.south)--++(0,-0.3);
\draw[quantum,enddots] (\x3.north)--++(0,0.3);
}
\end{tikzpicture}
\end{equation}
Note that this is not a causal *-tensor network though.
\end{myobs}

\begin{myobs}
The real-time evolution of a many-body quantum system has a notion of causality, so it would be natural to model it by a causal *-tensor network. This can be in principle obtained by changing the above procedure in the following way: Instead of cutting the trotterized time evolution into cubes of size $\Delta t$ times spacial unit cell as in step \ref{itm:block} above, we take rhombic patches. E.g. in $1+1$ we can cut the trotterized real-time evolution like the following:
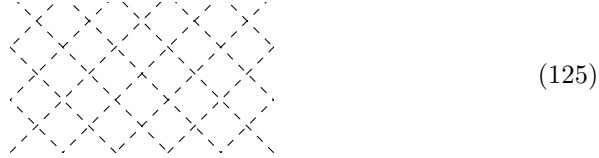
\begin{equation}
\begin{tikzpicture}
\clip (0,0) rectangle (3.5,2);
\foreach \d in {-7,...,9}{
\draw[dashed] (-\d*0.35,\d*0.35)--++(5,5);
\draw[dashed] (\d*0.35+2.5,\d*0.35-2.5)--++(-5,5);
}
\end{tikzpicture}
\end{equation}
Contracting the patches yields a tensor:
\begin{equation}
\begin{tikzpicture}
\draw[dashed] (0,0)--(0.5,0.5)--(0,1)--(-0.5,0.5)--cycle;
\draw[<->] (0.7,0)--node[midway,right]{$\Delta t$}++(0,1);
\draw[<->] (-0.5,-0.2)--node[midway,below]{$\Delta l$}++(1,0);
\end{tikzpicture}
\quad = \quad
\begin{tikzpicture}
\node[roundbox,rotate=45] (p) at (0,0){$P$};
\draw (p.north)edge[endlabel=$a$]++(135:0.3) (p.west)edge[endlabel=$b$]++(-135:0.3) (p.south)edge[endlabel=$c$]++(-45:0.3) (p.east)edge[endlabel=$d$]++(45:0.3);
\end{tikzpicture}
\end{equation}
In fact, it's not no important how precisely we cut the trotterized tensor network into patches. The important thing is that when we block indices, the indices of which each block consists should be either all from the lower all from the upper temportal half of the patch.

Again we consider filling this patch with smaller and smaller Trotterization steps $\Delta t/n$, and perform the same truncation as in Eq.~(\ref{eq:trotter_truncation}), just that now additionally try to achieve that $P$ is a unitary from the indices $b$ and $c$ to $a$ and $d$. We conjecture the following: If the time step $\Delta t$ is chosen sufficiently small (i.e. $\Delta l/\Delta t\gg c_R$, where $c_R$ is the Lieb-Robinson velocity of the quantum spin system), it will indeed be possible to choose $P$ to be unitary by on-site truncations. At least this will hold up to an approximation error exponentially small in $\Delta t$, due to the Lieb-Robinson bounds. In this case we get a normalized, positive *-tensor after doubling:
\begin{equation}
\begin{tikzpicture}
\node[froundbox,rotate=45] (p) at (0,0){$P$};
\draw[quantum] (p.north)edge[enddist=0.3cm,ndir=+,endlabel=$aa'$]++(135:0.4) (p.west)edge[enddist=0.3cm,ndir=-,endlabel=$bb'$]++(-135:0.4) (p.south)edge[enddist=0.3cm,ndir=-,endlabel=$cc'$]++(-45:0.4) (p.east)edge[enddist=0.3cm,ndir=+,endlabel=$dd'$]++(45:0.4);
\end{tikzpicture}
=
\begin{tikzpicture}
\node[roundbox,rotate=45] (p) at (0,0){$P$};
\draw (p.north)edge[endlabel=$a$]++(135:0.3) (p.west)edge[endlabel=$b$]++(-135:0.3) (p.south)edge[endlabel=$c$]++(-45:0.3) (p.east)edge[endlabel=$d$]++(45:0.3);
\node[roundbox,rotate=45] (p1) at (1.6,0){$P$};
\draw (p1.north)edge[endlabel=$a'$]++(135:0.3) (p1.west)edge[endlabel=$b'$]++(-135:0.3) (p1.south)edge[endlabel=$c'$]++(-45:0.3) (p1.east)edge[endlabel=$d'$]++(45:0.3);
\end{tikzpicture}
\end{equation}
The real-time evolution can then be approximated by the following causal *-tensor network:
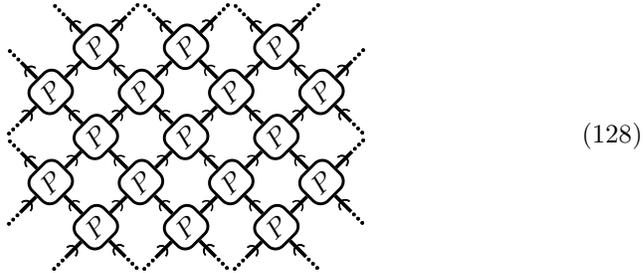
\begin{equation}
\begin{tikzpicture}
\foreach \x in {1,3,5}{
\foreach \y in {0,2,4}{
\node[froundbox,rotate=45] (\x\y) at (\x*0.6,\y*0.6){$P$};
}}
\foreach \x in {0,2,4,6}{
\foreach \y in {1,3}{
\node[froundbox,rotate=45] (\x\y) at (\x*0.6,\y*0.6){$P$};
}}
\foreach \x [evaluate=\x as \xx using int(\x+1),evaluate=\x as \xxx using int(\x-1)] in {1,3,5}{
\foreach \y [evaluate=\y as \yy using int(\y+1)] in {0,2}{
\draw[quantum,ndir=+] (\x\y.north)--(\xxx\yy.south);
\draw[quantum,ndir=+] (\x\y.east)--(\xx\yy.west);
}}
\foreach \x [evaluate=\x as \xx using int(\x+1),evaluate=\x as \xxx using int(\x-1)] in {1,3,5}{
\foreach \y [evaluate=\y as \yy using int(\y-1)] in {2,4}{
\draw[quantum,ndir=-] (\x\y.west)--(\xxx\yy.east);
\draw[quantum,ndir=-] (\x\y.south)--(\xx\yy.north);
}}
\draw[quantum] (01.north)edge[ndir=+,enddots]++(135:0.3) (01.west)edge[ndir=-,enddots]++(-135:0.3);
\draw[quantum] (03.north)edge[ndir=+,enddots]++(135:0.3) (03.west)edge[ndir=-,enddots]++(-135:0.3);
\draw[quantum] (61.east)edge[ndir=+,enddots]++(45:0.3) (61.south)edge[ndir=-,enddots]++(-45:0.3);
\draw[quantum] (63.east)edge[ndir=+,enddots]++(45:0.3) (63.south)edge[ndir=-,enddots]++(-45:0.3);
\draw[quantum] (10.west)edge[ndir=-,enddots]++(-135:0.3) (10.south)edge[ndir=-,enddots]++(-45:0.3);
\draw[quantum] (30.west)edge[ndir=-,enddots]++(-135:0.3) (30.south)edge[ndir=-,enddots]++(-45:0.3);
\draw[quantum] (50.west)edge[ndir=-,enddots]++(-135:0.3) (50.south)edge[ndir=-,enddots]++(-45:0.3);
\draw[quantum] (14.north)edge[ndir=+,enddots]++(135:0.3) (14.east)edge[ndir=+,enddots]++(45:0.3);
\draw[quantum] (34.north)edge[ndir=+,enddots]++(135:0.3) (34.east)edge[ndir=+,enddots]++(45:0.3);
\draw[quantum] (54.north)edge[ndir=+,enddots]++(135:0.3) (54.east)edge[ndir=+,enddots]++(45:0.3);
\end{tikzpicture}
\end{equation}
\end{myobs}

\begin{myobs}
Consider a process where we 1) prepare some product density state, 2) let the system evolve under a local Hamiltonian, and 3) perform some local measurements at different times and places. Using the Trotterization procedure from the previous observation, such a setup can be translated into a *-tensor network. E.g., consider the following setup where we measure some POVM $M$ involving two neighboring sites after $3$ units of time evolution in a $1+1$-dimensional system:
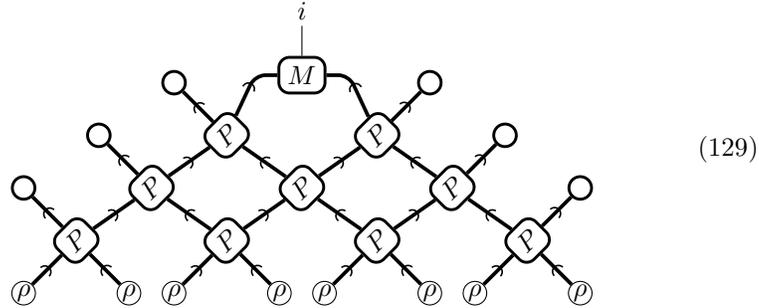
\begin{equation}
\begin{tikzpicture}
\foreach \x in {0,1,2,3}{
\node[froundbox,rotate=45] (a\x) at (\x*2,0){$P$};
};
\foreach \x in {0,1,2,3}{
\node[circle,draw,inner sep=0.01cm] (x\x) at (\x*2-0.7,-0.7){$\rho$};
\node[circle,draw,inner sep=0.01cm] (y\x) at (\x*2+0.7,-0.7){$\rho$};
\draw[quantum] (x\x)edge[ndir=+](a\x);
\draw[quantum] (y\x)edge[ndir=+](a\x);
};
\foreach \x in {0,1,2}{
\node[froundbox,rotate=45] (b\x) at (\x*2+1,0.7){$P$};
};
\foreach \x in {0,1}{
\node[froundbox,rotate=45] (c\x) at (\x*2+2,1.4){$P$};
};
\draw[quantum] (a0)edge[ndir=+](b0) (a1)edge[ndir=+](b0) (a1)edge[ndir=+](b1) (a2)edge[ndir=+](b1) (a2)edge[ndir=+](b2) (a3)edge[ndir=+](b2) (b0)edge[ndir=+](c0) (b1)edge[ndir=+](c0) (b1)edge[ndir=+](c1) (b2)edge[ndir=+](c1);
\pics{circ}{t0/-0.7,0.7|t1/0.3,1.4|t2/1.3,2.1};
\pics{circ}{s0/6.7,0.7|s1/5.7,1.4|s2/4.7,2.1};
\draw[quantum] (a0)edge[ndir=+](t0-c) (b0)edge[ndir=+](t1-c) (c0)edge[ndir=+](t2-c) (a3)edge[ndir=+](s0-c) (b2)edge[ndir=+](s1-c) (c1)edge[ndir=+](s2-c);
\node[froundbox] (m) at (3,2.2){$M$};
\draw[quantum,ndir=-,rounded corners] (m.west)--++(-0.3,0)--(c0);
\draw[quantum,ndir=-,rounded corners] (m.east)--++(0.3,0)--(c1);
\draw (m.north)edge[endlabel=$i$]++(0,0.4);
\end{tikzpicture}
\end{equation}
Note that due to the causal structure, we can restrict to a lightcone spreading in negative time direction from $M$.
\end{myobs}

\subsubsection*{Non-causal setups}
\begin{myobs}
Consider a (thermal) statistical quantum system with Hamiltonian $H$ over some Hilbert space. At inverse temperature $\beta$, the density matrix equals the following Boltzmann weight (after normalization):
\begin{equation}
W=e^{-\beta H}
\end{equation}
A measurement corresponds to a POVM $M$. The probability distribution corresponding to the measurement outcomes is the evaluation of the following *-tensor network (after normalization):
\begin{equation}
\begin{tikzpicture}
\node[froundbox] (r) at (0,0){$W$};
\node[froundbox] (p) at (1,0){$M$};
\draw[star,quantum] (r)--(p);
\draw[star,classical] (p.east)edge[endlabel=$i$]++(0.4,0);
\end{tikzpicture}
\end{equation}
\end{myobs}

\begin{myobs}
Consider a thermal quantum system at inverse temperature $\beta$ with degrees of freedom distributed over some lattice and local Hamiltonian terms. Also consider a set of local measurements, that is, a set of POVMs involving only degrees of freedom in a small region of the lattice. A *-tensor network for such a setup can be constructed in the following way:
\begin{enumerate}
\item Choose some spacial unit cell $C$, and take the tensor $P$ corresponding to the imaginary time evolution in the space-time volume $C\times \beta/2$ via Trotterization, as described in Obs.~(\ref{obs:hamiltonian_discretisation}).
\item Take two copies of $P$ to get a positive *-tensor.
\item Fill the space lattice by a tensor network of $P$ tensors, yielding a tensor network representing the doubled imaginary time evolution on the lattice for $\Delta t=i\beta/2$. It has $2$ open indices at every unit cell of the lattice, corresponding to the lower and upper time boundary.
\item Trace over all the indices at the lower time boundary.
\item For each measurement of the setup, contract the indices at the corresponding unit cells with the according quantum index of the POVM, transforming it into one open classical index.
\item For all unit cells without measurement, trace over the quantum index at the upper time boundary.
\end{enumerate}

E.g., consider a model on a $1$-dimensional square lattice, and one kind of measurement $M$ involving a single unit cell. The following shows a patch of the representing *-tensor network with $3$ measurements:
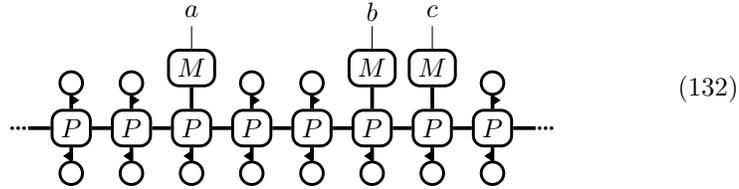
\begin{equation}
\begin{tikzpicture}
\foreach \x in {0,...,7}{
\node[froundbox] (\x) at (\x*0.8,0){$P$};
\pic (i\x) at (\x*0.8,-0.6){circ};
\draw[quantum] (i\x-c)edge[flags=+](\x.south);
}
\foreach \x in {0,1,3,4,7}{
\pic (j\x) at (\x*0.8,0.6){circ};
\draw[quantum] (j\x-c)edge[flags=+](\x.north);
}
\foreach \x/\l in {2/a,5/b,6/c}{
\node[froundbox] (m\x) at (\x*0.8,0.8){$M$};
\draw[quantum] (m\x.south)--(\x.north);
\draw[star,classical] (m\x.north)edge[endlabel=$\l$]++(0,0.3);
}
\foreach \x [count=\xx from 1] in {0,...,6}{
\draw[quantum] (\x)--(\xx);
}
\draw[quantum,enddots] (0.west)--++(-0.3,0);
\draw[quantum,enddots] (7.east)--++(0.3,0);
\end{tikzpicture}
\end{equation}

As another example, take a model on a $2$-dimensional square lattice with measurements involving $2$ neighboring cells. The tensor $P$ corresponding to a spacetime cube has $6$ indices, one for each side of the cube:
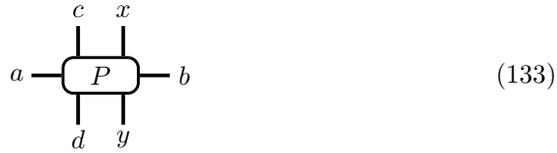
\begin{equation}
\begin{tikzpicture}
\node[froundbox,minimum width=1cm] (p) {$P$};
\draw[froundbox,quantum] (p.west)edge[endlabel=$a$]++(-0.4,0) (p.east)edge[endlabel=$b$]++(0.4,0) ([xshift=-0.3cm]p.north)edge[endlabel=$c$]++(0,0.4) ([xshift=0.3cm]p.north)edge[endlabel=$x$]++(0,0.4) ([xshift=-0.3cm]p.south)edge[endlabel=$d$]++(0,-0.4) ([xshift=0.3cm]p.south)edge[endlabel=$y$]++(0,-0.4);
\end{tikzpicture}
\end{equation}
Where $y$ and $x$ are the indices corresponding to the lower and upper time boundary, respectively. For cells where no measurement is taken we use the tensor:
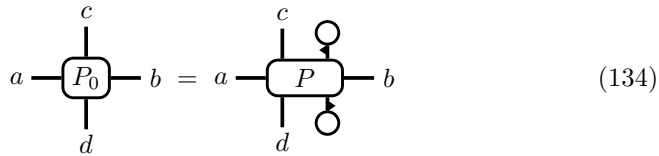
\begin{equation}
\begin{tikzpicture}
\node[froundbox] (p) {$P_0$};
\draw[quantum] (p.west)edge[endlabel=$a$]++(-0.4,0) (p.east)edge[endlabel=$b$]++(0.4,0) (p.north)edge[endlabel=$c$]++(0,0.4) (p.south)edge[endlabel=$d$]++(0,-0.4);
\end{tikzpicture}=
\begin{tikzpicture}
\node[froundbox,minimum width=1cm] (p) at (0,0) {$P$};
\pics{circ}{0/0.3,-0.6|1/0.3,0.6};
\draw[quantum] (p.west)edge[endlabel=$a$]++(-0.4,0) (p.east)edge[endlabel=$b$]++(0.4,0) ([xshift=-0.3cm]p.north)edge[endlabel=$c$]++(0,0.4) ([xshift=0.3cm]p.north)edge[flage=+](1-c) ([xshift=-0.3cm]p.south)edge[endlabel=$d$]++(0,-0.4) ([xshift=0.3cm]p.south)edge[flage=+](0-c);
\end{tikzpicture}
\end{equation}
For pairs of neighboring cells with a measurement take the following tensor:
\begin{equation}
\begin{tikzpicture}
\node[froundbox,minimum width=1cm] (p) {$P_M$};
\draw[quantum] (p.west)edge[endlabel=$a$]++(-0.3,0) (p.east)edge[endlabel=$b$]++(0.3,0) ([xshift=-0.3cm]p.north)edge[endlabel=$c$]++(0,0.3) ([xshift=0.3cm]p.north)edge[endlabel=$d$]++(0,0.3) ([xshift=-0.3cm]p.south)edge[endlabel=$e$]++(0,-0.3) ([xshift=0.3cm]p.south)edge[endlabel=$f$]++(0,-0.3);
\draw[star,classical] (p.north)edge[endlabel=$i$]++(0,0.4);
\end{tikzpicture}=
\begin{tikzpicture}
\node[froundbox,minimum width=1cm] (p0) at (0,0) {$P$};
\node[froundbox,minimum width=1cm] (p1) at (1.5,0) {$P$};
\node[froundbox,minimum width=0.7cm] (m) at (1,1.2) {$M$};
\pics{circ}{0/0.3,-0.6|1/1.8,-0.6};
\draw[quantum,rounded corners] (p0.west)edge[endlabel=$a$]++(-0.3,0) (p0.east)--(p1.west) (p1.east)edge[endlabel=$b$]++(0.3,0) ([xshift=-0.3cm]p0.north)edge[endlabel=$c$]++(0,0.2) ([xshift=0.3cm]p0.north)|-(m.west)  ([xshift=0.3cm]p1.north)|-(m.east) ([xshift=-0.3cm]p0.south)edge[endlabel=$e$]++(0,-0.2) ([xshift=0.3cm]p0.south)edge[flage=+](0-c) ([xshift=0.3cm]p1.south)edge[flage=+](1-c) ([xshift=-0.3cm]p1.south)edge[endlabel=$f$]++(0,-0.2) ([xshift=-0.3cm]p1.north)edge[endlabel=$d$]++(0,0.2);
\draw[star,classical] (m.north)edge[endlabel=$i$]++(0,0.3);
\end{tikzpicture}
\end{equation}
The following shows a patch of the representing tensor network with $2$ measurements:
\begin{equation}
\begin{tikzpicture}
\foreach \y/\x in {0/0,0/2,0/3,1/0,1/2,1/3,2/0,2/1,3/0,3/1,3/2,3/3}{
\node[froundbox] (\x\y) at (\x,\y) {$P_0$};
}
\node[froundbox,minimum height=1.4cm] (m0) at (1,0.5){$P_M$};
\node[froundbox,minimum width=1.4cm] (m1) at (2.5,2){$P_M$};
\draw[quantum] (00)--(01) (01)--(02) (02)--(03) (m0)--(12) (12)--(13) (20)--(21) (30)--(31) (02)--(12) (03)--(13) (13)--(23) (23)--(33) (20)--(30) (21)--(31) (12)--(m1) (00)--([yshift=-0.5cm]m0.west) (01)--([yshift=0.5cm]m0.west) (20)--([yshift=-0.5cm]m0.east) (21)--([yshift=0.5cm]m0.east) (21)--([xshift=-0.5cm]m1.south) (31)--([xshift=0.5cm]m1.south) (23)--([xshift=-0.5cm]m1.north) (33)--([xshift=0.5cm]m1.north);
\draw[quantum] (00.south)edge[enddots]++(-90:0.3) (m0.south)edge[enddots]++(-90:0.3) (20.south)edge[enddots]++(-90:0.3) (30.south)edge[enddots]++(-90:0.3) (00.west)edge[enddots]++(180:0.3) (01.west)edge[enddots]++(180:0.3) (02.west)edge[enddots]++(180:0.3) (03.west)edge[enddots]++(180:0.3) (03.north)edge[enddots]++(90:0.3) (13.north)edge[enddots]++(90:0.3) (23.north)edge[enddots]++(90:0.3) (33.north)edge[enddots]++(90:0.3) (30.east)edge[enddots]++(0:0.3) (31.east)edge[enddots]++(0:0.3) (m1.east)edge[enddots]++(0:0.3) (33.east)edge[enddots]++(0:0.3);
\draw[star,classical] (m0.east)edge[endlabel=$a$]++(0:0.2) (m1.north)edge[endlabel=$b$]++(90:0.2);
\end{tikzpicture}
\end{equation}
Evaluating this tensor network yields the (non-normalized) probability distribution over the measurement outcomes $a$ and $b$.
\end{myobs}

\section{*-tensor theories for other *-algebras}
\label{sec:other_star_algebras}
In the last two sections we have seen that both classical statistical and quantum physics can be formulated with *-tensor models, for the delta *-algebras and matrix times complex number *-algebras. If we allow for general *-algebras we can get different theories of physics. However, as we will see, quantum mechanics is enough to emulate all these theories.

\begin{mydef}
A \tdef{sub *-algebra}{sub_star_algebra} of a *-algebra $A$ is another *-algebra $B$ together with a tensor $R$ with one $A$-index and one $B$-index:
\begin{equation}
\begin{tikzpicture}
\piclab{r}{bwbox}{(0,0)}{(-90:0.4)}{$R$};
\draw (r-l)edge[endlabel=$b$]++(-0.3,0) (r-r)--++(0.3,0)node[right]{$a$};
\end{tikzpicture}
\end{equation}
We will omit the label $R$ if there is no ambiguity.

The following axioms have to hold:
\begin{itemize}
\item Consider contracting $n-1$ indices of a $n$-index $A$ *-algebra tensor with the $A$-indices of each of $n-1$ copies of $R$. This equals the contraction of the single remaining index of the according $n$-index $B$ *-algebra tensor with the $B$-index of a single copy of $R$. E.g.:
\begin{equation}
\label{eq:sub_star_algebra}
\begin{gathered}
a)\quad
\begin{tikzpicture}
\piclab{s}{circ}{(0,0)}{(-60:0.3)}{$A$};
\pics{bwbox}{r0/-0.9,0.5|r1/-0.9,-0.5};
\draw[rounded corners,flage=+] (r0-r)--++(0.3,0)--(s-c);
\draw[rounded corners,flage=+] (r1-r)--++(0.3,0)--(s-c);
\draw (r0-l)edge[endlabel=$a$]++(-0.4,0) (r1-l)edge[endlabel=$b$]++(-0.4,0) (s-c)edge[flags=+,endlabel=$c$]++(0.5,0);
\end{tikzpicture}=
\begin{tikzpicture}
\piclab{s}{circ}{(0,0)}{(-60:0.35)}{$B$};
\pics{bwbox}{r0/0.9,0};
\draw (s-c)edge[flags=+](r0-l) (r0-r)edge[endlabel=$c$]++(0.4,0);
\draw (s-c)edge[flags=-,endlabel=$a$]++(-0.5,0.5) (s-c)edge[flags=-,endlabel=$b$]++(-0.5,-0.5);
\end{tikzpicture}\\
b)\quad
\begin{tikzpicture}
\piclab{s}{circ}{(0,0)}{(180:0.3)}{$A$};
\draw (s-c)edge[flags=+,endlabel=$c$]++(0.5,0);
\end{tikzpicture}=
\begin{tikzpicture}
\piclab{s}{circ}{(0,0)}{(180:0.35)}{$B$};
\pics{bwbox}{r0/0.9,0};
\draw (s-c)edge[flags=+](r0-l) (r0-r)edge[endlabel=$c$]++(0.4,0);
\end{tikzpicture}
\end{gathered}
\end{equation}
\item Take two copies of $R$ and contract their $A$-indices. This yields the identity tensor:
\begin{equation}
\label{eq:sub_star_normalization}
\begin{tikzpicture}
\pics{bwbox}{r0/0,0.4|r1/0,-0.4};
\draw[rounded corners] (r0-r)--++(0.4,0)|-(r1-r);
\draw (r0-l)edge[endlabel=$a$]++(-0.3,0) (r1-l)edge[endlabel=$b$]++(-0.3,0);
\end{tikzpicture}=
\begin{tikzpicture}
\draw[rounded corners,startlabel=$a$,endlabel=$b$] (0,0.4)--++(0.4,0)|-(0,-0.4);
\end{tikzpicture}
\end{equation}
\end{itemize}
In other words, $R$ is an isometry (up to a prefactor) mapping one *-algebra to another. It is also fine if Eq.(\ref{eq:sub_star_algebra}) holds only up to a prefactor, as we can normalize either $A$ or $B$ such that it holds exactly. It seems to be the case that we've picked a compatible normalization if:
\begin{equation}
\begin{tikzpicture}
\piclab{s}{circ}{(0,0)}{(180:0.35)}{$A$};
\end{tikzpicture}=
\begin{tikzpicture}
\piclab{s}{circ}{(0,0)}{(180:0.35)}{$B$};
\end{tikzpicture}
\end{equation}
\end{mydef}
\begin{myobs}
\label{obs:matrix_algebra_homomorphism}
Every *-algebra $X$ with basis $B$ is contained as a sub *-algebra in the matrix *-algebra for the set $B$ (i.e. with basis $B\times B$). The corresponding tensor $R$ is given by:
\begin{equation}
\begin{tikzpicture}
\pics{bwbox}{r/0,0};
\draw (r-l)edge[endlabel=$x$]++(-0.3,0) (r-r)--++(0.3,0)node[right]{$ab$};
\end{tikzpicture}=
\begin{tikzpicture}
\piclab{s}{circ}{(0,0)}{(-120:0.4)}{$X$};
\draw (s-c)edge[flags=-,endlabel=$x$]++(-0.5,0) (s-c)edge[flags=+,endlabel=$a$]++(0.4,0.4) (s-c)edge[flags=-,endlabel=$b$]++(0.4,-0.4);
\end{tikzpicture}
\end{equation}

The axiom in Eq.~(\ref{eq:sub_star_algebra}) holds due to the fusion axiom:
\begin{equation}
\begin{gathered}
\begin{tikzpicture}
\piclab{s}{circ}{(0,0)}{(-60:0.4)}{$M$};
\pics{bwbox}{r0/-0.9,0.5|r1/-0.9,-0.5};
\draw[rounded corners,flage=+] (r0-r)--++(0.3,0)--(s-c);
\draw[rounded corners,flage=+] (r1-r)--++(0.3,0)--(s-c);
\draw (r0-l)edge[endlabel=$y$]++(-0.4,0) (r1-l)edge[endlabel=$x$]++(-0.4,0);
\draw[flags=+] (s-c)--(0.5,0) node[right]{$ab$};
\end{tikzpicture}
=
\begin{tikzpicture}
\piclab{0}{circ}{(0,0)}{(-120:0.4)}{$X$};
\piclab{1}{circ}{(0,1)}{(-120:0.4)}{$X$};
\draw[rounded corners,flags=-,flage=-] (1-c)--++(0.5,-0.3)--++(0,-0.4)--(0-c);
\draw (0-c)edge[endlabel=$y$,flags=-]++(-0.5,0) (1-c)edge[endlabel=$x$,flags=-]++(-0.5,0);
\draw[rounded corners,flags=-,endlabel=$b$] (0-c)--++(0.3,-0.3)--++(0.3,0)--++(0.3,0.5)--++(0.4,0);
\draw[rounded corners,flags=+,endlabel=$a$] (1-c)--++(0.3,0.3)--++(0.3,0)--++(0.3,-0.5)--++(0.4,0);
\end{tikzpicture}\\
=
\begin{tikzpicture}
\piclab{0}{circ}{(0,0)}{(-60:0.4)}{$X$};
\piclab{1}{circ}{(1,0)}{(-120:0.4)}{$X$};
\draw (0-c)edge[flags=+,flage=+](1-c);
\draw (0-c)edge[flags=-,endlabel=$x$]++(-0.5,0.5) (0-c)edge[flags=-,endlabel=$y$]++(-0.5,-0.5);
\draw[rounded corners,flags=+,endlabel=$a$] (1-c)--++(0.3,0.3)--++(0.4,0);
\draw[rounded corners,flags=-,endlabel=$b$] (1-c)--++(0.3,-0.3)--++(0.4,0);
\end{tikzpicture}
=
\begin{tikzpicture}
\piclab{s}{circ}{(0,0)}{(-60:0.4)}{$X$};
\pics{bwbox}{r0/0.9,0};
\draw (s-c)edge[flags=+](r0-l) (r0-r)--++(0.4,0) node[right]{$ab$};
\draw (s-c)edge[flags=-,endlabel=$x$]++(-0.5,0.5) (s-c)edge[flags=-,endlabel=$y$]++(-0.5,-0.5);
\end{tikzpicture}
\end{gathered}
\end{equation}

Eq.~(\ref{eq:sub_star_normalization}) is fulfilled if we pick a normalization such that the loop normalization from Eq.(\ref{eq:loop_normalization}) is fulfilled:
\begin{equation}
\begin{tikzpicture}
\pics{bwbox}{r/0,0};
\pic[rotate=180](r1)at (0.8,0){bwbox};
\draw (r-l)edge[endlabel=$x$]++(-0.3,0) (r1-l)edge[endlabel=$y$]++(0.3,0) (r-r)--(r1-r);
\end{tikzpicture}=
\begin{tikzpicture}
\piclab{0}{circ}{(0.1,0)}{(-120:0.4)}{$X$};
\piclab{1}{circ}{(1.4,0)}{(-60:0.4)}{$X$};
\draw[flags=+,flage=+,rounded corners] (0-c)--(0.5,0.5)--(1,0.5)--(1-c);
\draw[flags=-,flage=-,rounded corners] (0-c)--(0.5,-0.5)--(1,-0.5)--(1-c);
\draw (1-c)edge[flags=+,endlabel=$y$]++(0.5,0) (0-c)edge[flags=-,endlabel=$x$]++(-0.5,0);
\end{tikzpicture}=
\begin{tikzpicture}
\draw (0,0)edge[endlabel=$y$,startlabel=$x$](1,0);
\end{tikzpicture}
\end{equation}
\end{myobs}

\begin{myobs}
\label{obs:star_model_matrix}
Consider a *-tensor network with a contracted index pair equipped with an arbitrary *-algebra. By inserting a pair of the $R$-isomorphisms from Obs.~(\ref{obs:matrix_algebra_homomorphism}) we can turn the *-algebra of the contracted index pair into a matrix *-algebra, without changing the evaluation of the tensor network. Thus, every *-tensor model can be emulated by a *-tensor model using only matrix *-algebras (for contracted indices) and delta *-algebras (for open indices).
\end{myobs}

\begin{myobs}
A quantum *-algebra is a tensor product of the complex and a matrix *-algebra. Due to Obs.(\ref{obs:star_model_matrix}), every *-algebra is sub *-algebra of a matrix *-algebra, which is in turn a sub *-algebra of a quantum *-algebra. So using the same arguments as in Obs.~(\ref{obs:star_model_matrix}) we see that every *-tensor model is equivalent to a quantum model. However, directly using Obs.~(\ref{obs:star_model_matrix}), we see that real quantum mechanics (i.e. quantum mechanics with only real amplitudes) is already sufficient to emulate every *-tensor model including every quantum model.
\end{myobs}

\begin{mycom}
The fact that every quantum model can be emulated with a model using only real amplitudes is known in the context of circuit models and quantum computation. E.g. in \cite{Rudolph2002} they show how a real gate can emulate a complex universal quantum gate by introducing auxiliary qubits. In our formalism these auxiliary qubits come from the realification of the complex tensors.
\end{mycom}

\begin{myrem}
One could ask, if matrix *-algebras are sufficient, why bother with general *-algebras? As we have seen, in principle there is no reason for this. The only disadvantage from embedding general *-tensor models into matrix *-tensor models is that we get larger basis sets. Conversely one might be able to write models more compactly by using general *-algebras instead of matrix *-algebra. For example, quantum models that are genuinely complex are more compact than the corresponding real models.

So even if *-tensor models are equivalent to quantum mechanics they could be of potential use. In particular there could be known models that can be more compactly written down using the quaternion *-algebra.
\end{myrem}
\section*{Acknowledgments}
We thank R. Sweke and especially A. Nietner for lots of discussions and proof reading the document. Thanks to the Obst Office for providing a fruitful environment. We thank the DFG (CRC 183, B01) for support.

\bibliography{cstar_qmech_refs}{}

\providecommand{\href}[2]{#2}\begingroup\raggedright\begin{thebibliography}{1}

\bibitem{Abramsky2004}
S.~Abramsky and B.~Coecke, ``A categorical semantics of quantum protocols,''
  \href{http://dx.doi.org/10.1109/LICS.2004.1319636}{{\em Proceedings of the
  19th IEEE conference on Logic in Computer Science (LiCS'04)} (2004)
  415–425}, \href{http://arxiv.org/abs/quant-ph/0402130}{{\ttfamily
  arXiv:quant-ph/0402130}}.

\bibitem{Selinger2007}
P.~Selinger, ``Dagger compact closed categories and completely positive maps,''
  \href{http://dx.doi.org/10.1016/j.entcs.2006.12.018}{{\em Electronic Notes in
  Theoretical computer science} {\bfseries 170} (2007) 139–163}.

\bibitem{Coecke2013}
B.~Coecke, C.~Heunen, and A.~Kissinger, ``Categories of quantum and classical
  channels,'' \href{http://dx.doi.org/10.1007/s11128-014-0837-4}{{\em Quantum
  Inf Process} {\bfseries 15} (2016) 5179},
  \href{http://arxiv.org/abs/1305.3821}{{\ttfamily arXiv:1305.3821}}.

\bibitem{tensor_type}
A.~Bauer and A.~Nietner, ``Tensor types and their usage in physics.'' In
  preparation.

\bibitem{Li2003}
B.~Li, \href{http://dx.doi.org/10.1142/5284}{{\em Real operator algebras}}.
\newblock World Scientific Publishing Company, 2003.

\bibitem{tensor_lattice}
A.~Bauer, J.~Eisert, and C.~Wille, ``Towards a mathematical formalism for
  classifying phases of matter,''  (2019) ,
  \href{http://arxiv.org/abs/1903.05413}{{\ttfamily arXiv:1903.05413}}.

\bibitem{Rudolph2002}
T.~Rudolph and L.~Grover, ``A 2 rebit gate universal for quantum computing,''
  \href{http://arxiv.org/abs/quant-ph/0210187}{{\ttfamily
  arXiv:quant-ph/0210187}}.

\end{thebibliography}\endgroup
\bibliographystyle{utphys}
\end{document}